\documentclass[aps,prb,amsmath,amssymb,floatfix,twocolumn,amsmath,superscriptaddress,
twocolumn,nofootinbib,tighten,letterpaper]{revtex4-2}
\usepackage{lmodern}
\usepackage[T1]{fontenc}
\usepackage[latin9]{inputenc}
\usepackage{float}
\usepackage[normalem]{ulem}
\usepackage{amsmath, dsfont}
\usepackage{amssymb}
\usepackage{mathtools}
\usepackage{mathdots}

\usepackage{graphicx}
\usepackage{wasysym}
\usepackage{color}
\usepackage[table]{xcolor}
\usepackage{bm}
\usepackage{hyperref}
\hypersetup{pdfpagemode=UseNone}
\usepackage[many]{tcolorbox}
\newsavebox\MBox

\allowdisplaybreaks

\newcommand{\bsub}{\begin{subequations}}
\newcommand{\esub}{\end{subequations}}

\newcommand{\RR}[1]{{\bf r}_{#1}}

\newcommand\calphare{\cellcolor{green!10}}
\newcommand\calphaim{\cellcolor{green!30}}
\newcommand\cbetare{\cellcolor{red!10}}
\newcommand\cbetaim{\cellcolor{red!30}}
\newcommand\cgamma{\cellcolor{blue!20}}
\newcommand\caplus{\cellcolor{yellow!40}}
\newcommand\caminus{\cellcolor{yellow!20}}

\newcommand{\mfg}{\mathfrak g}

\newcommand{\mfn}{\mathfrak n}
\newcommand{\mfp}{\mathfrak p}
\newcommand{\mfk}{\mathfrak k}
\newcommand{\mfa}{\mathfrak a}

 \graphicspath{{Images/}}

\begin{document}

\title{Generalized multifractality at metal-insulator transitions and in metallic phases of 2D disordered systems}

\author{Jonas F.~Karcher}
\affiliation{{Institute for Quantum Materials and Technologies, Karlsruhe Institute of Technology, 76021 Karlsruhe, Germany}}
\affiliation{{Institut f\"ur Theorie der Kondensierten Materie, Karlsruhe Institute of Technology, 76128 Karlsruhe, Germany}}

\author{Ilya A.~Gruzberg}
\affiliation{Ohio State University, Department of Physics, 191 West Woodruff Ave, Columbus OH, 43210, USA}

\author{Alexander D.~Mirlin}
\affiliation{{Institute for Quantum Materials and Technologies, Karlsruhe Institute of Technology, 76021 Karlsruhe, Germany}}
\affiliation{{Institut f\"ur Theorie der Kondensierten Materie, Karlsruhe Institute of Technology, 76128 Karlsruhe, Germany}}

\date{May 27, 2022}

\begin{abstract}
We study generalized multifractality characterizing fluctuations and correlations of eigenstates in disordered systems of symmetry classes AII, D, and DIII.  Both metallic phases and Anderson-localization transitions are considered. By using the non-linear sigma-model approach, we construct pure-scaling eigenfunction observables. The construction is verified by numerical simulations of appropriate microscopic models, which also yield numerical values of the corresponding exponents. In the metallic phases, the numerically obtained exponents satisfy Weyl symmetry relations as well as generalized parabolicity (proportionality to eigenvalues of the quadratic Casimir operator). At the same time, the generalized parabolicity is strongly violated at critical points of metal-insulator transitions, signalling violation of local conformal invariance. Moreover, in classes D and DIII, even the Weyl symmetry breaks down at critical points of metal-insulator transitions. This last feature is related with a peculiarity of the sigma-model manifolds in these symmetry classes: they consist of two disjoint components. Domain walls associated with these additional degrees of freedom are crucial for ensuring Anderson localization and, at the same time, lead to the violation of the Weyl symmetry.
\end{abstract}

\maketitle

\section{Introduction}
\label{sec:intro}

Anderson localization in disordered systems keeps attracting much attention of researchers, both theoreticians and experimentalists~\cite{50_years_of_localization}.  Localized and delocalized phases are separated by critical points of Anderson transitions that have very intriguing properties~\cite{evers08}.  In a broader sense, Anderson transitions include also transitions between topologically distinct localized phases. The interest to Anderson localization has been additionally enhanced by the development of full symmetry classification of disordered fermionic systems~\cite{altland1997nonstandard, zirnbauer1996riemannian, heinzner2005symmetry} and by the advent of topological insulators and superconductors~\cite{chiu2016classification}.

Critical states at Anderson metal-insulator transitions exhibit a remarkable property---multifractality. When understood in a narrow sense, the Anderson-transition multifractality characterizes the scaling of moments of eigenfunction amplitudes (or, equivalently, of moments of the local density of states, LDOS)~\cite{evers08, rodriguez2011multifractal}. It has been recognized, however, that there is a much broader class of observables (described by gradientless composite operators in the sigma-model language)  characterizing the physics of critical eigenstates~\cite{gruzberg2013classification}. The scaling of such observables and associated correlation functions has been termed ``generalized multifractality''~\cite{karcher2021generalized}, and is characterized by an infinite set of scaling exponents $x_\lambda$, sometimes called ``multifractal spectra'',
where the index $\lambda$ labels different observables. In Refs.~\cite{mirlin2006exact, gruzberg2011symmetries, gruzberg2013classification} it was shown that the multifractal spectra $x_\lambda$ in five out of ten symmetry classes satisfy a certain exact Weyl symmetry that relates scaling exponents of seemingly unrelated multifractal observables.

In experimental studies of quantum transport, two-dimensional (2D) structures play a particularly prominent role. There is a vast variety of experimental realizations of 2D disordered electronic systems, which include interfaces in semiconductor heterostructures, surfaces of topological insulators and semiconductors, as well as 2D materials (graphene, transition-metal dichalcogenide monolayers, etc.)  Naively, one might expect, in view of an analogy with conventional second-order phase transitions with a continuous symmetry, that $d=2$ is the lowest critical dimension. This would imply that 2D systems are always in the localized phase. This is indeed true for the ``most conventional'' symmetry class AI  (also known as the Wigner-Dyson orthogonal class). Remarkably, in all the remaining nine symmetry classes, Anderson-localization critical points exist~\cite{evers08}. In the field-theory language, this is related to peculiar properties of the corresponding sigma-model manifolds: (i) supersymmetry (or $n \to 0$ replica limit in the replica formulation); (ii) combination of non-compact and compact degrees of freedom, with non-trivial topologies associated with the compact coordinates.

The most well-known example of a 2D Anderson-localization critical point is the celebrated integer quantum Hall (QH) plateau transition that belongs to the symmetry class A. It has counterparts in 2D disordered superconductors: the spin quantum Hall (SQH) transition (class C)~\cite{kagalovsky1999quantum, senthil1999spin,gruzberg1999exact, beamond2002quantum,mirlin2003wavefunction, evers2003multifractality, subramaniam2008boundary} as well as the thermal quantum Hall transition (class D)~\cite{read2000paired, senthil2000quasiparticle, bocquet2000disordered, chalker2001thermal, mildenberger2007density, medvedyeva2010effective, wang_multicriticality_2021}.
For class A, the construction of eigenstate observables was developed in Ref.~\cite{gruzberg2013classification}, and the generalized multifractality at the QH transition was studied numerically in Ref.~\cite{karcher2021generalized}.  For class C, a detailed analytical and numerical study of the generalized multifractality at the SQH transition was carried out in  Refs.~\cite{karcher2021generalized, karcher2022generalized}. A remarkable property of the SQH critical point is that exact analytical results for some of the critical exponents can be obtained by means of the mapping to classical percolation~\cite{gruzberg1999exact, beamond2002quantum}. In earlier works, several exponents $x_\lambda$ for the conventional multifractality were determined in this way~\cite{mirlin2003wavefunction, evers2003multifractality, subramaniam2008boundary}. More recently, we were able to extend the mapping to a broader subset of generalized-multifractality exponents $x_\lambda$~\cite{karcher2022generalized}. These analytical results are in excellent agreement with results of numerical simulations~\cite{karcher2022generalized}, which yield also exponents that cannot be found analytically. One of important implications of the analytical and numerical results of Refs.~\cite{karcher2021generalized, karcher2022generalized} is that the generalized parabolicity of the spectrum of exponents $x_\lambda$ (proportionality to eigenvalues of the quadratic Casimir operator) is strongly violated at the SQH transition. At the same time, it was shown in Ref.~\cite{karcher2021generalized} that, if the critical theory satisfies the local conformal invariance, the generalized-multifractality spectrum $x_\lambda$ must obey generalized parabolicity.  It follows that the local conformal invariance is violated at the SQH critical point. This striking result puts strong constraints on the form of the fixed-point theory of the SQH transition, excluding, in particular, models of Wess-Zumino-Novikov-Witten class.

Investigation of the generalized multifractality provides thus important ``fingerprints'' of an Anderson-transition critical point, which motivates an extension of the previous analysis to other 2D critical points. In this paper, we consider those three classes that are characterized by weak antilocalization:  AII, D, and DIII.  As a consequence, phase diagrams of the corresponding 2D systems feature a metallic phase. (In the superconducting classes D and DIII, this phase is called ``thermal metal''.)  These metallic phases are separated from the insulating phases by Anderson metal-insulator transitions, and we will explore the generalized multifractality at the corresponding critical points. Furthermore, while the antilocalization drives the system in the metallic phase to the ``supermetal''  (infinite conductivity) fixed point, the corresponding flow is logarithmically slow. Thus, at any realistic length scale the metallic systems exhibit generalized multifractality, which we will study as well in this work.

Two-dimensional systems of all three symmetry classes that we discuss here are of great physical interest, in particular, in connection with topological phenomena. The symmetry class AII is the Wigner-Dyson class for systems with strong spin-orbit interaction~\cite{asada2002anderson, Obuse-Multifractality-2007, obuse2007two-dimensional}. In particular, it includes 2D structures exhibiting quantum spin Hall effect as well as surfaces of 3D (weak or strong) topological insulators. Class D (already mentioned above in the context of the thermal quantum Hall effect) hosts, in particular, $p$-wave superconductors with broken time-reversal invariance; the corresponding excitations are Majorana fermions~\cite{read2000paired,senthil2000quasiparticle,bocquet2000disordered,chalker2001thermal, mildenberger2007density, medvedyeva2010effective, wang_multicriticality_2021}. Disordered systems of class D attract attention in context of paired states in the fractional quantum Hall effect with non-abelian statistics of excitations and also in connection with quantum spin liquids. Class DIII includes topological superconductors with broken spin symmetry (similarly to class D) but preserved time-reversal invariance~\cite{fulga2012thermal}. For all three classes, phase diagrams generically contain topologically distinct (thermal) insulator phases and a (thermal) metal phase.

The goal of this work is to address key questions related to the generalized multifractality in classes AII, D, and DIII:   How to construct the corresponding observables in terms of wave functions?  What are values of the exponents in 2D systems? Do  they satisfy the generalized parabolicity? (As explained above, this question is closely related to the presence of absence of local conformal invariance.) Do the exponents obey the Weyl symmetry? The paper answers all these questions; its most salient results are as follows:

\begin{enumerate}
	
	\item
	By using a renormalization-group (RG) analysis and the Iwasava decomposition, we derive the pure-scaling observables in terms of sigma-model composite operators and in terms of eigenfunction observables. We find that the eigenfunction observable construction follows one of two patterns. If the symmetry class has a symmetry of Kramers type, i.e., either a time reversal symmetry $T$ satisfying $T^2=-1$ or a particle-hole symmetry $P$ satisfying $P^2=-1$, the construction is of ``spinful'' type, as we have derived for class C in Ref.~\cite{karcher2022generalized}. This is the case for classes AII and DIII.  In the opposite case (in particular, in class D), the ``spinless'' construction applies, as obtained earlier  for class A~\cite{gruzberg2013classification}.
	
	\item
	Using appropriate microscopic models, we confirm numerically this construction for classes AII, D, and DIII,  both in metallic phases and at the metal-insulator transition points.
	(The validity of the derivation based on the sigma model is not entirely trivial at these transition points, in view of the importance of topological defects for localization.)
	These simulations also allow us to find numerical values of the generalized-multifractality exponents.
	
	\item
	We find that, in the metallic phases, the Weyl symmetry and the generalized parabolicity are fulfilled (at least, within the numerical accuracy), as expected analytically.
	
	\item
	At the metal-insulator transitions, the generalized parabolicity is strongly violated, which also implies the violation of the local conformal invariance.

	\item
	Moreover, for metal-insulator transitions in classes D and DIII, even the Weyl symmetry is violated.  We attribute this to the topology of the sigma-model manifolds in this symmetry classes, which contain two disconnected components, i.e., additional $\mathbb{Z}_2$ degrees of freedom. ``Jumps'' between these two components (or, equivalently, domain walls) are responsible for Anderson localization and, at the same time, lead to the violation of the Weyl symmetry.
	
\end{enumerate}

\section{General considerations}
\label{sec:general}

\subsection{Non-linear $\sigma$-model and composite operators. }
\label{sec:NLSM}

Field theories of Anderson localization are non-linear supersymmetric $\sigma$-models~\cite{efetov1983supersymmetry, efetov1997supersymmetry, zirnbauer1994towards, mirlin00, evers08, wegner2016supermathematics}. Many properties of the theory (including, in particular, the perturbative analysis and the symmetry classification) can be equivalently understood within a replica version of the $\sigma$-model~\cite{wegner1979the, hoef1986calculation, wegner1987anomalous1, wegner1987anomalous2, dellanna_anomalous_2006}. The target spaces of the $\sigma$-models are symmetric (super)spaces $G/K$, see the review~\cite{evers08} for target spaces corresponding to all ten symmetry classes of disordered systems.

Within the $\sigma$-model field theory, observables characterizing the generalized multifractality are represented by gradientless composite operators $\mathcal{P}(Q)$. Here the $\sigma$-model field $Q \in G/K$ is a matrix, $Q = g\Lambda g^{-1}$, where $\Lambda$ is a matrix that commute with all $k \in K$ (a standard choice is $\Lambda = \text{diag}(I_{m}, -I_{m})$, where the identity blocks are in the retarded-advanced space), and $g \in G$. Since $Q$ does not change when $g$ is multiplied on the right ($g \to gk$) by any element $k \in K$, the set of matrices $Q$ realizes the symmetric space $G/K$.

The pure-scaling observables $\mathcal{P}_\lambda(Q)$ are labeled by a tuple $\lambda = (q_1,\ldots,q_n)$ of numbers $q_i$, which is the highest weight of the corresponding irreducible representation. In general, $q_i$ may be complex but one usually focusses on real $q_i$. The composite operators belonging to the representation $\lambda$ show at criticality a power-law scaling characterized by an exponent (scaling dimension) $x_\lambda$. There are many ways to choose a representative $\mathcal{P}_\lambda(Q)$ of a given representation $\lambda$. One important choice is provided by the Iwasawa decomposition~\cite{Helgason-Differential-1978, Helgason-Groups-1984}. (In the supersymmetric approach we need a generalization to Lie supergroups that was worked out in Ref.~\cite{Alldridge-The-Harish-Chandra-2012}.) Here is a brief description of the method in the classical setting.

Any connected non-compact semisimple Lie group $G$ has a global Iwasawa decomposition $G = NAK$, where $N$ is a nilpotent group, $A$ is an Abelian group, and $K$ is the maximal compact subgoup of $G$. This factorization provides a very useful parametrization of the target space $G/K$. An element $a \in A$ is fully specified by $n$ real numbers $x_i$, which play the role of radial coordinates on $G/K$. In terms of the radial coordinates, the pure-scaling operators $\phi_\lambda(Q)$ are simply ``plane waves'':
\begin{equation}
\phi_\lambda(Q) = \phi_\lambda(x_1,x_2, \ldots, x_n) = e^{-2 \sum_i q_i x_i}.
\label{eq:phi-lambda}
\end{equation}
(Note that, in the supersymmetric formulation, $x_i$ are the Iwasawa radial coordinates in the boson sector.)

To construct the pure-scaling operators explicitly as combinations of matrix elements of $Q$, we use the key fact that there exists a choice of basis in which elements of $a \in A$ are diagonal matrices, while elements of $n \in N$ are upper triangular with units on the diagonal. This has immediate consequences for the matrix $Q \Lambda$: since elements of $K$ commute with $\Lambda$, the Iwasawa decomposition $g = nak$ leads to $Q \Lambda = n a^2 \Lambda n^{-1} \Lambda$, which is a product of an upper triangular, a diagonal, and a lower triangular matrices. In this form the lower principal minors of the advanced-advanced block of $Q \Lambda$ are simply products of diagonal elements of $a^2$, which are exponentials of the radial coordinates $x_i$ on $G/K$. These minors are basic building blocks, which can be raised to arbitrary powers and multiplied to produce the most general exponential functions~\eqref{eq:phi-lambda}.

A great advantage of this choice is that the functions $\phi_\lambda(x_1,x_2, \ldots, x_n)$ are positive (and thus can be raised to any power) and satisfy Abelian fusion:
\begin{equation}
\phi_{\lambda_1} \phi_{\lambda_2} = \phi_{\lambda_1 + \lambda_2},
\qquad \phi_{c\lambda} = (\phi_\lambda)^c.
\label{eq:phi-lambda-fusion}
\end{equation}
Furthermore, they are in direct correspondence with the pure-scaling eigenfunction observables satisfying analogous properties (to be discussed below). The Iwasawa decomposition was explicitly performed for class A in Ref.~\cite{gruzberg2013classification} and for class C in Ref.~\cite{karcher2021generalized}. In Appendix~\ref{app:Iwasawa} we provide detials of the Iwasawa construction for classes AII, D, and DIII.

Alternatively, one can use the Cartan decomposition $G=KAK$, which naturally leads to a definition of $K$-invariant  (or $K$-radial) eigenfunctions $\mathcal{P}_\lambda(Q)$. A very convenient way  to find them is to use the one-loop RG. Details of this approach and the results for all ten symmetry classes are presented in Appendix~\ref{appendix:rg}.

\subsection{Scaling dimensions and Weyl symmetry}
\label{sec:weyl}

\subsubsection{Scaling dimensions}

At the point of an Anderson transition, the observables $ \langle \mathcal{P}_\lambda(Q) \rangle$ exhibit a power-law scaling with the system size $L$,
\begin{equation}
\langle \mathcal{P}_\lambda(Q) \rangle \sim L^{-x_\lambda},
\label{eq:P-lambda-Q-scaling}
\end{equation}
where we set the ultraviolet scale (lattice constant in numerical simulations) to be unity. The angular brackets $\langle \ldots \rangle$ in Eq.~\eqref{eq:P-lambda-Q-scaling} denote the integration over $Q$ with the corresponding $\sigma$-model action.
In terms of eigenfunctions, the pure-scaling observables $P_\lambda[\psi]$ are homogeneous functions of degree $2q$ with respect to eigenfunction amplitudes, where
\begin{equation}
q = q_1 + q_2 + \ldots + q_n \equiv |\lambda|.
\label{eq:q}
\end{equation}
The explicit construction of $P_\lambda[\psi]$ is presented in Sec.~\ref{sec:pure_scaling} below. These observable show at criticality the scaling
\begin{equation}
L^{qd} \langle P_\lambda[\psi] \rangle \sim L^{-\Delta_\lambda},
\label{eq:P-lambda-psi-scaling}
\end{equation}
where $d$ is the spatial dimensionality and  $\langle \ldots \rangle$ denotes the disorder averaging. (In this paper, we focus on the case of $d=2$.) The factor $L^{qd}$ in Eq.~\eqref{eq:P-lambda-psi-scaling} takes into account the conventional metallic scaling, $|\psi^2| \sim L^{-d}$, so that the exponent $\Delta_\lambda$ characterizes the anomalous scaling. The correspondence between $\mathcal{P}_\lambda(Q)$ and
$P_\lambda[\psi]$, obtained earlier for class A in Ref.~\cite{gruzberg2013classification} and for class C in Refs.~\cite{karcher2021generalized, karcher2022generalized}, is as follows:
\begin{equation}
\mathcal{P}_\lambda(Q) \ \ \longleftrightarrow \ \   \bar{\nu}^q L^{qd} P_\lambda[\psi].
\label{eq:PQ-Ppsi}
\end{equation}
The meaning of the correspondence is that the disorder averages of (products of) $\bar{\nu}^q L^{qd} P_\lambda[\psi]$ map to $\sigma$-model correlation functions involving (products of) the corresponding  composite operators $\mathcal{P}_\lambda(Q)$.

The average density of states $\bar{\nu}$ that appears in Eq.~\eqref{eq:PQ-Ppsi} corresponds to the representation $\lambda = (1)$ and thus scales with $L$ as
\begin{equation}
\bar{\nu} (L) \sim L^{-x_{(1)}}.
\label{eq:DOS}
\end{equation}
This yields the relation between the exponents $x_\lambda$ and $\Delta_\lambda$:
\begin{equation}
x_{(q_1,\ldots,q_n)} = \Delta_{(q_1,\ldots, q_n)}  + qx_{(1)},
\label{eq:Deltaq}
\end{equation}
with $q$ given by Eq.~\eqref{eq:q}.

\subsubsection{Density of states}
\label{sec:DOS}

It is worth briefly commenting on the scaling  of the average density of states  $\bar{\nu}$, Eq.~\eqref{eq:DOS}. In the Wigner-Dyson classes (including class AII considered in this paper), $\nu$ does not have any critical behavior (i.e., is essentially a constant), so that $x_{(1)} = 0$ and $x_\lambda = \Delta_\lambda$.  On the other hand, for the unconventional (non-Wigner-Dyson) symmetry classes---in particular, classes D and DIII considered here---the average density of states exhibits at criticality a power-law scaling  with energy $\epsilon$ near $\epsilon = 0$:
\begin{equation}
\bar{\nu}(\epsilon) \sim |\epsilon|^{\kappa}.
\label{eq:power_ldos}
\end{equation}
The connection between this formula and  Eq.~\eqref{eq:DOS} is as follows.  The typical position of the $n$-th lowest eigenstate with $\epsilon >0$ is found from   Eq.~\eqref{eq:power_ldos} by using $L^2\int_0^{\epsilon_n} d\epsilon \,  \bar{\nu}(\epsilon) = n$ (we set the spatial dimensionality $d$ to be $d=2$), which yields
\begin{align}
\epsilon_n &= \left(\dfrac{n}{L^2}\right)^{\frac{1}{1+\kappa}}.
\end{align}
As we are interested in a system at criticality, we set $n \sim 1$ here and substitute the resulting $\epsilon_n$ in  Eq.~\eqref{eq:power_ldos}, which gives Eq.~\eqref{eq:DOS} with
\begin{align}
x_{(1)} &= \dfrac{2\kappa}{1+\kappa}, &  \kappa&= \dfrac{x_{(1)} }{2-x_{(1)}}.
\label{eq:exprel}
\end{align}

\subsubsection{Weyl symmetry}

As was shown in Refs.~\cite{gruzberg2011symmetries, gruzberg2013classification}, the scaling dimensions $x_\lambda$ in classes A, AI, AII, C and CI should be invariant under the action of Weyl symmetry,
\begin{equation}
x_{w\lambda} = x_\lambda, \qquad w \in W.
\end{equation}
Here $W$ is the Weyl group that acts in the space of weights $\lambda$ and is generated by the operations of the following two types:
\begin{itemize}
	\item  Weyl reflections: $q_i\rightarrow -c_i - q_i$,
	\item  Weyl permutations: \\ $q_i\rightarrow q_j+(c_j-c_i)/2$, \qquad $q_j\rightarrow q_i+(c_i-c_j)/2$.
\end{itemize}

In classes D and DIII, the situation is more subtle since the $\sigma$-model target space in either class consists of two parts: the group $G$ has two connected components, and the homotopy group $\pi_0(G/K) = \mathbb{Z}_2$. The corresponding domain walls associated with jumps between the two components of the manifold spoil the proof of the Weyl symmetry. On a technical level, the derivation of the Weyl symmetry relations in Refs.~\cite{gruzberg2011symmetries, gruzberg2013classification} was based on the supersymmetric generalization~\cite{Alldridge-The-Harish-Chandra-2012} of the classical Iwasawa decomposition of the group $G$, and the related Harish-Chandra integral formula and Harish-Chandra isomorphism~\cite{Helgason-Differential-1978, Helgason-Groups-1984}. All these exist only for {\it connected} noncompact groups, which is {\it not} the case for classes D and DIII.

Thus, the Weyl symmetry is expected to hold in classes D and DIII only when the domain walls are suppressed, i.e., the $\sigma$-model field $Q$ stays---at least approximately---in a single component of the manifold. As we discuss below, this condition is fulfilled in the thermal-metal phase. The constants $c_j$ (with $j=1,2, \ldots$) are determined by (the bosonic part of) the half sum of positive roots, $\rho_b = \sum_j c_j x_j$, and are known for all the symmetry classes. In particular, they read for the three classes that are in the focus of this paper:
\begin{align}
c_j & = 3-4j, \qquad & \text{class AII}, \label{eq:cj-AII}  \\
c_j & = 1-j, \qquad & \text{class D}, \label{eq:cj-D} \\
c_j & = 2-2j, \qquad & \text{class DIII}. \label{eq:cj-DIII}
\end{align}

\subsection{Conformal invariance and generalized parabolicity}
\label{sec:conformal}

It was shown in Ref.~\cite{karcher2021generalized} that, if the following two assumptions are met:
\begin{itemize}
	\item the theory is invariant under local conformal transformations generated by the Virasoro algebra,
	\item  there exists a set of pure-scaling operators $\phi_\lambda(Q)$ (covering all, continuously varying, $\lambda$) that are Virasoro primaries satisfying the abelian fusion rules,
\end{itemize}
\noindent
the scaling dimensions $x_{(q_1,q_2,\ldots)}$ are quadratic functions with respect to the set of $q_i$:
\begin{align}
x_{(q_1,q_2,\ldots)}^{\rm CFT}&= \sum_i A_i q_i  + \sum_{ij} B_{ij} q_i q_j.
\label{eq:Deltaq_CFT}
\end{align}
The assumption of the abelian fusion is explicitly verified by the Iwasawa construction, see Sec.~\ref{sec:NLSM}.  Thus, the only remaining assumption behind Eq.~\eqref{eq:Deltaq_CFT} is that of local conformal invariance, which is indicated by the superscript ``CFT'' (conformal field theory). This quadratic dependence of $x_{(q_1,q_2,\ldots)}$ was called in Ref.~\cite{karcher2021generalized} ``the generalized parabolicity''. In the presence of Weyl symmetry (i.e., for classes A, AI, AII, C, and CI, as well as for classes D and DIII with suppressed $\sigma$-model domain walls), the generalized parabolicity implies a very strong restriction on the spectrum of scaling dimensions $x_\lambda$.  Specifically, the spectrum is then characterized by a single parameter $b$:
\begin{align}
x_{(q_1,q_2,\ldots)}^{\rm para}&=  - b\sum_i q_i(q_i + c_i) \equiv - b \lambda \cdot (\lambda + \rho_b).
\label{eq:xq}
\end{align}
Equation~\eqref{eq:xq} means, that, in the presence of Weyl symmetry and local conformal invariance,  the generalized-multifractality spectrum is given by $x^{\rm para}_\lambda = -bz_\lambda$,  where
\begin{align}
z_\lambda &=  \sum_i q_i(q_i + c_i) \equiv  \lambda \cdot (\lambda + \rho_b).
\label{eq:laplace-eigenvalues}
\end{align}
are eigenvalues of the Laplace-Beltrami operator on the $\sigma$-model target space.
The generalized parabolicity of the spectrum $x_\lambda$ represents thus a stringent test of the local conformal invariance of the theory. In particular, it was found in Refs.~\cite{karcher2021generalized,karcher2022generalized} that the generalized parabolicity is strongly violated at the critical point of the SQH transition, thus implying a strong violation of the local conformal invariance. The numerical simulations in this paper (see Sections~\ref{sec:AII},~\ref{sec:D}, and~\ref{sec:DIII} below) allow us to test the local conformal invariance at metal-insulator transitions in these symmetry classes.

\subsection{Pure-scaling eigenfunction observables}
\label{sec:pure_scaling}

\subsubsection{Classes without (pseudo)spin degree of freedom}
\label{eq:spinless_classes}

We consider first the symmetry classes for which the following conditions are met: (i) time-reversal invariance $T$ is either absent or satisfies $T^2=1$,  and (ii) particle-hole symmetry $P$ is either absent or satisfies $P^2=1$.  This means that there is no Kramers degeneracy associated with $T^2=-1$ and no Kramers-like near-degeneracy (at $\epsilon \to 0$) associated with $P^2=-1$. We will thus loosely call this situation ``spinless''.  This case is realized for the following five symmetry classes: A, AI, BDI, AIII, and D.

For class A, pure-scaling eigenfunction combinations were derived  in Ref.~\cite{gruzberg2013classification}. The building blocks for the constructions are the eigenfunction observables corresponding to the representations $\lambda = (1,1, \ldots, 1) \equiv (1^m)$:
\begin{align}
P_{(1^m)}[\psi] &= \left|\det
\left(M_m[\psi]\right)\right|^2
= \left|\det
(\psi_{i}(\RR{j}))_{m \times m} \right|^2
\nonumber\\[0.2cm]
& \equiv \left|\det
\left(\begin{array}{cccc}
\psi_{1}(\RR1) & \psi_{2}(\RR1) & \ldots & \psi_{m}(\RR1) \\
\psi_{1}(\RR2) & \psi_{2}(\RR2) & \ldots & \psi_{m}(\RR2) \\
\vdots & \vdots & \ddots & \vdots \\
\psi_{1}(\RR{m}) & \psi_{2}(\RR{m}) & \ldots & \psi_{m}(\RR{m})
\end{array}\right)\right|^2.
\label{eq:det}
\end{align}
For a generic $\lambda$ with $n$ entries, $\lambda = (q_1,\ldots, q_n)$, the scaling observable is obtained by raising the blocks~\eqref{eq:det} to the corresponding powers and multiplying them:
\begin{align}
P_{\lambda}[\psi] &=  (P_{(1^1)}[\psi])^{q_1-q_2} (P_{(1^2)}[\psi])^{q_2-q_3}\cdots\nonumber\\
& \times \cdots(P_{(1^{n-1})}[\psi])^{q_{n-1}-q_n}(P_{(1^n)}[\psi])^{q_n}.
\label{eq:abelianA}
\end{align}
For a $\lambda$ with $n$ entries, one therefore needs $k$ wavefunctions $\psi_1, \ldots, \psi_n$, each evaluated at $n$ points, $\RR1, \ldots \RR{n}$.  The factors $P_{(1^m)}[\psi]$ with $m < n$ in Eq.~\eqref{eq:abelianA} are understood as calculated on the first $m$ wavefunctions $\psi_1, \ldots, \psi_m$ and at the first $m$ points, $\RR1, \ldots \RR{m}$.
For integer positive $q_i$ satisfying $q_1\geq q_2\geq\ldots\geq q_n>0$, the observable~\eqref{eq:abelianA} is characterized by a Young diagram $\lambda$ with $n$ rows and $q_i$ elements in the $i$-th row. The pure-scaling character of $P_{\lambda}[\psi]$ was proven in Ref.~\cite{gruzberg2013classification} by a mapping to the $\sigma$-model, which yields a pure-scaling composite operator $\mathcal{P}_\lambda(Q)$. More specifically, one obtains the radial eigenfunctions $\phi_\lambda(Q)$ of the Iwasawa decomposition,
Eq.~\eqref{eq:phi-lambda}. After this, an analytical continuation permits to prove that Eq.~\eqref{eq:abelianA} represents a pure-scaling observable for arbitrary complex $q_i$.
Note that the construction of Eq.~\eqref{eq:abelianA} out of its building blocks $P_{(1^m)}[\psi]$ is in a clear analogy with the abelian fusion rules~\eqref{eq:phi-lambda-fusion} of the $\sigma$-model functions $\phi_\lambda(Q)$.

While the derivation sketched above is rather technical, one can presented a transparent physical argument clarifying this construction of eigenfunction observables. Specifically, let us focus on the central element of the construction, Eq.~\eqref{eq:det}  for $P_{(1^m)}[\psi]$.  We know that amplitudes of critical eigenstates that are nearby in energy are strongly correlated: adjacent-in-energy eigenstates look locally almost the same. Further, the observable $P_{(1^m)}[\psi]$ should be a polynomial of degree $2m$ with respect to eigenstate amplitudes. Finally, it is the most irrelevant out of all such observables with given $m$. To have a non-zero average, each eigenfunction should enter twice (once as $\psi$, and once as $\psi^*$). With these arguments, it is clear that  $P_{(1^m)}[\psi]$ should involve the maximal possible antisymmetrization, i.e., a Slater determinant built on $k$ functions, which implies Eq.~\eqref{eq:det}.  This argumentation is equally applicable to other ``spinless'' symmetry classes (AI, BDI, AIII, and D). Out of this group of classes, we consider in detail in this work the class D.

In Sec.~\ref{sec:D}, we will explicitly verify, by performing numerical simulations at the Anderson transition and in the thermal-metal phase, that the Eq.~\eqref{eq:abelianA} indeed provides the correct form of pure-scaling observables in class D. An analytical derivation of Eq.~\eqref{eq:abelianA} proceeds by mapping to the $\sigma$-model, using the Iwasawa decomposition and establishing a one-to-one correspondence to pure-scaling composite operators~\eqref{eq:phi-lambda}. This program was carried out for class A in Ref.~\cite{gruzberg2013classification}. In Appendix~\ref{app:Iwasawa}, we show that the same procedure works in class D.

\subsubsection{Classes with (pseudo)spin degree of freedom}
\label{eq:spinful_classes}

The remaining five classes---AII, C, CI, DIII and CII---possess either time reversal symmetry with $T^2=-1$, or particle-hole symmetry with $P^2=-1$, or both. We call this situation ``spinful'': in addition to the spatial coordinate, we necessarily have in these classes a spin-type index corresponding to the space where $T$ or $P$ invariance acts. In particular, both classes from this group that we consider in detail in this paper---AII and DIII---are characterized by the time-reversal symmetry with $T^2=-1$.  This implies the Kramers degeneracy:  together with each critical state $\psi_i$, we have its partner state $T\psi_i$ which we denote for brevity by $\psi_{\bar{\imath}}$. A linear combination of $\psi_i$ and $\psi_{\bar{\imath}}$ is then also an eigenstate with the same energy. Thus, the physical argument in Sec.~\ref{eq:spinless_classes} motivating Eq.~\eqref{eq:det} has to be modified: instead of an eigenstate $\psi_i$, one should consider a two-dimensional linear space spanned by $\psi_i$ and $\psi_{\bar{\imath}}$.  Via the same token, the coordinate $\RR{j}$ should be supplemented by the spin index that takes two values $\uparrow$ and $\downarrow$. Then, instead of Eq.~\eqref{eq:det} of the spinless sutiation, we get in the spinful case the following Slater determinant:
\onecolumngrid
\begin{align}
P^{\rm sp}_{(1^m)}[\psi] &
\equiv \det
\left(
\begin{array}{c|c}
(\psi_{i,\uparrow}(\RR{j}))_{m \times m}
&
(\psi_{\bar{\imath},\uparrow}(\RR{j}))_{m \times m}\\
\hline
(\psi_{i,\downarrow}(\RR{j}))_{m \times m}
&
(\psi_{\bar{\imath},\downarrow}(\RR{j}))_{m \times m}
\end{array}
\right)
= \det
\left(\begin{array}{ccc|ccc}
\psi_{1,\uparrow}(\RR1)  & \ldots & \psi_{m,\uparrow}(\RR1)
&
\psi_{\bar{1},\uparrow}(\RR1) & \ldots & \psi_{\bar{m},\uparrow}(\RR1)
\\
\vdots & \ddots & \vdots &\vdots & \ddots & \vdots
\\
\psi_{1,\uparrow}(\RR{m}) & \ldots  & \psi_{m,\uparrow}(\RR{m})
&
\psi_{\bar{1},\uparrow}(\RR{m}) & \ldots & \psi_{\bar{m},\uparrow}(\RR{m})
\\
\hline
\psi_{1,\downarrow}(\RR1) & \ldots & \psi_{m,\downarrow}(\RR1)
&
\psi_{\bar{1},\downarrow}(\RR1) & \ldots & \psi_{\bar{m},\downarrow}(\RR1)
\\
\vdots & \ddots & \vdots & \vdots & \ddots & \vdots
\\
\psi_{1,\downarrow}(\RR{m}) & \ldots & \psi_{m,\downarrow}(\RR{m})
&
\psi_{\bar{1},\downarrow}(\RR{m}) & \ldots & \psi_{\bar{m},\downarrow}(\RR{m})
\end{array}\right).
\label{eq:det-sp}
\end{align}
\twocolumngrid
\noindent It is not difficult to prove that that the Slater determinant~\eqref{eq:det-sp} is real and positive.

This argumentation also applies to classes C and CI, which possess either no $T$-invariance (class C) or $T$-invariance with $T^2=+1$ (class CI), but are instead characterized but particle-hole symmetry with $P^2=-1$. In this case, the partner state $P\psi_i \equiv \psi_{\bar{\imath}}$ has an energy opposite to that of $\psi_i$. However, for distances between the points $\RR{j}$ much smaller than the correlation (localization) length (which is of the order of the system size $L$ for eigenstates closest to zero energy), this is immaterial, and the above argumentation leading to Eq.~\eqref{eq:det-sp} fully applies. Indeed, Eq.~\eqref{eq:det-sp} has been derived for class C  in Ref.~\cite{karcher2022generalized} by using the mapping to the $\sigma$-model and the Iwasawa decomposition of Ref.~\cite{karcher2021generalized}. Furthermore, it was shown that the observable $P_{\lambda}[\psi]$ with generic $\lambda$ has the same form~\eqref{eq:abelianA} as for spinless case; one only has to use Eq.~\eqref{eq:det-sp} for the building blocks.

We thus argue that the construction given by Eqs.~\eqref{eq:det-sp} and~\eqref{eq:abelianA} yields the pure-scaling eigenstate observables in all five ``spinful'' classes, including classes AII and DIII studied in this paper. We will  verify this below by numerical simulations in Sec.~\ref{sec:AII} (class AII) and Sec.~\ref{sec:DIII} (class DIII).
Furthermore, the connection~\eqref{eq:PQ-Ppsi} that was analytically established for class C in Refs.~\cite{karcher2021generalized} and~\cite{karcher2022generalized}, works for other spinful classes as well, and we have explicitly verified this for classes AII and DIII, see Appendix~\ref{app:Iwasawa}.

In Appendix~\ref{appendix:rg}, we present a complementary approach to derivation of pure-scaling observables. This is the one-loop RG analysis for the $\sigma$-model, which allows us to determine $K$-invariant polynomial composite operators ${\cal P}_\lambda (Q)$ as eigenfunctions of the RG. For completeness, this is done in Appendix~\ref{appendix:rg} for all ten symmetry classes. The results are very instructive, as they clearly demonstrate relations between different symmetry classes. In particular, they show that the classes split into two groups---``spinless'' and ``spinful''---in agreement with the above physical arguments based on the presence or absence of Kramers degeneracy and Kramers-like near-degeneracy and with the analysis based on the Iwasawa construction (Appendix~\ref{app:Iwasawa}).

\subsection{Scaling at a metal-insulator transition}
\label{sec:scaling-MIT}

In numerical simulations below, we study eigenfunction observables $P_{\lambda}[\psi]$ as defined by Eq.~\eqref{eq:abelianA}  in combination with Eq.~\eqref{eq:det} for class D or with Eq.~\eqref{eq:det-sp} for classes AII and DIII. For $\lambda= (q_1, \ldots, q_n)$, we use $n$ eigenstates $\psi_i$ closest in energy to criticality (with their time-reversal partners for classes AII and DIII) and $n$ spatial points $\RR{j}$ that are separated by distances $\sim r$ from each other. For minimal distances of order of lattice spacing, $r \sim 1$, the correspondence~\eqref{eq:PQ-Ppsi} implies the scaling (with spatial dimensionality $d=2$)
\begin{equation}
\bar{\nu}\,^q(L) L^{2q} \langle P_\lambda[\psi] (r,L) \rangle  \sim L^{-x_\lambda}, \qquad r \sim 1.
\label{eq:Plambda-scaling-r1}
\end{equation}
In the opposite limit of the largest possible separation, $r \sim L$, the wave functions at distance $\sim r$ become uncorrelated, yielding
\begin{equation}
\bar{\nu}\,^q(L) L^{2q} \langle P_\lambda[\psi] (r,L) \rangle  \sim L^{-x_{(q_1)} - x_{(q_2)} - \ldots - x_{(q_n)} }, \ \ r \sim L.
\label{eq:Plambda-scaling-rL}
\end{equation}
Since the dependences on $r$ and $L$ are of power-law type at criticality, we find the result for arbitrary $r$  ($ 1 \lesssim r \lesssim L$):
\begin{equation}
\bar{\nu}\,^q(L) L^{2q} \langle P_\lambda[\psi] (r,L) \rangle  \sim L^{-x_\lambda}  r^{x_\lambda-x_{(q_1)} - x_{(q_2)} - \ldots - x_{(q_n)} }.
\label{eq:Plambda-scaling-all-r}
\end{equation}
A more formal way to obtain Eq.~\eqref{eq:Plambda-scaling-all-r} is to use the $\sigma$-model RG analysis. We briefly sketch it. In the first step of RG, which proceeds from the ultraviolet scale until $r$, the operators $\mathcal{P}_{(q_i)}(Q)$ at each point $\RR{i}$ are renormalized independently, with the scaling dimensions $x_{(q_i)}$. After this, they fuse to the operator $\mathcal{P}_\lambda(Q)$. In the second step of the renormalization, from $r$ to $L$, this composite operator is renormalized with the scaling dimension $x_\lambda$.

Using Eqs.~\eqref{eq:DOS} and~\eqref{eq:Deltaq}, the result~\eqref{eq:Plambda-scaling-all-r} can be rewritten as
\begin{equation}
L^{2q} \langle P_\lambda[\psi] (r,L) \rangle  \sim L^{-\Delta_\lambda}  r^{\Delta_\lambda-\Delta_{(q_1)} - \Delta_{(q_2)} - \ldots - \Delta_{(q_n)} }.
\label{eq:Plambda-scaling-all-r-Delta}
\end{equation}

\subsection{Scaling in the (thermal) metal regime}
\label{sec:metal_scaling}

In addition to critical points of metal-insulator transitions, we explore the generalized multifractality in the (thermal) metal phase in all the three classes AII, D and DIII. In this phase, the dimensionless conductivity $g$ is large or, in other words, the coupling constant $t= 1/\pi g$ of the $\sigma$-model is small. In the limit $L\to \infty$ the conductivity $g$ renormalizes to infinity due to weak antilocalization: the system becomes a ``supermetal''.  However, we are interested in the behavior of a system at a finite $L$, when $g(L)$ is finite. As discussed below, the system then is characterized by the generalized multifractality, with effective exponents proportional to $t(L) \sim g^{-1}(L)$.  Since the renormalization $g(L)$ is only logarithmic, the generalized multifractality remains rather significant even for large system sizes (up to $L =1024$) used in our simulations.

The one-loop RG equation for the coupling constant $t$ reads~\cite{efetov1997supersymmetry, bocquet2000disordered, senthil2000quasiparticle, evers08, ostrovsky2010diffusion, wegner2016supermathematics}
\begin{align}
\dfrac{\mathrm{d} \ln t}{\mathrm{d} \ln \ell } &= - \alpha t, \qquad \alpha = \left\{ \begin{array}{ll} 1, \qquad & \text{AII and D}, \\  2, \qquad & \text{DIII}.
\end{array} \right.
\label{eq:RG-thermal-metal}
\end{align}
It is supplemented by an equation describing renormalization of the energy $\epsilon$ that determines the scaling of the averaged density of states (cf. Sec.~\ref{sec:DOS}):
\begin{equation}
\dfrac{\mathrm{d} \ln \epsilon}{\mathrm{d} \ln \ell } = \left\{ \begin{array}{ll} 2+t,
\qquad &  \text{D and DIII}, \\ 2, \qquad & \text{AII}.
\end{array} \right.
\label{eq:RG-thermal-metal-energy}
\end{equation}
Here the term 2  is the normal dimension corresponding to the spatial integral $\int d^2r$. Class AII is one of conventional (Wigner-Dyson) classes, where the energy does not exhibit any anomalous renormalization, so that the density of states is constant. In classes D and DIII, the energy does show a non-trivial renormalization, implying an anomalous scaling of the density of states.

Since $t$ is small in the (thermal) metal phase, the one-loop RG is controllable. Furthermore, it is worth emphasizing that, for classes D and DIII, the one-loop RG equations~\eqref{eq:RG-thermal-metal} discard not only higher-loop contributions but also jumps between two components of the $\sigma$-model manifold. This is fully justified since such topological excitations (domain walls) are exponentially suppressed at small $t$.

Integrating Eq.~\eqref{eq:RG-thermal-metal} from the ultraviolet scale (lattice constant) $a$ to a running scale $\ell$, we get
\begin{align}
t(\ell) &= \left[ t_0^{-1}+ \alpha \ln \left(\dfrac{\ell}{a}\right)\right]^{-1},
\qquad t_0 \equiv t(a).
\label{eq:one-loop-RG-running-t}
\end{align}
Below we set $a \equiv 1$.

To obtain the scaling of averages of $\sigma$-model composite operators  $\langle \mathcal{P}_\lambda(Q) \rangle$  within the RG scheme, one perturbs the $\sigma$-model action by the term $\mathcal{C}_\lambda^{(0)} \mathcal{P}_\lambda(Q)$. The elimination of fast degrees of freedom leads to a flow of the coupling constant $\mathcal{C}_\lambda(\ell)$ as a function of the RG scale $\ell$.  It was shown in Ref.~\cite{karcher2021generalized, Friedan-Nonlinear-1980, Friedan-Nonlinear-1985} that the one-loop RG acts on gradientless composite operators on the $\sigma$-model manifold as the Laplace-Beltrami operator (times a constant). The pure-scaling operators $\mathcal{P}_\lambda(Q)$ are eigenoperators of the Laplacian by construction. The eigenvalues of the RG describing this flow are thus proportional to eigenvalues of the Laplacian (quadratic Casimir invariant) $z_\lambda$:
\begin{align}
\dfrac{\mathrm{d} \ln \mathcal{C}_\lambda}{\mathrm{d} \ln \ell }&=  \gamma z_\lambda t, \qquad \gamma = \left\{ \begin{array}{ll} 1/2, \qquad & \text{AII}, \\  1, \qquad & \text{D and DIII}.
\end{array} \right.
\label{eq:one-loop-RG-C-lambda}
\end{align}
For class AII, the value of $\gamma$ in Eq.~\eqref{eq:one-loop-RG-C-lambda} [with the definition of coupling $t$ used in Eq.~\eqref{eq:RG-thermal-metal}] follows from the analysis in Refs.~\cite{wegner1987anomalous1, wegner1987anomalous2}, where the renormalization of polynomial composite operators in Wigner-Dyson classes was studied up to four-loop order. For classes D and DIII, the value $\gamma = 1$ can be restored from Eq.~\eqref{eq:RG-thermal-metal-energy}. Indeed,  the energy couples to the representation $\lambda = (1)$ and  $z_{(1)}^{\rm D} = z_{(1)}^{\rm DIII} = 1$.

Substituting here the running coupling~\eqref{eq:one-loop-RG-running-t}  and integrating Eq.~\eqref{eq:one-loop-RG-C-lambda} from the ultraviolet scale to $L$, we obtain
\begin{align}
\mathcal{C}_\lambda(L) &= \left(1+ \alpha t_0\ln L \right)^{ \frac{\gamma}{\alpha} z_\lambda}\mathcal{C}_\lambda^{(0)},
\label{eq:Lscal}
\end{align}
where $\mathcal{C}_\lambda^{(0)}$ is the bare value of $\mathcal{C}_\lambda$. This implies the following scaling of the averaged $\sigma$-model composite operator  $\mathcal{P}_\lambda(Q)$ with the system size $L$:
\begin{equation}
\langle \mathcal{P}_\lambda(Q; L) \rangle \sim  \left(1+ \alpha t_0\ln L \right)^{ \frac{\gamma}{\alpha} z_\lambda}.
\label{eq:one-loop-renormalization-comp-op}
\end{equation}
Setting here $\lambda = (1)$, we get the behavior of the averaged density of states $\bar{\nu}(L)$:
\begin{equation}
\bar{\nu}(L) \sim \left(1+ \alpha t_0\ln L \right)^{ \frac{\gamma}{\alpha} z_{(1)}}.
\label{eq:one-loop-renorm-DOS}
\end{equation}
Equation ~\eqref{eq:one-loop-renorm-DOS} implies the known logarithmic increase of the density of states at $\epsilon\to 0$ in classes D and DIII~\cite{bocquet2000disordered, senthil2000quasiparticle,mildenberger2007density}
\begin{align}
\bar{\nu}(\epsilon) & \sim 1+ \frac12 t_0\ln \frac{1}{\epsilon}, \qquad & \text{D},
\label{eq:one-loop-renorm-DOS-e-D}
\\
\bar{\nu}(\epsilon) & \sim \left( 1+  t_0\ln \frac{1}{\epsilon} \right)^{1/2}, \qquad & \text{DIII}.
\label{eq:one-loop-renorm-DOS-e}
\end{align}
Using the correspondence~\eqref{eq:PQ-Ppsi}, we find from Eqs. ~\eqref{eq:one-loop-renormalization-comp-op} and ~\eqref{eq:one-loop-renorm-DOS}   the scaling of the eigenfunction observables
\begin{equation}
L^{2q} \langle P_\lambda[\psi] (L) \rangle  \sim \left(1+ \alpha t_0\ln L \right)^{ \frac{\gamma}{\alpha} [ z_\lambda - q z_{(1)} ] },
\label{eq:metal-scaling-P-lambda-L}
\end{equation}
where, as before, $q = |\lambda|$.

In analogy with the discussion of the Anderson-transition point in Sec.~\ref{sec:scaling-MIT}, we can extend  Eq.~\eqref{eq:metal-scaling-P-lambda-L}, which is derived for distances $r \sim 1$ between the points, to arbitrary $r$.  In analogy with the critical point, we perform the RG in two steps: first from the ultraviolet scale till $r$, where the fusion takes place, and then from $r$ till $L$. The result reads
\begin{align}
L^{2q} \langle P_\lambda[\psi] (r, L) \rangle  &\sim \left(1+ \alpha t_0\ln L \right)^{ \frac{\gamma}{\alpha} [ z_\lambda - q z_{(1)} ] }  \nonumber \\
& \times \left(1+ \alpha t_0\ln r \right)^{ \frac{\gamma}{\alpha} \left[ - z_\lambda + \sum_i z_{(q_i)} \right] }.
\label{eq:metal-scaling-P-lambda-Lr}
\end{align}

For a small bare coupling $t_0$, there is an exponentially broad range of $L$ for which $\alpha t_0 \ln L \ll 1$. In this situation,  we can approximate
\begin{equation}
1+\alpha t_0\ln L \simeq e^{\alpha t_0 \ln L} = L^{\alpha t_0}.
\label{eq:metal-log-to-power-approx}
\end{equation}
Upon this approximation, Eqs.~\eqref{eq:one-loop-renormalization-comp-op} and~\eqref{eq:metal-scaling-P-lambda-Lr} take the form of Eqs.~\eqref{eq:P-lambda-Q-scaling} and~\eqref{eq:Plambda-scaling-all-r-Delta}, with the exponents
\begin{equation}
x_\lambda  = -  \gamma t_0 z_\lambda,  \qquad \Delta_\lambda = - \gamma t_0 (z_\lambda - q z_{(1)}).
\label{eq:metal-x-lambda}
\end{equation}
For a generic $L$, one can define running exponents
\begin{align}
x_\lambda(L) & =  - \frac{d \ln \langle \mathcal{P}_\lambda(Q; L)  \rangle}{d \ln L}, \\[0.1cm]
\Delta_\lambda(L) & = - \frac{d \ln \left[ L^{2q} \langle P_\lambda[\psi] (L) \rangle \right]} {d \ln L} = x_\lambda(L) - q x_{(1)}(L).
\label{eq:metal-running-exp}
\end{align}
Substituting here Eqs.~\eqref{eq:one-loop-renormalization-comp-op} and~\eqref{eq:metal-scaling-P-lambda-Lr}, we obtain
\begin{equation}
x_\lambda (L)  = - \gamma t(L) z_\lambda,  \qquad \Delta_\lambda (L) = - \gamma t(L) (z_\lambda - q z_{(1)}),
\label{eq:metal-x-lambda-run}
\end{equation}
where $t(L)$ is the running coupling~\eqref{eq:one-loop-RG-running-t}.

In the numerical analysis, we will plot $L^{2q} \langle P_\lambda[\psi] (L) \rangle$ as a function of $L$ on the log-log scale. We see that, as long as the condition $\alpha t_0 \ln L \ll 1$ is reasonably well met, this plot is expected to give an approximately straight line, with a slope $-\Delta_\lambda$ given by Eq.~\eqref{eq:metal-x-lambda}. With increasing $L$, the line should exhibit a curvature, and a slope will be given by $-\Delta_\lambda(L)$, Eq.~\eqref{eq:metal-x-lambda-run}. We also emphasize that the spectrum~\eqref{eq:metal-x-lambda} satisfies the single-parameter generalized parabolicity~\eqref{eq:xq}, with $b =  \gamma t_0$.  Moreover, the generalized parabolicity~\eqref{eq:xq}  holds also for the running spectrum~\eqref{eq:metal-x-lambda-run}, with $b = \gamma t(L)$.  Below, we will confront these analytical predictions with results of numerical simulations.

\section{Class AII}
\label{sec:AII}

\subsection{Model and generalities}

\begin{table*}
	\centering
	\begin{tabular}{cc||cc|cc||c}
		& rep. $\lambda$ & $x_\lambda^{\rm MIT}$  & $x_\lambda^{\rm MIT}/b$ & $x_\lambda^{\rm metal}$ & $x_\lambda^{\rm metal}/b$ & $x^{\rm para}_\lambda $\\[5pt]
		\hline
		\hline
		&&&&&&\\[-5pt]
		$q=2$ & (2) & $-0.361 \pm 0.001$ & $-2.08\pm 0.01$ & $-0.0551\pm 0.0001$ & $-2.017\pm 0.005$ & $-2b$\\[3pt]
		& (1,1) & $0.489\pm 0.001$ & $2.83\pm 0.01$ & $0.1095\pm 0.0001$ & $4.012\pm 0.005$ & $4b$
		\\[5pt]
		\hline
		&&&&&&\\[-5pt]
		$q=3$ & (3) & $-1.14\pm 0.01$ & $-6.57\pm 0.06$ & $-0.1659\pm 0.0004$ & $-6.08\pm 0.02$ & $-6b$\\[3pt]
		& (2,1) & $0.225\pm 0.001$ & $1.30\pm 0.01$ & $0.0547\pm 0.0002$ & $2.04\pm 0.01$ & $2b$\\[3pt]
		& (1,1,1)  & $1.333 \pm 0.001$ & $7.70\pm 0.01$ & $0.3278\pm 0.0003$ & $12.01\pm 0.01$ & $12b$
		\\[5pt]
		\hline
		&&&&&&\\[-5pt]
		$q=4$ & (4) & $-2.27\pm 0.05$ & $-13.13\pm 0.29$ & $-0.334\pm 0.001$ & $-12.21\pm 0.04$ & $-12b$\\[3pt]
		& (3,1) & $-0.36\pm 0.01$ & $-2.06\pm 0.06$ & $-0.0557\pm 0.0005$ & $-2.04\pm 0.02$ & $-2b$\\[3pt]
		& (2,2)  & $0.493\pm 0.005$ & $2.85\pm 0.03$ & $0.1095 \pm 0.0005$ & $4.01\pm 0.02$ & $4b$\\[3pt]
		& (2,1,1) & $1.111\pm 0.003$ & $6.42\pm 0.02$ & $0.2728\pm 0.0005$ & $9.99\pm 0.02$ & $10b$\\[3pt]
		& (1,1,1,1) & $2.515\pm 0.002$ & $14.54\pm 0.01$ & $0.6545 \pm 0.0003$ & $23.97\pm 0.01$ & $24b$\\[3pt]
		\hline		
		&&&&&&\\[-5pt]
		$q=5$ & (5) & $-3.52 \pm 0.09$ & $-20.37\pm 0.17$ & $-0.559\pm 0.003$ & $-20.48\pm 0.52$ & $-20b$\\[3pt]
		& (4,1) & $-1.35\pm 0.07$ & $-7.82\pm 0.40$ & $-0.223\pm 0.001$ & $-8.16\pm 0.04$ & $-8b$\\[3pt]
		& (3,2)  & $0.02 \pm 0.02$ & $0.08\pm 0.12$ & $-0.0006 \pm 0.0009$ & $0.02\pm 0.03$ & $0$\\[3pt]
		& (3,1,1) & $0.64\pm 0.01$ & $3.67\pm 0.06$ & $0.1623\pm 0.0008$ & $5.95\pm 0.03$ & $6b$\\[3pt]
		& (2,2,1) & $1.333\pm 0.005$ & $7.70\pm 0.03$ & $0.327 \pm 0.0008$ & $11.97\pm 0.03$ & $12b$\\[3pt]
		& (2,1,1,1) & $2.316\pm 0.004$ & $13.39\pm 0.02$ & $0.5997\pm 0.0005$ & $21.99\pm 0.02$ & $22b$\\[3pt]
		& (1,1,1,1,1) & $4.031\pm 0.004$ & $23.30\pm 0.02$ & $1.0895\pm 0.0004$ & $39.91\pm 0.02$ & $40b$
	\end{tabular}
	\caption{Scaling exponents $x_\lambda$ of the generalized multifractality in class AII (Ando model) for polynomial eigenstate observables with $q\equiv |\lambda|\leq 5$. The exponents $x_\lambda^{\rm MIT}$ and $x_\lambda^{\rm metal}$ are obtained numerically at the metal-insulator transition (critical disorder $W= 5.84$) and in the metallic phase ($W=3$), respectively.  The Weyl symmetry implies the relations $x_{(2,2)} = x_{(1,1)}$, $x_{(2,2,1)} = x_{(1,1,1)}$, $x_{(3,1)} = x_{(2)}$, and $x_{(3,2)} = 0$; all of them are nicely fulfilled both at the MIT and in the metallic phase. The last column displays the exponents $x_\lambda^{\rm para}$ corresponding to the generalized parabolic spectrum~\eqref{eq:xq} with a single parameter $b$. To check whether the generalized parabolicity holds, we choose $b=0.0273$ in the metal and $b=0.173$ at the transition
		and show the values of $x_{\lambda}/b$. A comparison with $x_\lambda^{\rm para}$ shows that the generalized parabolicity holds nearly perfectly in the metallic regime but is strongly violated at the transition. In the metallic regime, the parameter $b$ of the spectrum flows logarithmically with scale: $b= \frac12 t(L)$, where $t(L)$ is the running coupling~\eqref{eq:one-loop-RG-running-t}.}
	\label{tab:lAII}
\end{table*}

The symmetry class AII is the Wigner-Dyson class with time-reversal invariance $T$ satisfying $T^2=1$. At variance with classes D and DIII considered below, the density of states is not critical in class AII, i.e., $x_{(1)} = 0$, so that $x_\lambda = \Delta_\lambda$. Thus, exploring numerically the scaling of the eigenfunction observables
$\langle P_\lambda[\psi] (L) \rangle$ in class AII, we will obtain directly the exponents $x_\lambda$.  Another important feature of class AII is the absence of two-loop and three-loop corrections to the beta- and zeta-functions~\cite{wegner1987anomalous1, wegner1987anomalous2},
\begin{align}
\dfrac{\mathrm{d} \ln t}{\mathrm{d} \ln \ell } = - t + O (t^4) ; \qquad
\dfrac{\mathrm{d} \ln \mathcal{C}_\lambda}{\mathrm{d} \ln \ell }=  \frac{1}{2} z_\lambda t + O (t^4).
\label{eq:AII-four-loop}
\end{align}
with $t= 1/\pi g$. In view of this, the one-loop formulas of Sec.~\ref{sec:metal_scaling} are expected to be especially accurate in the metallic phase of class AII.

For the numerical analysis, we use the Ando model~\cite{ando1989numerical} defined by the following Hamiltonian:
\begin{align}
H&= \sum_{i\sigma}\epsilon_i c_{i,\sigma}^\dagger c_{i,\sigma}+\sum_{\langle i,j\rangle \sigma}V_{i,\sigma;j,\sigma'} c_{i,\sigma}^\dagger c_{i,\sigma}\label{eq:ham_ando}.
\end{align}
Here  the random on-site potentials $\epsilon_i$ are uniformly distributed on the interval $[-W/2,W/2]$ and the nearest-neighbor hopping $V_{i,\sigma;i+k,\sigma'}= V_0\exp(i\theta_k\sigma_k)$ depends on the direction $k=x,y$.  The time-reversal symmetry acts as $\sigma_2\mathcal{K}$, where $\mathcal{K}$ is the complex conjugation.
The Ando model  (as well as its variation known as the SU(2) model) exhibits the metal-insulator transition characteristic for class AII, and the corresponding phase diagram has been extensively studied~\cite{ando1989numerical, kawarabayashi1996diffusion, yakubo1998finite-size, asada2002anderson, ando2004anderson, markos2006critical, Mildenberger-Wave-2007}.
We set $V_0 = 1$ and $\theta_x=\theta_y=\pi/6$. It is known from previous studies that the Anderson transition at the band center, $\epsilon=0$, takes place at disorder strength
$W=5.84$, and we use this value of $W$ to explore the system at criticality.  In addition, we perform the numerical analysis for a considerably weaker disorder, $W = 3.00$, for which the system is deeply in the metallic phase.

We verify below that the construction~\eqref{eq:abelianA},~\eqref{eq:det-sp} correctly yields the scaling observables. To this end, we will consider the polynomial observables up to the order $q=5$. The Weyl symmetry, which is expected to hold exactly in class AII, implies a number of relations between the corresponding scaling exponents:
\begin{align}
x_{(1,1)} &= x_{(2,2)}, & x_{(1,1,1)} &= x_{(2,2,1)},\nonumber \\
x_{(3,1)} &= x_{(2)}, & x_{(3,2)} &= 0.
\label{eq:weyl_aii}
\end{align}
The generalized parabolic spectrum~\eqref{eq:xq} has in class AII the following form:
\begin{align}
x_{(q_1,q_2,\ldots)}^{\rm para}&= b\left[ q_1(1-q_1) + q_2(5-q_2) + q_3(9-q_3) +\ldots\right].
\label{eq:xqaii}
\end{align}
As discussed above, violation of this form of the generalized-multifractality spectrum at the metal-insulator transition would imply that the local conformal invariance does not hold.

We proceed now by presenting numerical results first for the metallic phase (Sec.~\ref{sec:AII-metallic}) and then for the metal-insulator transition in Sec.~\ref{sec:AII-MIT}.

\subsection{Metallic phase}
\label{sec:AII-metallic}

\begin{figure*}
	\centering
	\includegraphics[width =.46\textwidth]{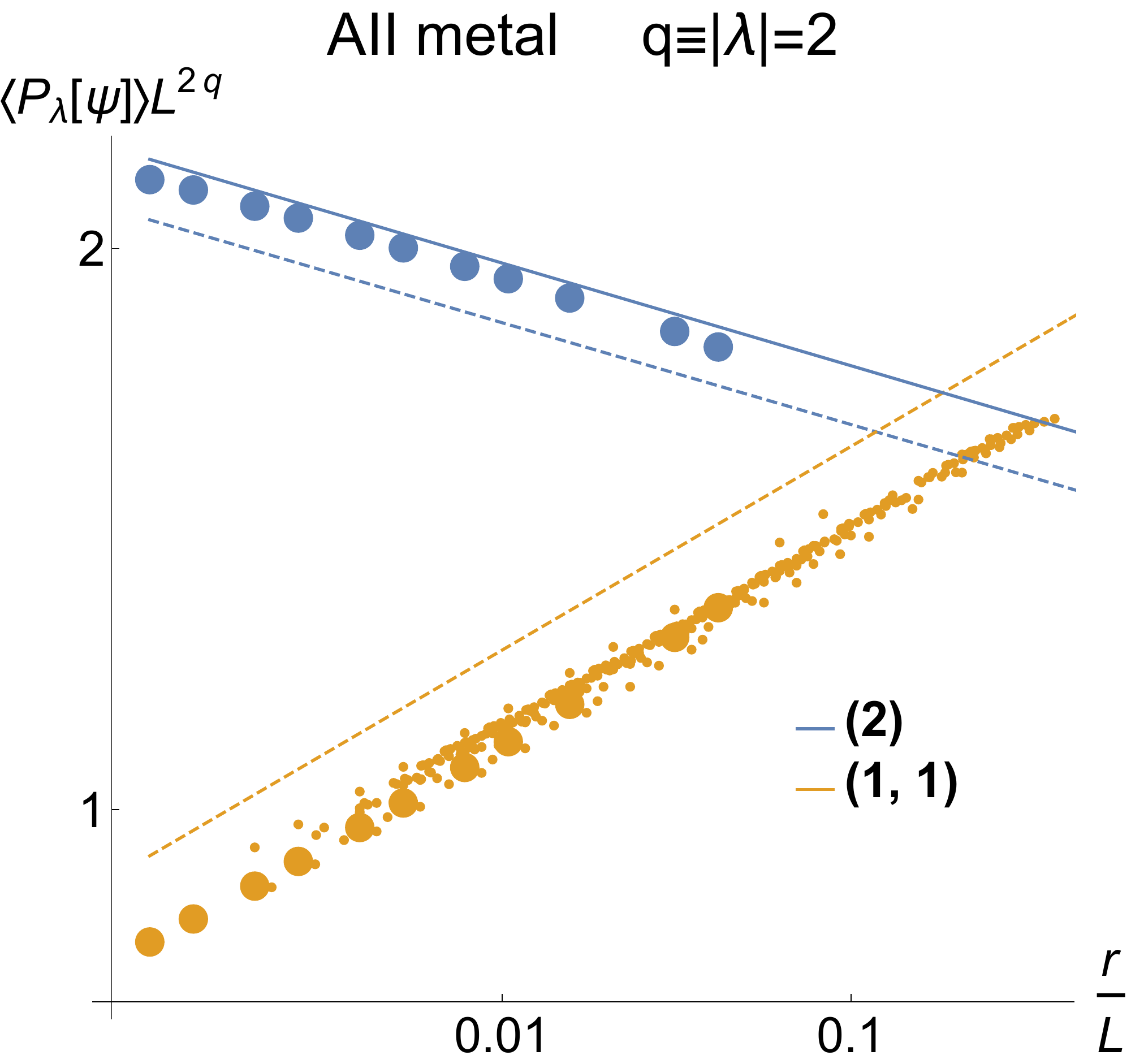}
	\includegraphics[width =.46\textwidth]{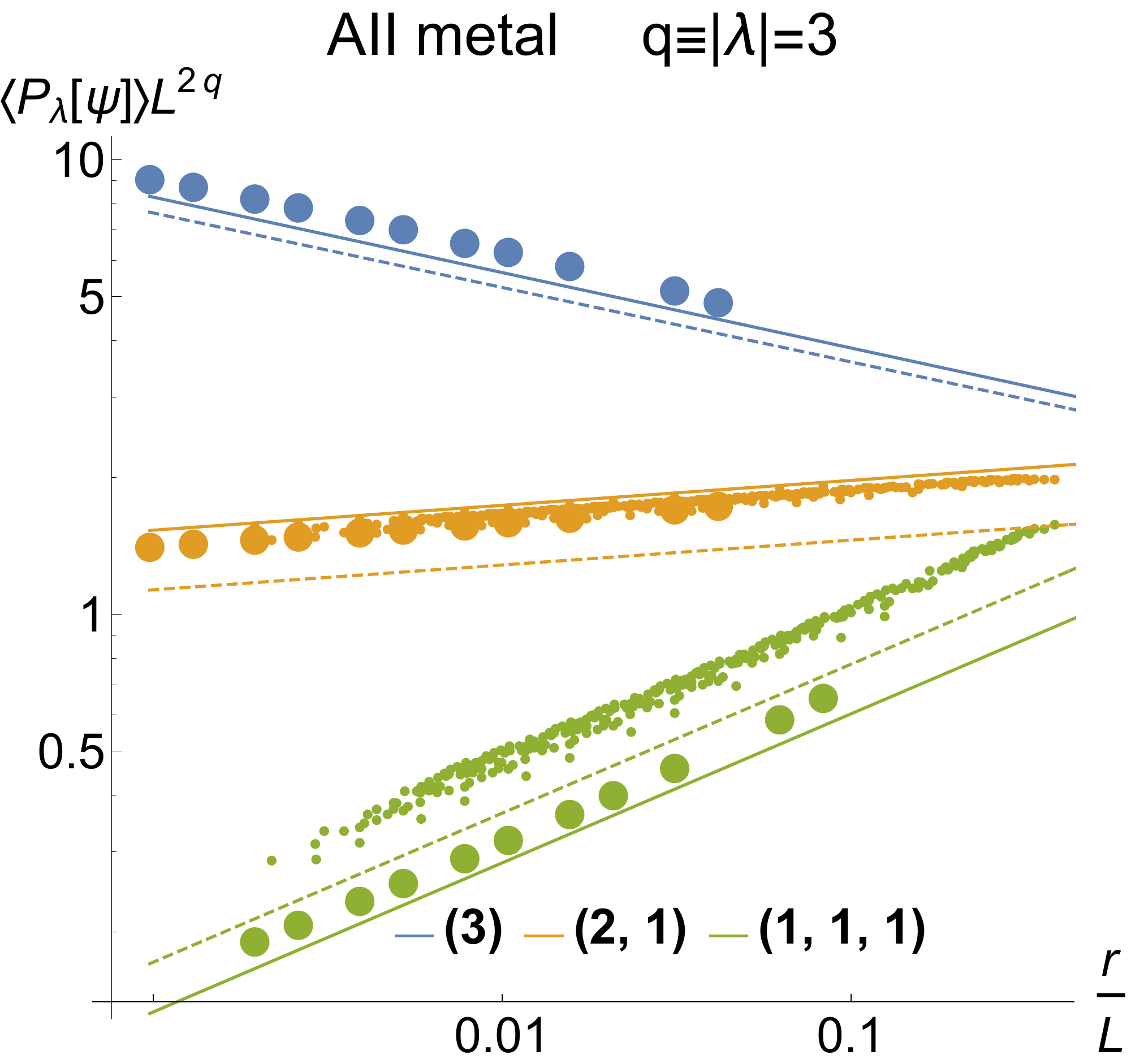}
	\includegraphics[width =.46\textwidth]{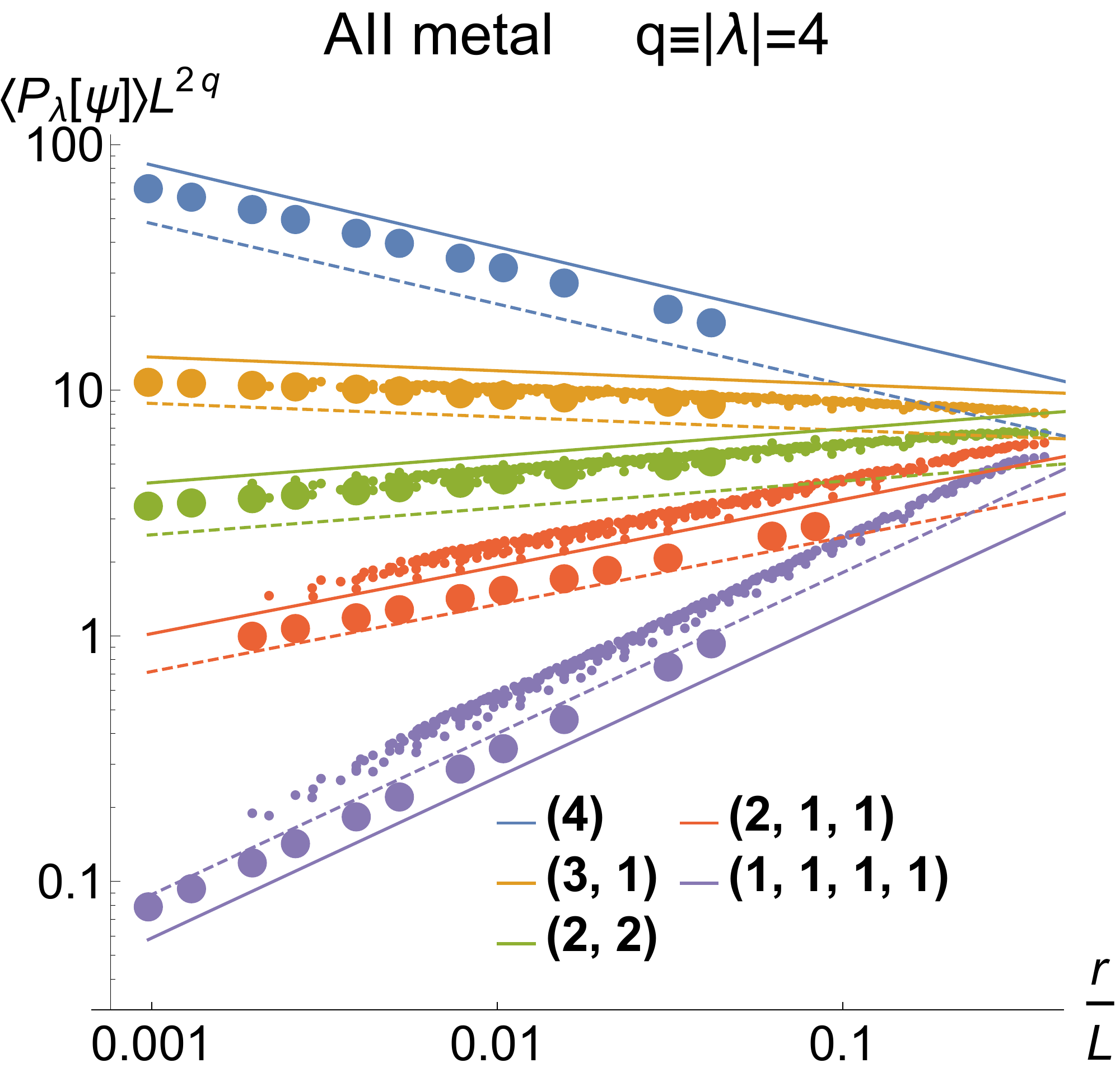}
	\includegraphics[width =.46\textwidth]{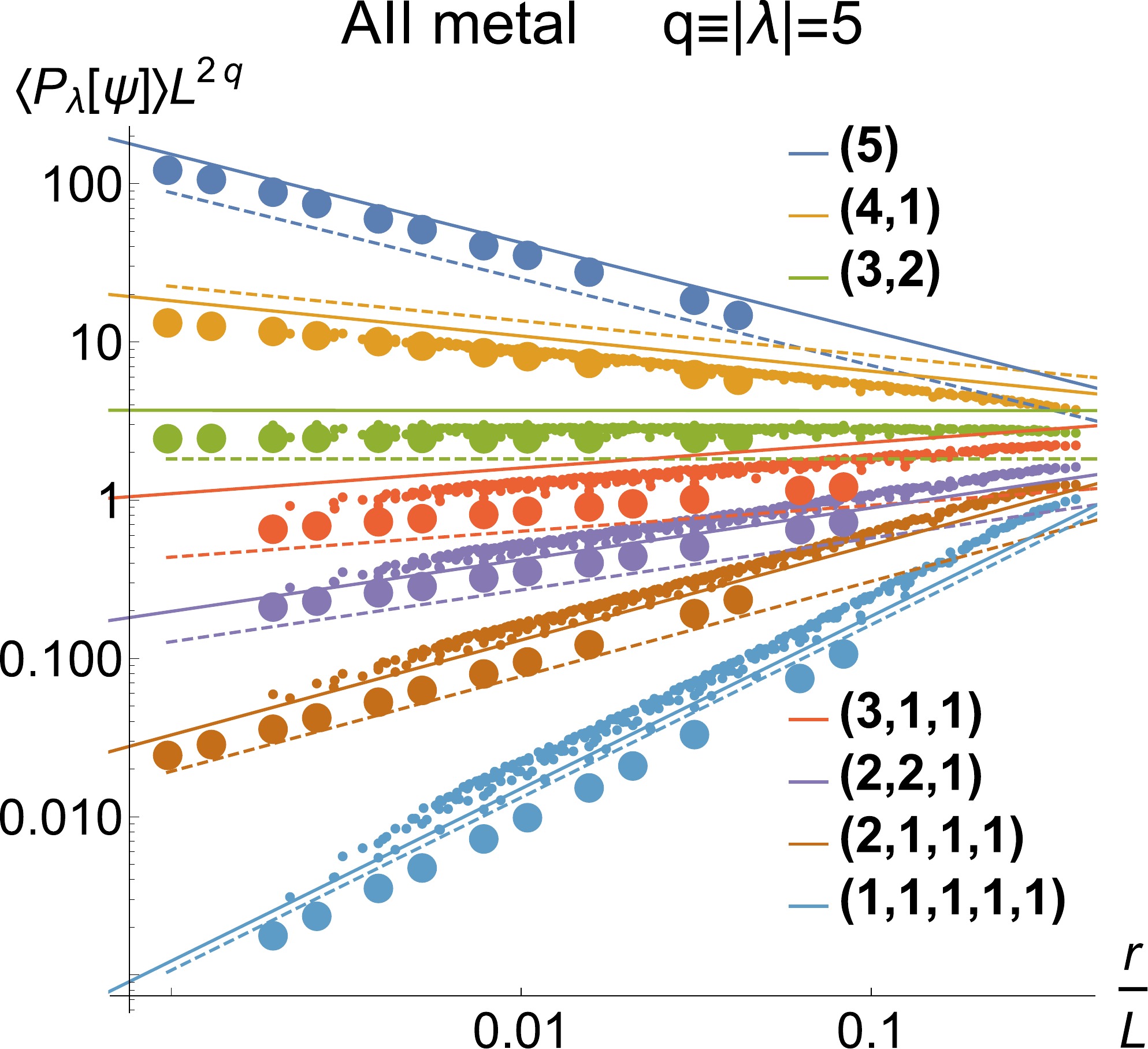}
	\caption{Numerical determination of the generalized multifractality in the metallic phase of class AII [Ando model~\eqref{eq:ham_ando}, disorder $W=3$, energy $\epsilon = 0$], for $q=2$ (top left), $q=3$ (top right), $q=4$ (bottom left) and $q=5$ (bottom right) eigenstate observables. The spinful pure-scaling combinations~\eqref{eq:abelianA},~\eqref{eq:det-sp} are computed, with averaging over the system area and $10^4$ realizations of disorder. The data are scaled with $r^{\Delta_{q_1}+ \ldots + \Delta_{q_n}}$, which yields an
		expected collapse as functions of $r/L$. For each $\lambda$, data points corresponding to the smallest $r \sim 1$ are highlighted as large dots, visualizing the $L$-dependence at a fixed $r$. The full lines are fits to these data points; the resulting exponents $x^{\rm metal}_\lambda$ are given in Table~\ref{tab:lAII}. The dashed lines corresponds to the generalized parabolic spectrum~\eqref{eq:xq} with $b=0.0273$; see the column $x^{\rm para}_\lambda$ in  Table~\ref{tab:lAII}.  The slopes of full and dashed lines are nearly identical for all $\lambda$, which means that the generalized parabolicity holds to a great accuracy in the metallic phase.}
	\label{fig:AII-metal}
\end{figure*}

In Fig.~\ref{fig:AII-metal}, we show the numerical results for polynomial observables $L^{2q} \langle P_\lambda[\psi] (r,L) \rangle$ with $q = |\lambda|$ equal to 2, 3, 4, and 5.
We take the required number (up to five) of eigenstates closest to criticality, together with their Kramers partners, evaluate the observables defined
by Eqs.~\eqref{eq:abelianA},~\eqref{eq:det-sp}, and perform the averaging over all points in the sample ($\sim L^2$) and over $10^4$ realizations of disorder.  The system size is varied from $L=32$ to $L = 1024$.

The points are separated by distances $\sim r$, with $r \lesssim 10$  (see Ref.~\cite{karcher2022generalized} for an analogous study for SQH transition). The data are scaled by the factor $r^{\Delta_{q_1}+ \ldots + \Delta_{q_n}}$.  As expected from Eq.~\eqref{eq:Plambda-scaling-all-r-Delta}, this leads to a collapse of the data for any given $\lambda$ onto a straight  line corresponding to a power-law scaling with $r/L$. The slope of this line yields $\Delta_\lambda = x_\lambda$.  For each $\lambda$, the data points for the smallest $r \sim 1$ are highlighted by larger symbols in order to visualize the $L$ dependence at fixed small $r$. These data points are used for power-law fits (full lines); the resulting exponents $x^{\rm metal}_\lambda$ are  presented in Table~\ref{tab:lAII}.  The error bars of the exponents $x_\lambda$ in Table~\ref{tab:lAII} (and in the analogous tables for classes D and DIII below) are determined using the same method as for the SQH transition in Appendix C of Ref. \cite{karcher2022generalized}. As seen from the Table~\ref{tab:lAII}, the Weyl-symmetry relations~\eqref{eq:weyl_aii} are fulfilled by the numerically found exponents $x^{\rm metal}_\lambda$ with an excellent accuracy.

In order to analyze whether the spectrum $x^{\rm metal}_\lambda$ satisfies the generalized parabolicity~\eqref{eq:xqaii}, we fix the parameter $b$
to $b=0.0273$  and present the values of $x^{\rm metal}_{\lambda} / b $ in Table~\ref{tab:lAII}.
(We choose the optimal value of $b$ such that  the corresponding parabolic approximations describes in the best possible way the exponents $x^{\rm metal}_{(q_1^n)}$ with $n=1,2,3$ and not too large $q$, see below.)
The values of $x^{\rm metal}_{\lambda} / b $  should be compared to exponents $x^{\rm para}_\lambda $ for the strictly parabolic spectrum, Eq.~\eqref{eq:xqaii}, that are also included in the Table. An excellent agreement is observed, in full consistency with the analytical prediction~\eqref{eq:metal-x-lambda} for the metallic phase.  The generalized parabolicity of the spectrum $x^{\rm metal}_\lambda$ is illustrated also in Fig.~\ref{fig:AII-metal}, where the dashed lines correspond to a parabolic spectrum~\eqref{eq:xqaii} with $b=0.0273$. The slopes of full and dashed lines are essentially indistinguishable.

\begin{figure*}
	\centering
	\vspace*{5cm}
	\includegraphics[width = \textwidth]{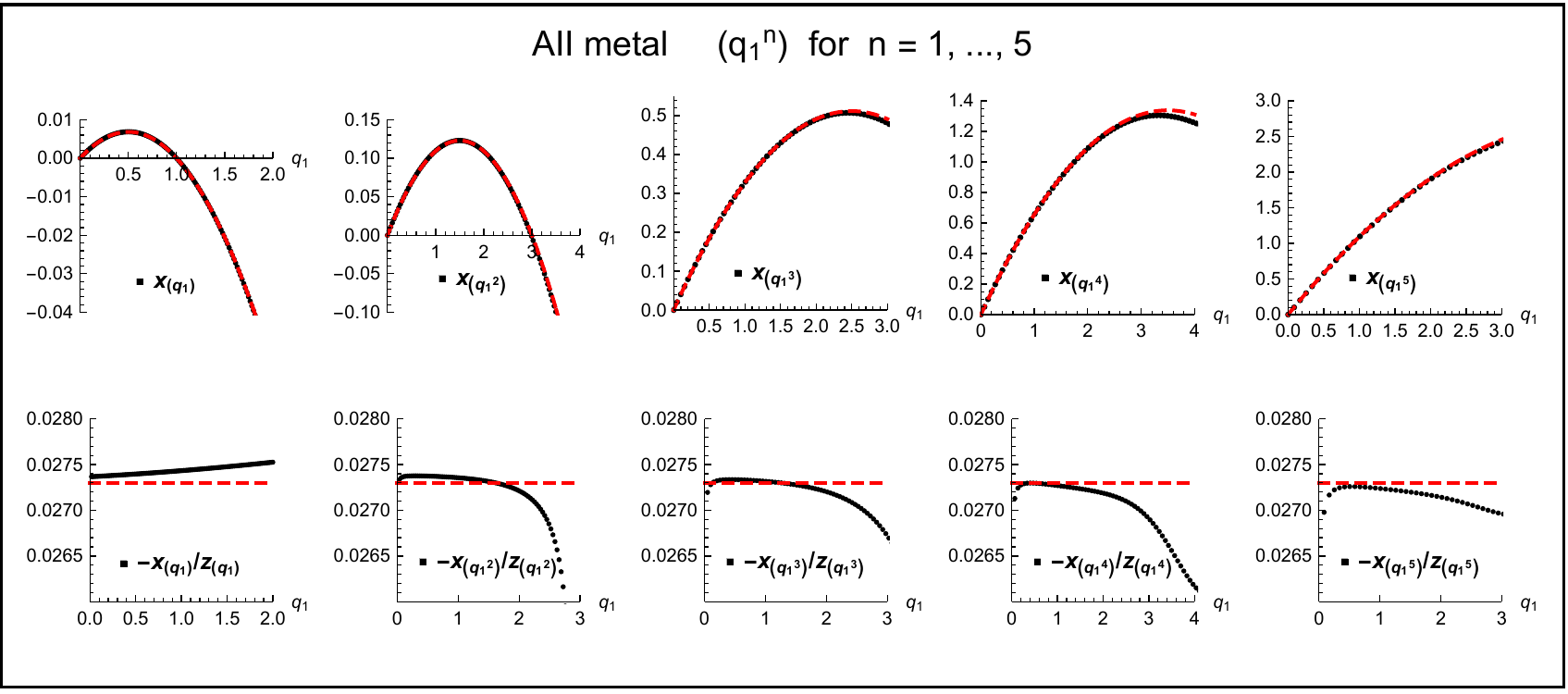}
	\caption{Exponents $x_{(q_1^n)}$ with $n=1, \ldots, 5$ for the metallic phase of class AII (same parameters as in Fig.~\ref{fig:AII-metal}). Data points in the top panels are  numerically obtained exponents $x_\lambda$. In the bottom panels, the same data are shown in the form $- x_\lambda / z_\lambda$, where $z_\lambda$ is the quadratic Casimir invariant.  The red dashed lines in all panels correspond to the generalized parabolicity~\eqref{eq:xqaii} with $b=0.0273$.  It is seen that the generalized parabolicity holds to a great accuracy. At large $q$, deviations related to insufficient ensemble averaging are observed. 	
	}
	\label{fig:AII-metal-2}
\end{figure*}

Let us emphasize that we have fitted the data by power-law dependences~\eqref{eq:Plambda-scaling-all-r-Delta}.  As discussed in Sec.~\ref{sec:metal_scaling}, this holds approximately in the metallic phase ($t_0 \ll 1$), as long as the system size is not too big, so that $t_0 \ln L$ is sufficiently small.  Our results for the multifractality spectrum imply $t_0 \simeq 0.076$,  implying that $t_0 \ln L$ indeed remains quite small in the whole range of considered system sizes. This explains why the data in Fig.~\ref{fig:AII-metal} are fitted well by straight lines.  More accurately, the slope changes with $L$ according to Eqs.~\eqref{eq:metal-x-lambda-run},~\eqref{eq:one-loop-RG-running-t}, but this change is logarithmically slow.  Indeed, a closer inspection of the data in Fig.~\ref{fig:AII-metal} shows the predicted reduction of the slope with increasing $L$.
As we have shown, the scaling in the whole range of $L$ should be described by Eq.~\eqref{eq:metal-scaling-P-lambda-L}, which yields for class AII
\begin{equation}
L^{2q} \langle P_\lambda[\psi] (L) \rangle  \sim \left(1+t_0\ln L \right)^{z_\lambda/2}.
\label{eq:metal-scaling-P-lambda-L-AII}
\end{equation}
We have verified that a fit according to  Eq.~\eqref{eq:metal-scaling-P-lambda-L-AII} with $t = 0.076$  indeed describes all the data excellently.  We also note that reaching the asymptotic regime, where the logarithmic term in brackets  of Eq.~\eqref{eq:metal-scaling-P-lambda-L-AII} is dominant, so that
$L^{2q} \langle P_\lambda[\psi] (L) \rangle  \propto (\ln L)^{z_\lambda/2}$, would require unrealistically large system sizes ($\ln L$ considerably exceeding $1/t_0 \approx 15$).

We complement the numerical analysis of the multifractality by investigating the exponents $x_\lambda$ for $\lambda = (q_1, \ldots, q_1)$ (tuple of length $n$), with a continuously changing $q_1 = q/n$ and with $n=1,2,\ldots$. We use a short notation $\lambda = (q_1^n)$ for these representations.  For a given $n$, such exponents $x_\lambda$ characterize the distribution of the determinant~\eqref{eq:det-sp}. In Fig.~\ref{fig:AII-metal-2}, we show the results for $n=1$, 2, 3, 4, and 5.  In the lower panels, the data are presented in the form $-x_\lambda / z_\lambda$, which is a constant $b$ in the case of a generalized parabolic spectrum~\eqref{eq:xqaii}. The lines correspond to the generalized parabolicity~\eqref{eq:xqaii} with $b=0.0273$. We see again that the generalized parabolicity in the metallic phase is very well fulffilled. At high $q$, the ensemble averaging becomes insufficient (as is always the case with numerical studies of spectra of multifractality), which leads to observed increase of deviations.

\subsection{Metal-insulator transition}
\label{sec:AII-MIT}
\begin{figure*}
	\centering 
	\includegraphics[width=.3\textwidth]{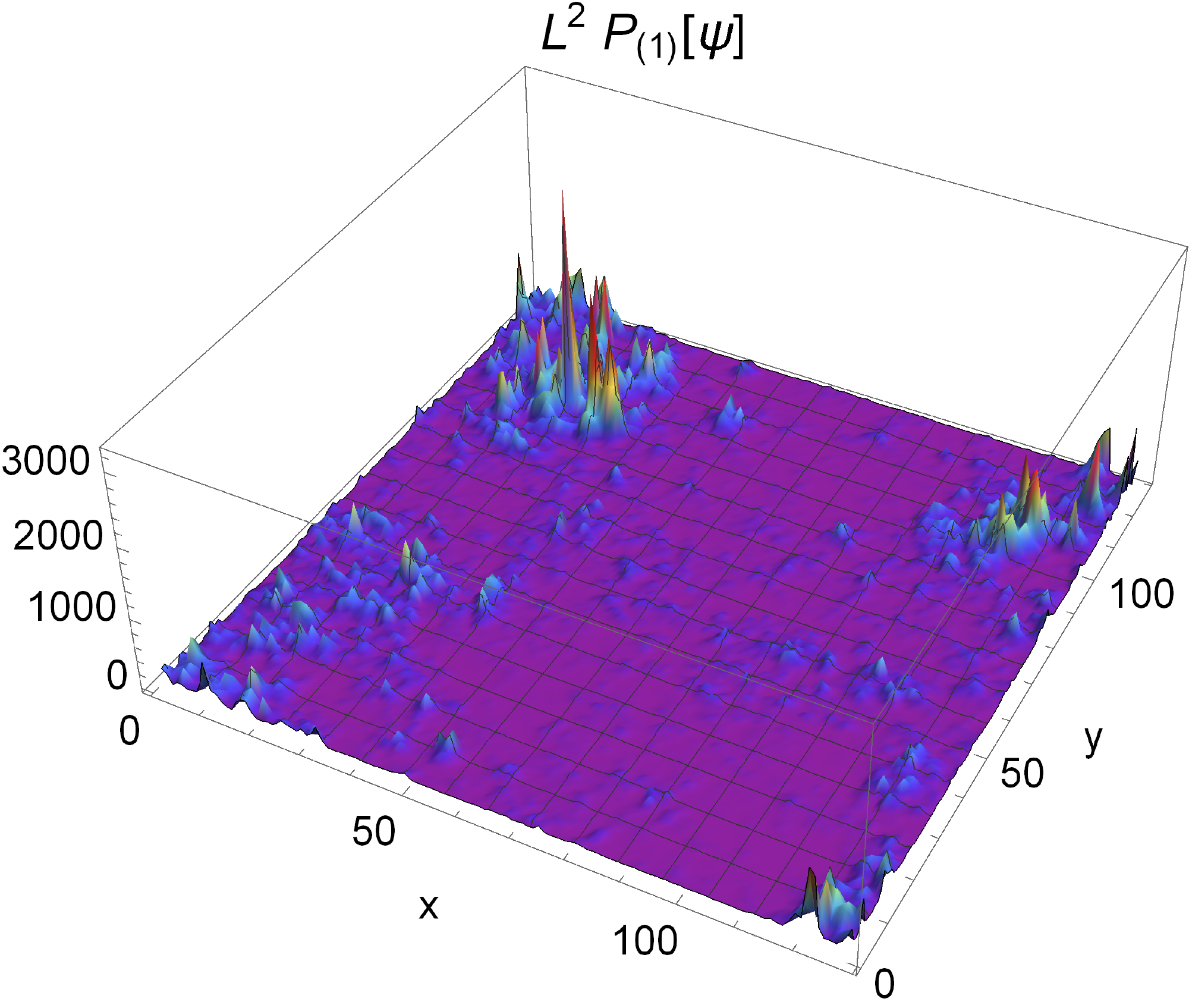}
	\includegraphics[width=.3\textwidth]{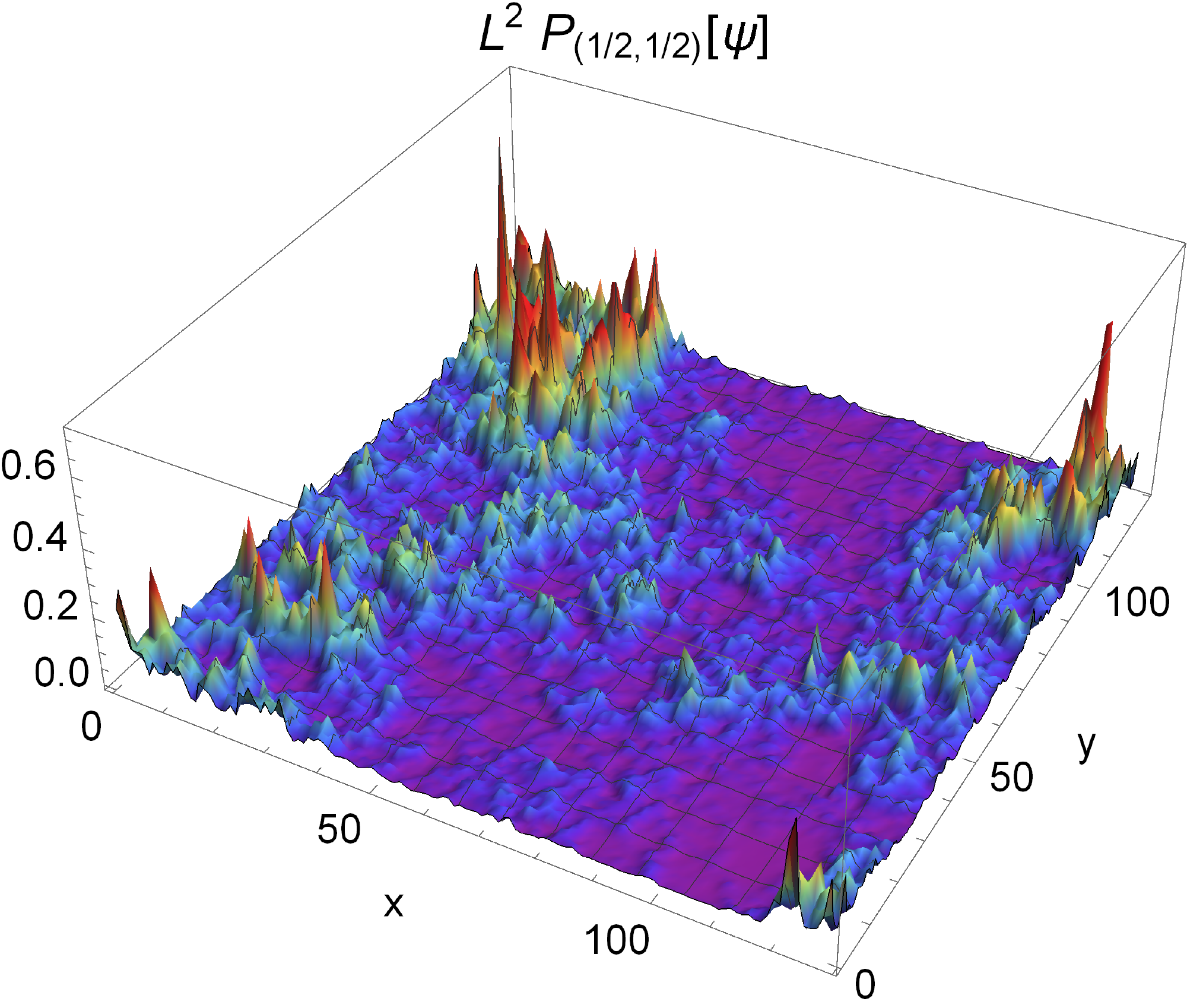}
	\includegraphics[width=.3\textwidth]{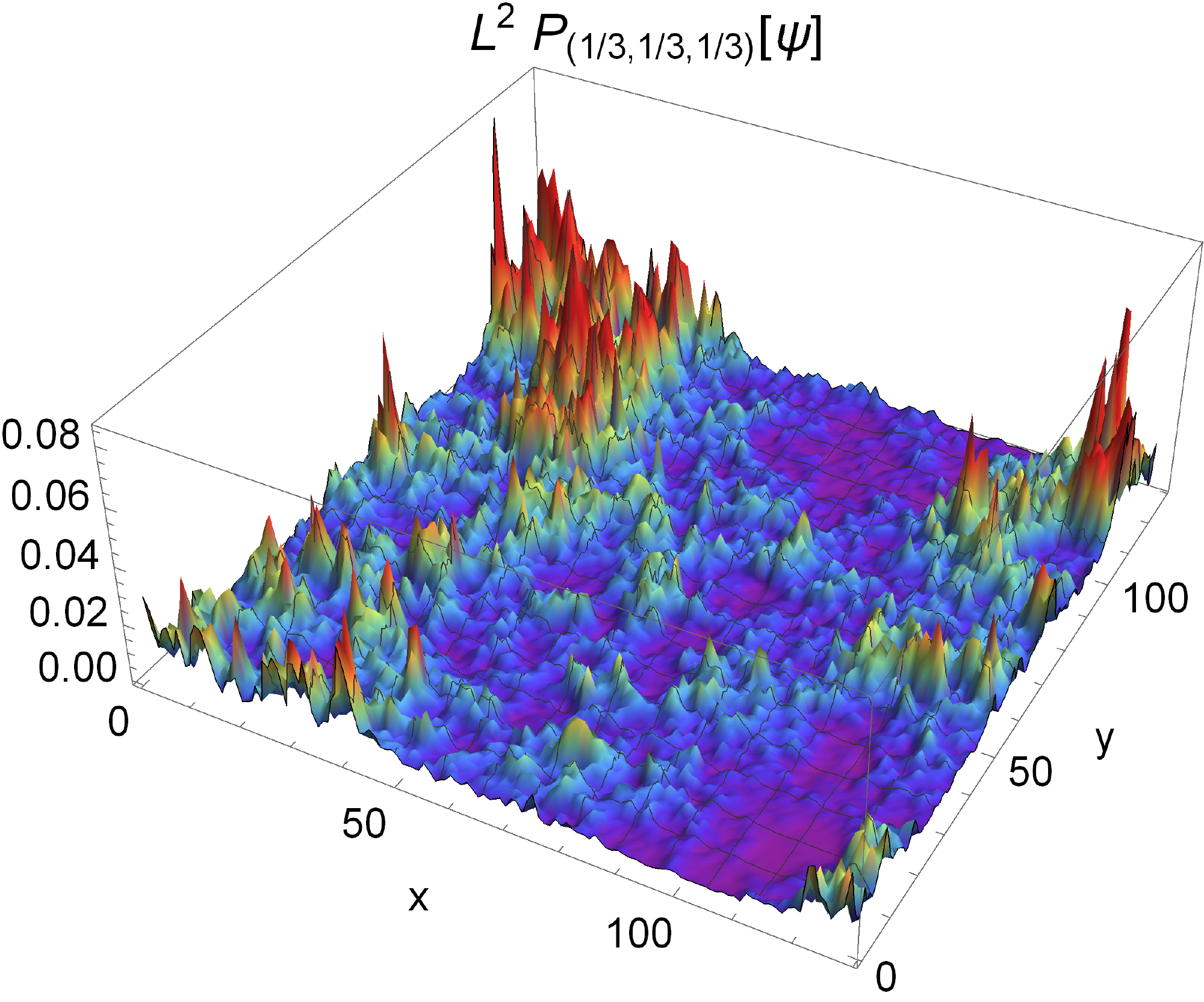}
	\caption{Illustration of spatial structure of building blocks of
		generalized multifractality, Eq. (21). Pure-scaling observables
		$L^2 P_{(1)}[\psi]$ (left panel), $L^2 P_{(1,1)}^{1/2}$ (middle panel),
		and $L^2 P_{(1,1,1)}^{1/3}$ (right panel) evaluated for a randomly
		chosen disorder realization at the metal-insulator transition in
		class AII. } 
	\label{fig:ill}
\end{figure*}

\begin{figure*}
	\centering
	\includegraphics[width =.46\textwidth]{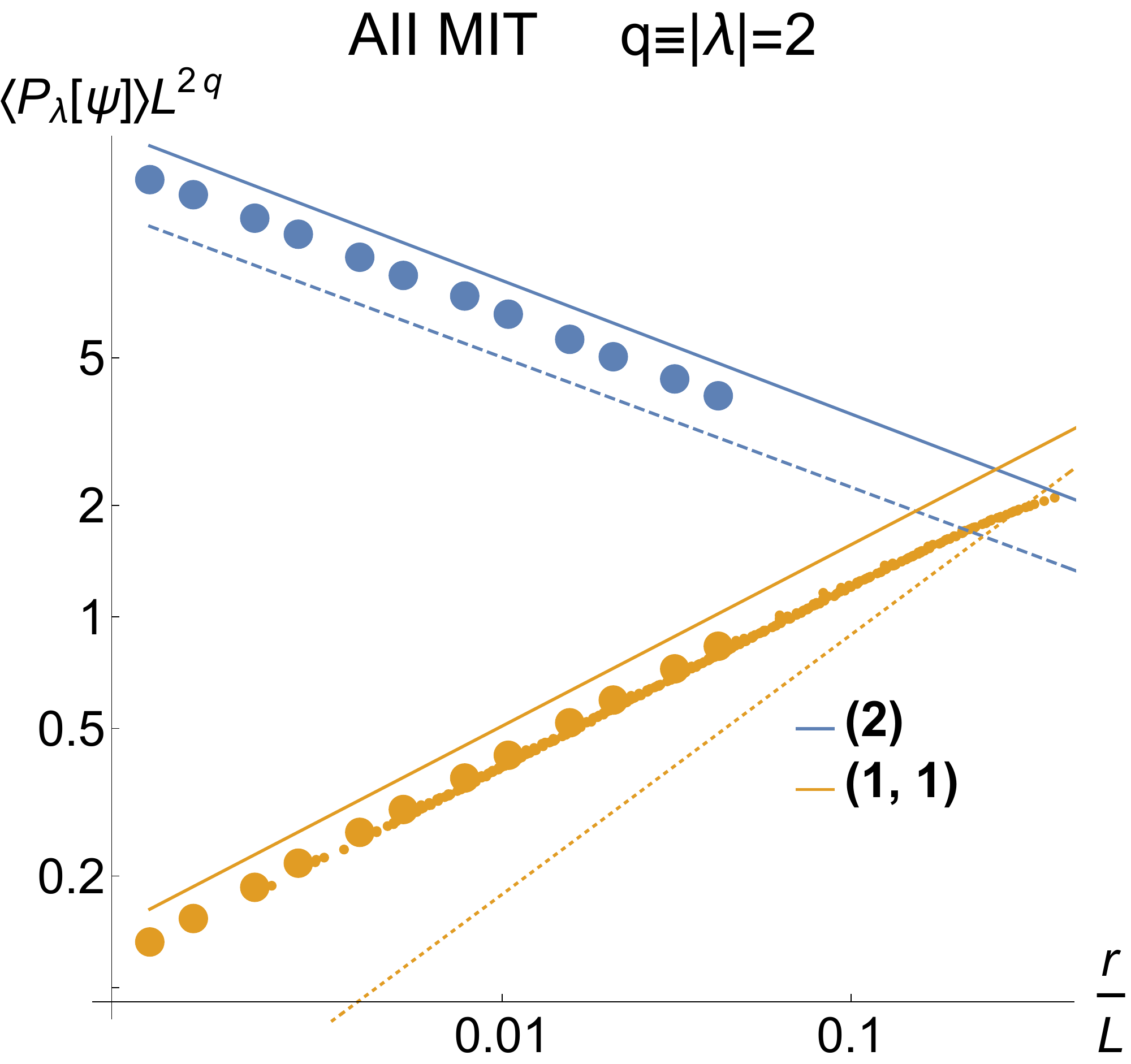}
	\includegraphics[width =.46\textwidth]{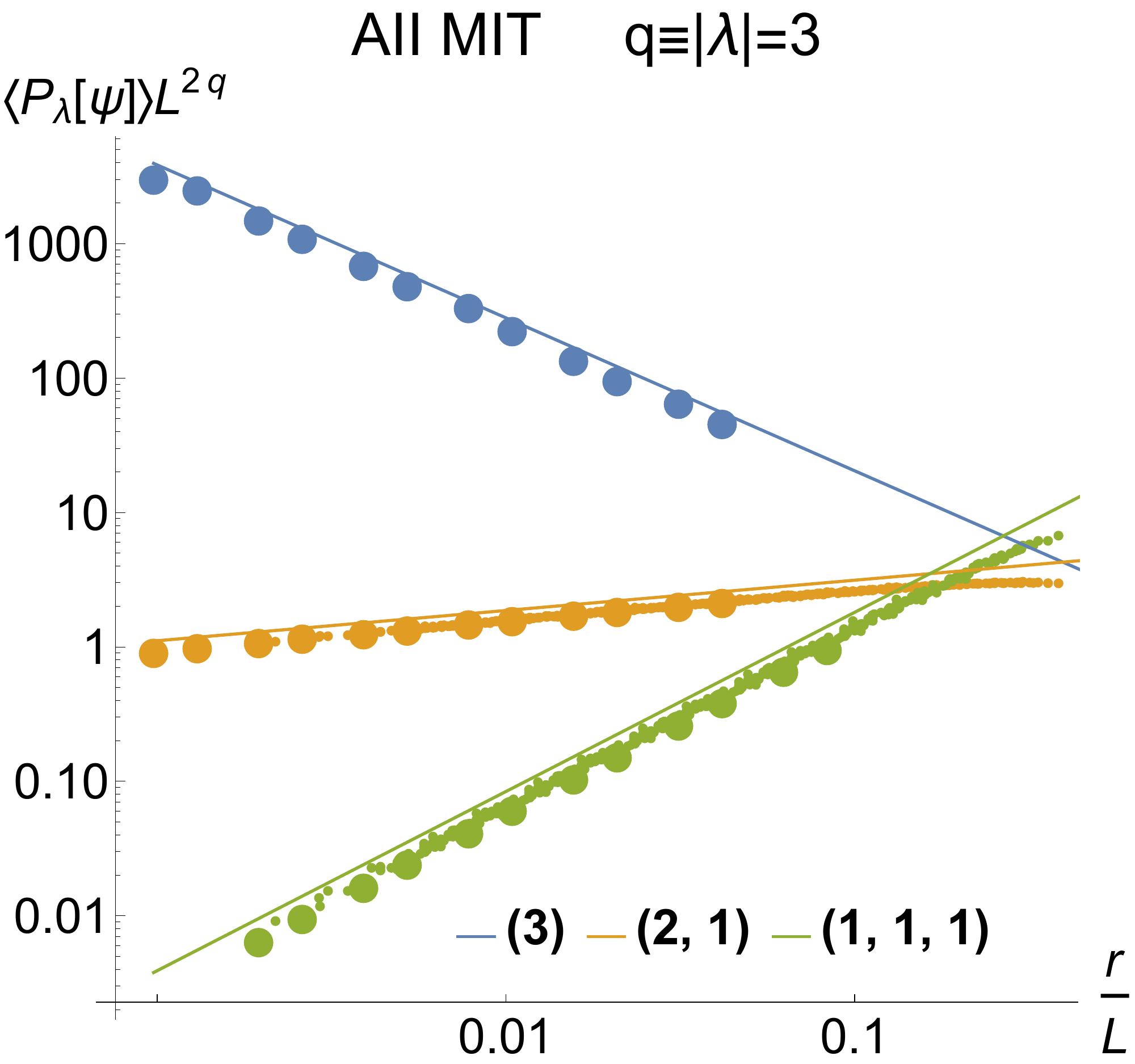}
	\includegraphics[width =.46\textwidth]{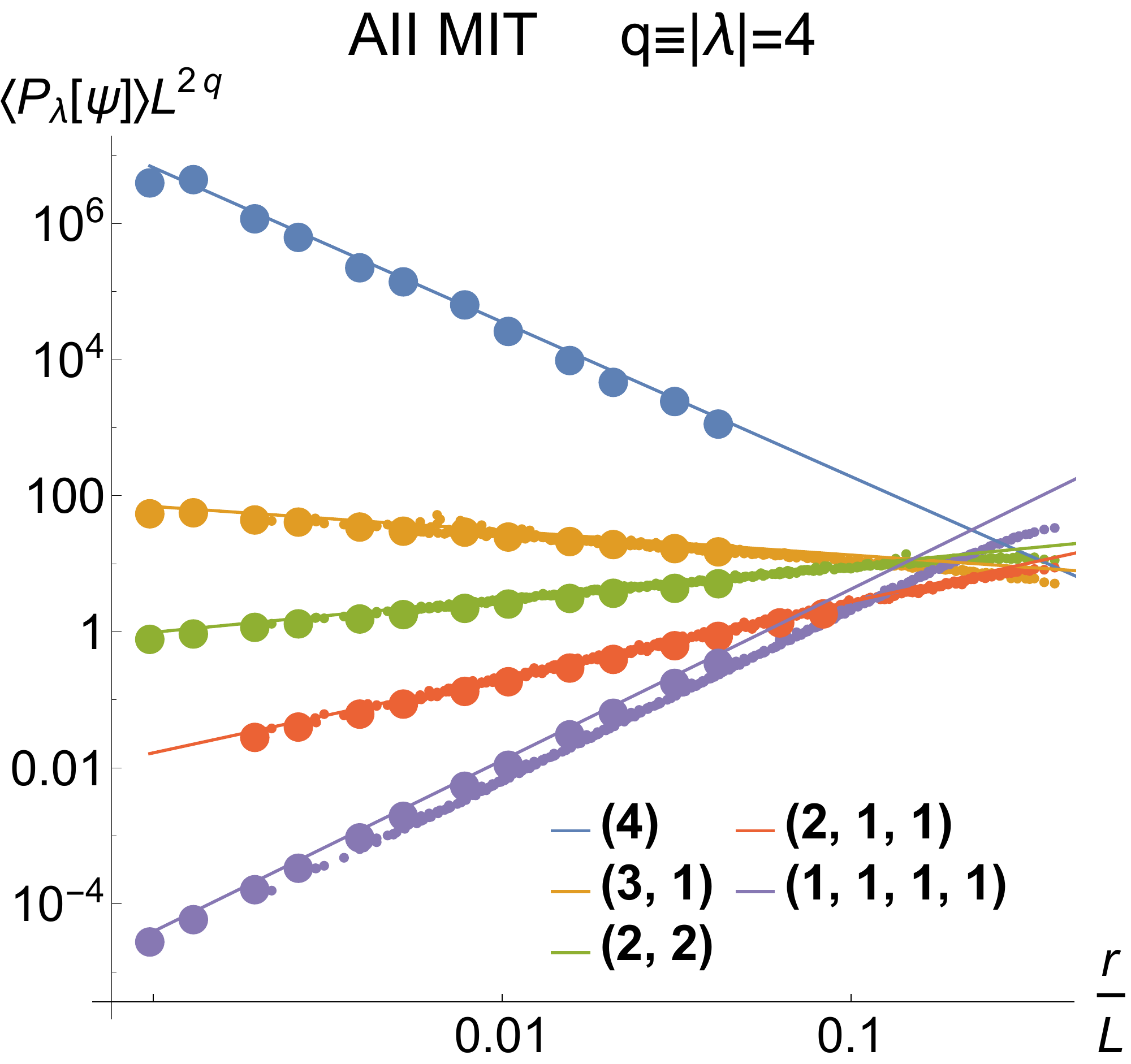}
	\includegraphics[width =.46\textwidth]{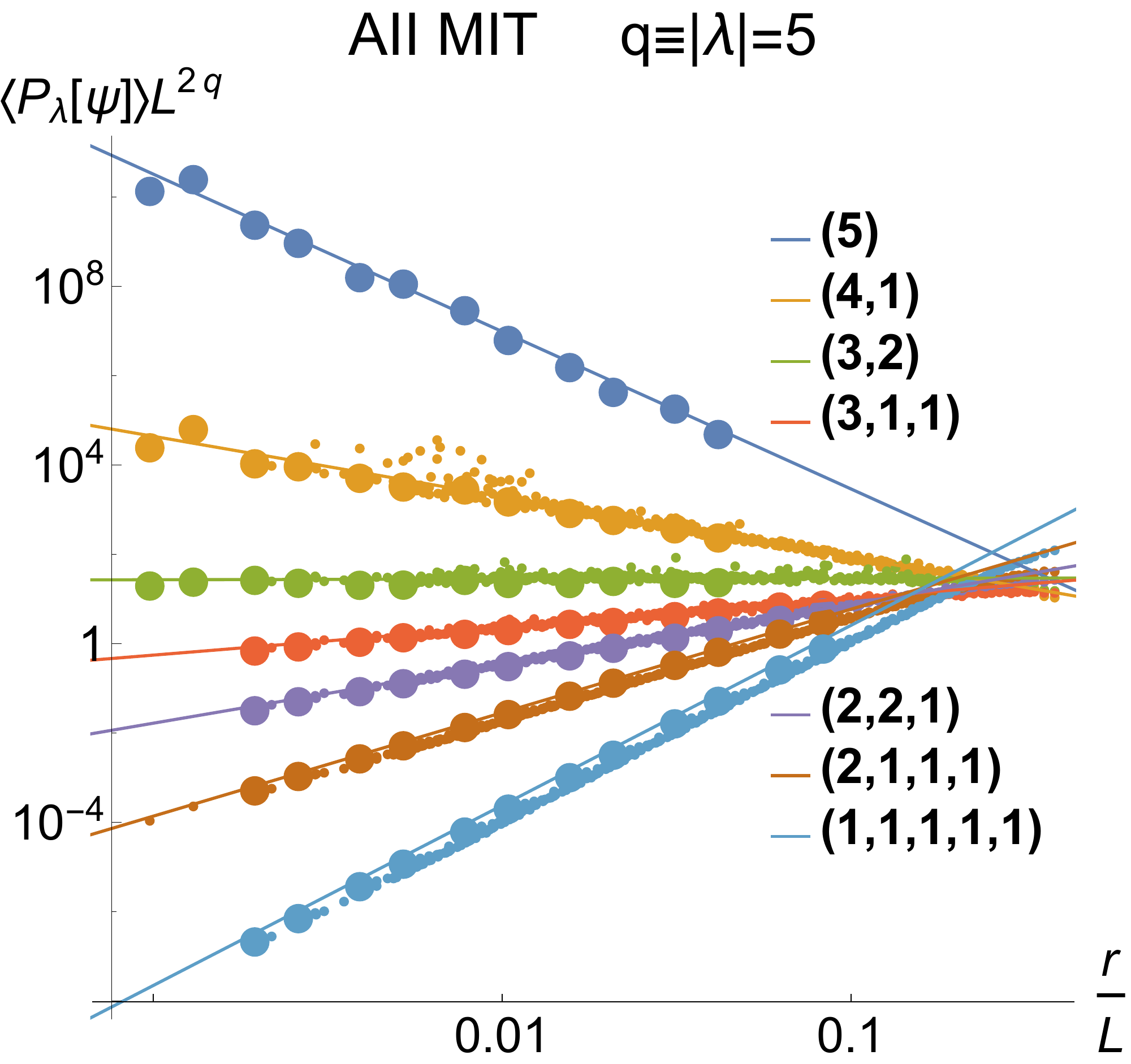}
	\caption{Numerical determination of the generalized multifractality at the metal-insulator transition in class AII
		[Ando model~\eqref{eq:ham_ando}, energy $\epsilon = 0$, critical disorder $W=5.84$], for $q=2$ (top left), $q=3$ (top right), $q=4$ (bottom left) and $q=5$ (bottom right) eigenstate observables. Numerical data are shown in the same way as in Fig.~\ref{fig:AII-metal}.
		The obtained exponents are shown in Table~\ref{tab:lAII} in the column $x^{\rm MIT}_\lambda$. In the $q=2$ panel, the yellow dotted line shows the generalized-parabolicity value $x_{(1,1)}$, with the prefactor $b$ fixed according to $b = 0.173$.  A large difference between the slope of this line and that of the full yellow line (actual value of $x_{(1,1)}$) demonstrates strong violation of the generalized parabolicity at the metal-insulator transition; see also Table~\ref{tab:lAII} and Fig.~\ref{fig:AII-sym}.}
	\label{fig:AII-MIT}
\end{figure*}

We turn now to the numerical study of the generalized multifractality at the critical point of the metal-insulator transition in the Ando model, $W = 5.84$. For an illustration of a spatial structure of the spinful Slater
determinants from Eq. (21), see Fig. \ref{fig:ill} which shows the scaling
combinations $L^2 P_\lambda[\psi]$ for $\lambda=(1)$,
$\lambda=(1/2,1/2)$, and $\lambda=(1/3,1/3,1/3)$, computed for a
fixed disorder realization at the metal-insulator transition in
class AII. The results for the scaling of the polynomial observables with $q \le 5$ are shown in Fig.~\ref{fig:AII-MIT} (in the same way as the data for the metallic phase were plotted in Fig.~\ref{fig:AII-metal}, see Sec.~\ref{sec:AII-metallic}). We see again that the construction~\eqref{eq:abelianA},~\eqref{eq:det-sp} yields correctly the pure-scaling observables. For every $\lambda$, the data show nice straight lines on the log-log scale, i.e., exhibit a power-law scaling with $L$, as expected at criticality.  Obtained values of the scaling exponents are presented in Table~\ref{tab:lAII} in the column $x_\lambda^{\rm MIT}$.

Analyzing the results for the exponents $x_\lambda^{\rm MIT}$ in Table~\ref{tab:lAII}, we first observe that the Weyl-symmetry relations~\eqref{eq:weyl_aii} are nicely satisfied.
It is worth emphasizing that the exact Weyl symmetry at the class-AII metal-insulator transition is rather non-trivial. As we have already pointed out in Sec.~\ref{sec:intro}, the Weyl symmetry has in general a character of a hidden symmetry.  Furthermore, the conductance renormalization in a 2D system of class AII is essentially affected by vortices associated with the homotopy group $\pi_1(G/K) = \mathbb{Z}_2$~\cite{koenig2012metal-insulator,fu2012topology}. Moreover, it was shown that these vortices are crucial for establishing localization~\cite{fu2012topology}. One could thus ask whether the vortices might invalidate the proof of the Weyl symmetry based on the $\sigma$-model. The crucial point is that the symmetry of the $\sigma$-model remains intact under RG also in the presence of vortices, so that the classification of the observables and the Weyl-symmetry relations hold also at the strong-coupling fixed point of the Anderson transition. The $G$-invariance of the RG transformation leads to the Weyl symmetry for critical exponents~\cite{gruzberg2013classification}. Numerical verification of the Weyl-symmetry relations, as carried out here, thus constitutes an important confirmation of the validity of our analytical understanding (based on the $\sigma$-model field theory) of the Anderson-transition physics.

Having verified that the Weyl symmetry holds at the critical point, we inspect whether the exponents $x_\lambda$ satisfy the generalized parabolicity~\eqref{eq:xqaii}.
A quick inspection of Table~\ref{tab:lAII} tells us that this is clearly not the case. In the column $x_\lambda^{\rm MIT}$ we present the corresponding ratio, with $b=0.173$ chosen in such a way that the parabolic approximation~\eqref{eq:xqaii} optimally describes the exponent $x_{(q)}$ in the range $0 \le q \le 1.5$ (see below discussion of these data).
Comparing entries in this column with those in the column $x_\lambda^{\rm para}$ corresponding to a parabolic spectrum, we see a strong (reaching a factor of two) violation of the generalized  parabolicity. As an example, one can look at $\lambda= (1,1)$, in which case the numerical result is $x_{(1,1)}^{\rm MIT} / b = 2.8$, which should be compared with the value $x_{(1,1)}^{\rm para} = 4b$ for a generalized parabolic spectrum.
This is also illustrated by a dotted yellow line in the $q=2$ panel of Fig.~\ref{fig:AII-MIT}.

\begin{figure}
	\centering
	\includegraphics[width =.48\textwidth]{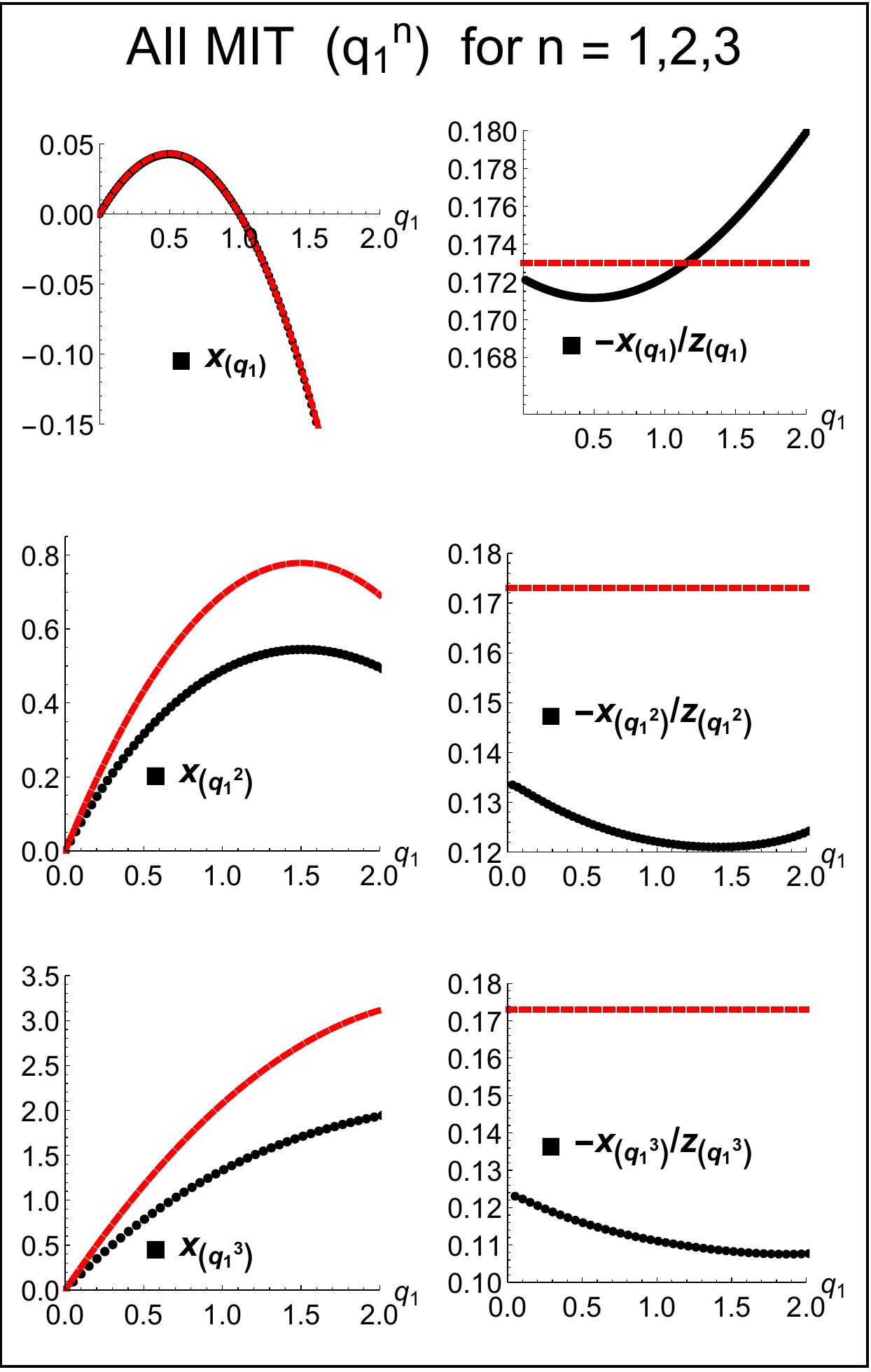}
	\caption{Exponents $x_{(q_1^n)}$ with $n=1, 2, 3$ for the metal-insulator transition in class AII (same parameters as in Fig.~\ref{fig:AII-MIT}).
		Data points in the left panels are  numerically obtained exponents $x_\lambda$. In the right panels, the same data are shown in the form $- x_\lambda / z_\lambda$, where $z_\lambda$ is the quadratic Casimir invariant.  The red dashed lines in all panels correspond to the generalized parabolicity~\eqref{eq:xqaii} with $b=0.173$. 	 It is seen that the generalized parabolicity is strongly violated.  }
	\label{fig:AII-sym}
\end{figure}

The strong violation of the generalized parabolicity is also evident from  Fig.~\ref{fig:AII-sym}, where we show the spectra $x_\lambda$ for $\lambda = (q_1^n)$ with $q_1 = q/n$ and and $n =1,2,3$. (The presentation of data in this figure is analogous to Fig.~\ref{fig:AII-metal-2} for the metallic phase.)  For a generalized parabolic spectrum, we would have $-x_\lambda / z_\lambda = b$, i.e., the same constant in all three right panels of this figure. In each of these panels, we see sizeable deviations of $-x_\lambda / z_\lambda$ from a constant, signalling a modest violation of parabolicity of $x_\lambda$.  For $n=1$, these results are in agreement with Refs.~\cite{Mildenberger-Wave-2007, Obuse-Multifractality-2007}. Comparing results for different $n$, we see that the violation  of the generalized parabolicity is in fact strong: $- x_\lambda / z_\lambda$ increases above 0.18 for $n=1$ and drops below $0.11$ for $n=3$. This strong violation is also evident in left panels of Fig.~\ref{fig:AII-sym}, where red dashed lines correspond to Eq.~\eqref{eq:xqaii} with $b=0.173$.

We note that the properties of the generalized multifractality at the metal-insulator transition in class AII studied here are largely analogous to those
at the SQH transition (class C) explored in our recent works~\cite{karcher2021generalized, karcher2022generalized}: the Weyl symmetry nicely holds, and the generalized parabolicity is strongly violated.  As emphasized above, the former property confirms the validity of the $\sigma$-model approach to the problem, while the latter one implies a violation of the local conformal invariance.

\section{Class D}
\label{sec:D}

\subsection{Model and generalities}
\label{sec:D-model-gener}

Superconductors without further symmetries are in the Bogolyubov-de-Gennes class D with particle-hole symmetry $P$ satisfying $P^2=1$.  At variance with class AII discussed above in Sec.~\ref{sec:AII}, the beta-function has a two-loop correction to the leading (one-loop) term:
\begin{align}
\dfrac{\mathrm{d} \ln t}{\mathrm{d} \ln \ell } = - t + 2t^2 + O (t^3),
\label{eq:D-beta}
\end{align}
with $t= 1/\pi g$. Thus, in the metallic phase of class D, one expects larger deviations from the asymptotic one-loop formulas of Sec.~\ref{sec:metal_scaling} than in class AII.
The one-loop zeta-functions for scaling dimensions of the operators read  [see Eq.~\eqref{eq:one-loop-RG-C-lambda}]
\begin{align}
\dfrac{\mathrm{d} \ln \mathcal{C}_\lambda}{\mathrm{d} \ln \ell }=  z_\lambda t + O (t^2).
\label{eq:D-zeta}
\end{align}
Since class D is not of Wigner-Dyson type, the average density of states is critical, i.e., $x_{(1)} \neq 0$. To one-loop order we have $x_{(1)} = t + O (t^2)$
[see Eq.~\eqref{eq:RG-thermal-metal-energy}], which is Eq.~\eqref{eq:D-zeta} for $\lambda = (1)$. To convert the exponents $\Delta_\lambda$ characterizing the eigenfunction observables into $x_\lambda$, we will need the numerical value of $x_{(1)}$, see Eq.~\eqref{eq:Deltaq}, that should be determined separately. This substantially reduces the accuracy of numerical determination of the exponents $x_\lambda$ in class D (and also in class DIII studied below) in comparison with class AII.

In our numerical analysis, we use the Cho-Fisher model~\cite{cho1997criticality}, which is defined as a network model of the Chalker-Coddington type \cite{chalker1988percolation, kramer2005random} with the following distribution of scattering angles $\alpha_i$ at nodes (labeled by $i$):
\begin{align}
\mathcal{P}(\alpha_i) &= (1-p)\delta(\alpha_i-\alpha)+\frac12p\,\delta(\alpha_i+\alpha) \nonumber \\
& +\frac12p\,\delta(\alpha_i+\alpha-\pi) \label{eq:dcf}.
\end{align}
The model is characterized by two parameters, $\alpha$ and $p$. For $p=0$, the network is fully regular, with the scattering angle $\alpha$. At every node, the particle turns right or left with the amplitudes $\pm \cos \alpha$ or $\pm \sin \alpha$, respectively. The parameter $p$ is the concentration of defects inserted at some nodes of the network. A defect corresponds to the change of the scattering angle $\alpha$ at a node $i$ to $-\alpha$ or to $\pi-\alpha$, with equal probabilities. This changes signs of either both $\cos \alpha$ or both $\sin \alpha$ associated with the node $i$ and thus can be viewed as an insertion of a pair of vortices into two plaquettes meeting at the node $i$ and belonging to the same sublattice (either cosine or sine). The phase diagram of the Cho-Fisher model and the behavior of the density of states were studied in Refs.~\cite{chalker2001thermal, mildenberger2007density}. The phase diagram contains two topologically distinct insulating phases and a metallic phase, with metal-insulator transition lines in the $\alpha$-$p$ plane separating the metallic phase from each of the insulating phases.  The model is dual under the exchange $p\leftrightarrow 1-p$, with $p=1/2$ being the maximal disorder. Here, we will use the Cho-Fisher model to study the generalized multifractality at this metal-insulator transition. Specifically, we will focus on the point $\sin^2\alpha = 0.19$, $p=0.19$, which is known to belong to the metal-insulator transition line~\cite{chalker2001thermal}.

We also study numerically the generalized multifractality in the thermal-metal phase of class D.  One way to do this is to consider the metallic phase of the Cho-Fisher model. A slightly different way is to use the so-called $\mathrm{O}(1)$ model, in which the sign disorder is randomly distributed over the links of the network~\cite{chalker2001thermal}. A defect in the $\mathrm{O}(1)$ model thus inserts two vortices into the two plaquettes bordering the corresponding link (and thus belonging to different sublattices). For the maximal disorder, $p= 1/2$, the $\mathrm{O}(1)$ model and the Cho-Fisher model were found to be equivalent~\cite{chalker2001thermal}. (In fact, there is a subtle difference related to the random-matrix-theory (RMT) behavior at the lowest energies, see below.) It is known that the $\mathrm{O}(1)$ model exhibits only the metallic phase since this type of disorder suppresses jumps between the components of the $\sigma$-model manifold and thus prohibits localization~\cite{read2000paired, bocquet2000disordered, chalker2001thermal}. Similar effect occurs in one-dimensional or quasi-one-dimensional wires in class D studied within the scattering theory formalism in Refs.~\cite{Brouwer-Localization-2000, chalker2001thermal, Gruzberg-Localization-2005}. In this formalism, the presence of uncorrelated random $\pi$-fluxes (sign changes of the transfer matrix) in a wire prevents localization~\cite{chalker2001thermal, Gruzberg-Localization-2005}.

As discussed in Sec.~\ref{sec:weyl}, the Weyl symmetry is expected to be violated at the metal-insulator transition of class D because of jumps (domain walls) between the connected components of the $\sigma$-model. At the same time, the Weyl symmetry is expected to be restored in the metallic phase where such jumps are suppressed (approximately in the Cho-Fisher model and exactly in the $\mathrm{O}(1)$ model), see Sec.~\ref{sec:D-metallic}.

We show below that the construction~\eqref{eq:abelianA},~\eqref{eq:det} correctly yields the scaling observables and determine the exponents in the thermal-metal phase (Sec.~\ref{sec:D-metallic}) and at the metal-insulator transition (Sec.~\ref{sec:D-MIT}). In the metallic phase, we find that the Weyl symmetry and the single-parameter generalized parabolicity~\eqref{eq:xq} hold with a good accuracy. On the other hand, at the metal-insulator transition, both the Weyl symmetry and the weak form~\eqref{eq:Deltaq_CFT} of the generalized parabolicity are strongly violated. The non-parabolic character of the spectrum implies that the local conformal invariance does not hold at the class-D metal-insulator transition.

\subsection{Metallic phase}
\label{sec:D-metallic}

\begin{figure}
	\centering
	\includegraphics[width=0.45\textwidth]{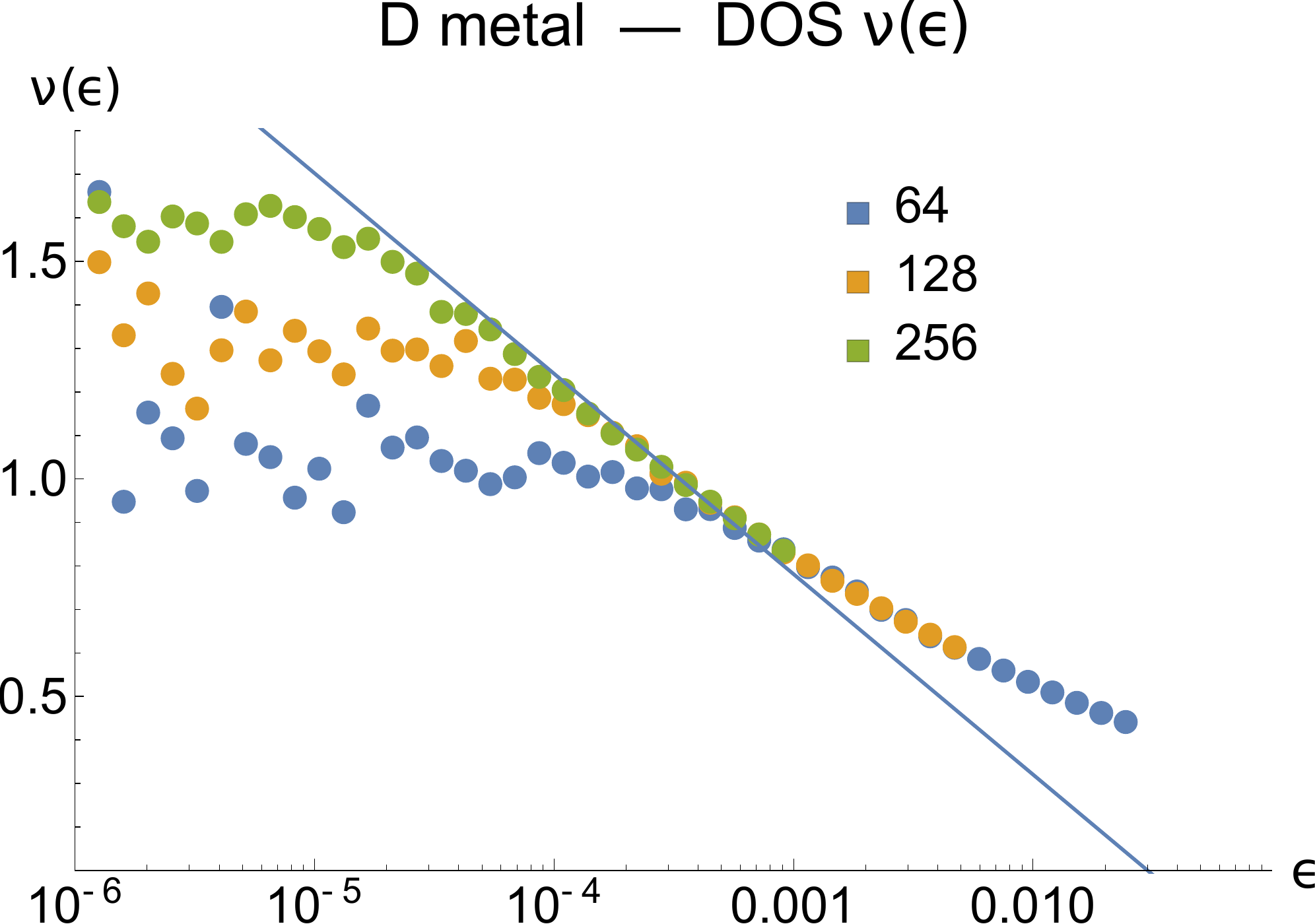}
	\caption{Average density of states $\bar{\nu}(\epsilon)$ in the thermal-metal phase of class D [O(1) network model, with parameters $\alpha=\pi/4$ and $p = 1/2$]. The low-energy behavior exhibited by $\bar{\nu}(\epsilon)$ is logarithmic, in agreement with the analytical prediction ~\eqref{eq:one-loop-renorm-DOS-e-D}. At the lowest energies, below the level spacing at zero energy, a saturation of the density of states is observed, which is the RMT behavior corresponding to class BD.
	}
	\label{fig:D-metal-DOS}
\end{figure}

\begin{figure*}
	\centering
	\includegraphics[width=0.4\textwidth]{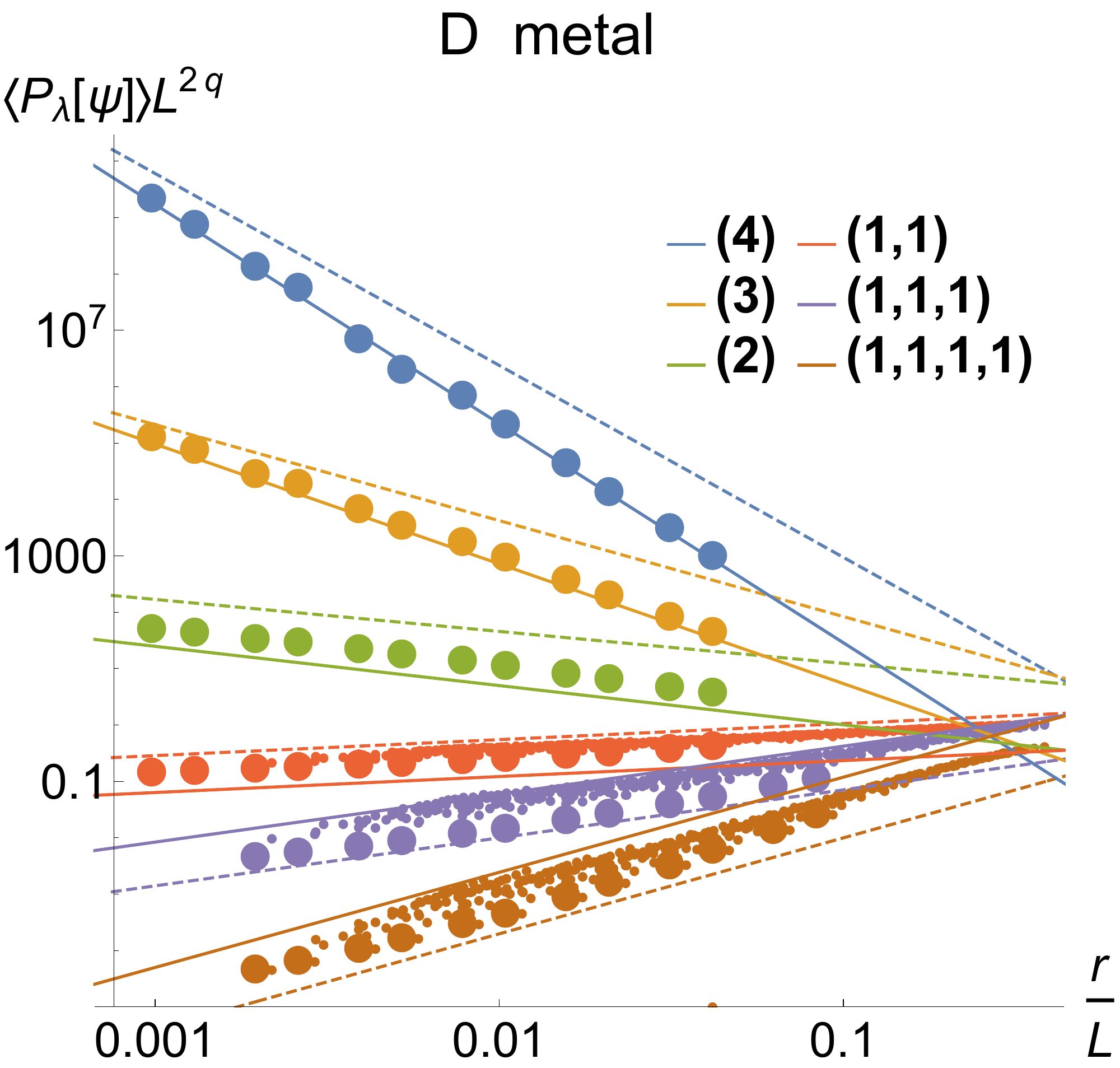} \hspace{0.05\textwidth}
	\includegraphics[width=0.48\textwidth]{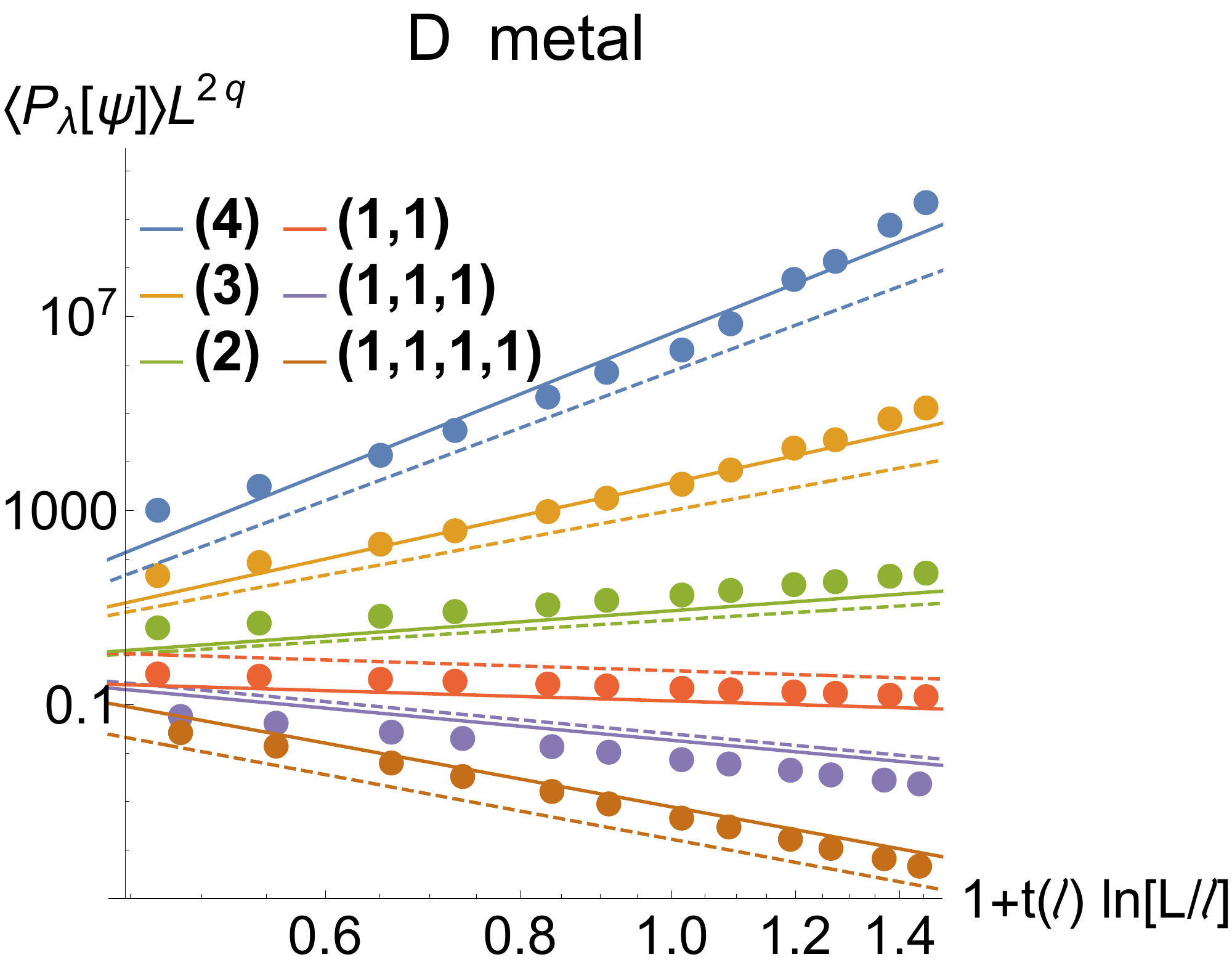}
	\caption{
		Numerical determination of generalized multifractality in the class-D metallic phase [O(1) model]. The spinless pure-scaling combinations~\eqref{eq:abelianA},~\eqref{eq:det} are computed, with averaging over the system area and over $10^4$ disorder realizations. \textit{Left panel:} Data shown on a log-log scale, with straight lines corresponding to power-law scaling~\eqref{eq:Plambda-scaling-all-r-Delta}. The corresponding slopes yield the running exponents~\eqref{eq:metal-running-exp}.
		The data are scaled with $r^{\Delta_{q_1}+ \ldots + \Delta_{q_n}}$, which yields an expected collapse as functions of $r/L$. For each $\lambda$, data points corresponding to the smallest $r \sim 1$ are highlighted as large dots, visualizing the $L$-dependence at a fixed $r$. The full lines are fits to these data points; the resulting exponents $\Delta^{\rm metal}_\lambda$ are given in Table~\ref{tab:lD}. The dashed lines corresponds to the generalized parabolic spectrum~\eqref{eq:xq} with $b=0.283$; see the column $\Delta^{\rm para}_\lambda$ in  Table~\ref{tab:lD}.   The slopes of full and dashed lines are close for all $\lambda$, which means that the generalized parabolicity is a good approximation in the metallic phase. At the same time, there are pronounced deviations for $\lambda= (3)$ and especially for $\lambda= (4)$.
		\textit{Right panel:} Same data as in the left panel (only with the smallest $r \sim 1$) plotted as function of $1+t(\ell) \ln(L/\ell)$
		[with $\ell =180$ and $t(\ell) = 0.263$] on the log-log scale,  according to the asymptotic scaling form~\eqref{eq:metal-scaling-P-lambda-L_2} in the metallic phase.
	}
	\label{fig:D-metal-2}
\end{figure*}

We study numerically the polynomial observables up to the order $q=4$. The Weyl symmetry, which is expected to hold in the metallic phase of class D, implies a number of relations between the corresponding scaling exponents:
\begin{align}
x_{(1)} &= x_{(1,1)}, & x_{(1,1,1)} &= 0,\nonumber \\
x_{(2,1)} &= x_{(2)}, & x_{(3,1)} &= x_{(3)}.
\label{eq:weyl_d}
\end{align}
The generalized parabolic spectrum~\eqref{eq:xq} (which combines parabolicity with the Weyl symmetry) has in class D the following form:
\begin{align}
x_{(q_1,q_2,\ldots)}^{\rm para}&= b\left[ -q_1^2 + q_2(1-q_2) + q_3(2-q_3) +\ldots\right].
\label{eq:xqd}
\end{align}
In analogy with class AII, we will also study numerically the exponents $x_\lambda$ for $\lambda = (q_1^n)$ with $q_1 = q/n$ and $n =1,2,3$. The Weyl symmetry for such exponents reads
\begin{align}
& x_{(q_1)} = x_{(- q_1)}, \quad x_{(q_1, q_1)} = x_{(\frac12 - q_1, \frac12-q_1)}, \nonumber \\
& x_{(q_1, q_1, q_1)} = x_{(1- q_1, 1-q_1, 1-q_1)}.
\label{eq:weyl-D-q1n}
\end{align}
As for class AII, we take the required number (up to four) of eigenstates closest to criticality, evaluate the observables defined
by Eqs.~\eqref{eq:abelianA},~\eqref{eq:det}, and perform the averaging over all points in the sample ($\sim L^2$) and over $10^4$ realizations of disorder.  The system size is varied from $L=24$ to $L = 1024$.

In Fig.~\ref{fig:D-metal-DOS}, we show numerical results for the average density of states $\bar{\nu}(\epsilon)$. We observe a logarithmic increase of the density of states at low energies, in agreement with the  analytical prediction~\eqref{eq:one-loop-renorm-DOS-e-D} and with previous numerical studies~\cite{mildenberger2007density, wang_multicriticality_2021}.

{At the lowest energies, i.e., on the scale set by the level spacing at zero energy in a system of given spatial size, we observe a saturation of the density of states. In the previous work~\cite{mildenberger2007density}, where the Cho-Fisher model was studied, it was found that the density of states shows, in the thermal-metal phase, an oscillatory behavior in this range of energies, as predicted by the RMT of class D~\cite{altland1997nonstandard, bocquet2000disordered}. A difference is that, in the case of O(1) model that we consider, the number of O(1) defects can be either even or odd.  If this number is odd, the determinant of the scattering matrix defined by the network is $-1$ rather than $+1$, and one of its eigenvalues is strictly unity. At the level of a Hamiltonian, this corresponds to a Majorana zero mode. From the RMT point of view, the system is said to belong to class D for an even number of defects and to class B for an odd number. As these two classes of network realizations have equal probabilities, the averaged density of states is an arithmetic mean of those for classes D and B. In such a class BD, the RMT density of states has a constant behavior: the oscillations characteristic for classes D and B  exactly cancel each other~\cite{bocquet2000disordered, ivanov2002supersymmetric}. }

\begin{table*}
	\centering
	\begin{tabular}{cc||c|ccc||c}
		& rep. $\lambda$ & $\Delta_\lambda^{\rm MIT}$ &  $\Delta_\lambda^{\rm metal}$ & $ \displaystyle \frac{\Delta_\lambda^{\rm metal}}{b}$ & $\tilde{\Delta}_\lambda^{\rm metal}$ & $\Delta^{\rm para}_\lambda $\\[5pt]
		\hline
		\hline
		&&&&&&\\[-5pt]
		$q=2$ & (2) & $-1.546\pm 0.004$ &  $-0.695\pm 0.04$ & $-2.45\pm 0.14$ & $-2.34\pm 0.01$ & $-2b$\\[3pt]
		& (1,1) & $0.44\pm 0.02$ &    $0.278\pm 0.005$    &$0.98\pm 0.02$ & $0.955\pm 0.003$ & $b$
		\\[5pt]
		\hline
		&&&&&&\\[-5pt]
		$q=3$ & (3)& $-3.55\pm 0.09$ &   $-2.11\pm 0.03$ & $-7.44\pm 0.11$ & $-7.04\pm 0.10$ & $-6b$\\[3pt]
		& (2,1)& $-0.71\pm 0.10$ &  $-0.45\pm 0.01$ & $-1.57\pm 0.04$ & $-1.49\pm 0.04$ & $-b$\\[3pt]
		& (1,1,1)& $1.18\pm 0.03$ &  $0.852\pm 0.002$  & $3.003\pm 0.007$ & $2.988\pm 0.007$ & $3b$
		\\[5pt]
		\hline
		&&&&&&\\[-5pt]
		$q=4$ & (4) & $-5.56\pm 0.01$ &  $-3.86\pm 0.07$ & $-13.61\pm 0.25$  & $-12.84\pm 0.22$  & $-12b$\\[3pt]
		& (3,1) & $-2.35\pm 0.18$ &  $-1.85\pm 0.03$  & $-6.50\pm 0.11$  & $-6.16\pm 0.10$ & $-5b$\\[3pt]
		& (2,2) & $-1.7\pm 0.5$ &  $-0.83\pm 0.05$ & $-2.93\pm 0.18$  & $-2.79\pm 0.14$ &   $-2b$\\[3pt]
		& (2,1,1) & $0.32\pm 0.04$ &  $0.16\pm 0.02$ & $0.57\pm 0.07$ & $0.58\pm 0.05$ &   $b$\\[3pt]
		& (1,1,1,1) & $2.39\pm 0.03$ &  $1.691\pm 0.004$ & $5.96\pm 0.01$  & $5.93\pm 0.02$ & $6b$
	\end{tabular}
	\caption{Scaling exponents of generalized multifractality in class D for eigenstate observables with $q\equiv |\lambda|\leq 4$.
		The exponents $\Delta_\lambda$ shown in the table are related to the exponents $x_\lambda$ via $\Delta_\lambda = x_\lambda - qx_{(1)}$.
		The exponents $\Delta_\lambda^{\rm MIT}$ are found numerically from the Cho-Fisher model at the transition point, $\sin^2\alpha = 0.19$, $p=0.19$.  The thermal-metal exponents $\Delta_\lambda^{\rm metal}$ are obtained from the O(1) model with $\sin^2\alpha = 0.5$, $p=0.5$.
		The last column displays the exponents $\Delta_\lambda^{\rm para} = b(q- z_\lambda)$ corresponding to the generalized parabolic spectrum, Eq.~\eqref{eq:xq}, with a single parameter $b$.  In the metallic phase, the exponents are reasonably close to the generalized parabolicity, as can be seen from the comparison of the column $\Delta_\lambda^{\rm metal} / b$ (where $b=0.283$, see text) with  $\Delta_\lambda^{\rm para}$. The deviations are, however, quite substantial, which is expected since $b$ is not so small.
		The exponent $x_{(1)}^{\rm metal}$ as obtained from a power-law fit of the density of states is $x_{(1)} \approx -b = -0.283$ as expected.
		The Weyl symmetry relations~\eqref{eq:weyl_d} (that can be easily translated to $\Delta_\lambda$) are approximately satisfied in the metallic phase, which can be seen by inspection of $\Delta_\lambda^{\rm metal} / b$.
		The column $\tilde{\Delta}_\lambda^{\rm metal}$ contains the exponents obtained by a fit of the thermal-metal data to the asymptotic form ~\eqref{eq:metal-scaling-P-lambda-L_3}.
		It is seen that $\tilde{\Delta}_\lambda^{\rm metal}$ is rather close to $\Delta_\lambda^{\rm metal} / b$, indicating that both types of fits are quite similar in the considered range of $L$ in the metallic phase.
		The metal-insulator transition exponents $\Delta_\lambda^{\rm MIT}$, in combination with $x_{(1)}^{\rm MIT} = -0.85$ obtained from the fit of the density of states, strongly violate the Weyl symmetry, which is a manifestation of the effect of topological excitations (domain walls between two connected components) in the $\sigma$ model. }
	\label{tab:lD}
\end{table*}

In Fig.~\ref{fig:D-metal-2} we show the data for $L^{2q} \langle P_\lambda[\psi] \rangle$ with $q=2$, 3, and 4. We do not consider all $\lambda$ in each order but rather restrict ourselves to the most relevant (in the RG sense) observables $(q)$ (that correspond to the conventional multifractality) and the most irrelevant $(1^q)$ that serve as building blocks for all generalized-multifractality observables, Eq.~\eqref{eq:abelianA}. Note that, quite generally, for each order $q$, the statistical fluctuations turn out to be the smallest for $(1^q)$. The left panel presents the conventional log-log plot, so that the slopes yield the exponents $\Delta_\lambda^{\rm metal}$, in analogy with Fig.~\ref{fig:AII-metal} for class AII. As a first key observation, we notice that the numerics confirms that  Eqs.~\eqref{eq:abelianA},~\eqref{eq:det} properly yield pure-scaling observables. In the left panel of Fig.~\ref{fig:D-metal-2}, the data are shown on the log-log scale, with straight lines corresponding to the power-law scaling~\eqref{eq:Plambda-scaling-all-r-Delta}.
The corresponding slopes yield the running exponents~\eqref{eq:metal-running-exp} (i.e., effective exponents for the given range of $L$). We recall that these exponents should slowly (logarithmically) decrease with increasing $L$,  see Sec.~\ref{sec:metal_scaling}. The corresponding curvature is indeed noticeable in the data for $\lambda = (2)$, $(1,1)$, $(1,1,1)$, and $(1,1,1,1)$, which are least affected by statistical fluctuations. The change of the slope in the considered range of $L$ is, however, not so big, so that the power-law fit is meaningful. The corresponding exponents $\Delta_\lambda^{\rm metal}$ are presented in Table~\ref{tab:lD}. To convert $\Delta_\lambda$ into $x_\lambda$, we still need the value of $x_{(1)}$. A nice way to find it is to use the Weyl symmetry relations~\eqref{eq:weyl_d}
(which are expected to hold in the metallic phase of class D, see Sec.~\ref{sec:D-model-gener}). Indeed, we find that, if we set $x_{(1)} = - 0.283$, then all these symmetry relations are well satisfied.  Note that the value $x_{(1)} =  - 0.283$ translates into the value $\kappa = -0.13$ of the density-of-states exponent,   Eq.~\eqref{eq:power_ldos}, which is consistent with the density-of-states behavior, Fig.~\ref{fig:D-metal-DOS} (if one fits it with a power law). At the same time, finding $x_{(1)}$ directly from the data for the density of states is difficult: the corresponding error turns out to be rather large.

The thermal-metal data are not too far from the single-parameter generalized  parabolicity, Eq.~\eqref{eq:xq}. To illustrate this, we include in Table~\ref{tab:lD} the column
$\Delta_\lambda^{\rm metal} / b$,  where $b=0.283$ in consistency with the above value of  $x_{(1)}$.  One sees that the values of $\Delta_\lambda^{\rm metal} / b$ are reasonably close to  $\Delta_\lambda^{\rm para} / b \equiv q - z_\lambda$ (last column) corresponding to the generalized parabolic spectrum~\eqref{eq:xq}. At the same time, deviations from the generalized parabolicity are quite substantial. In particular, they considerably exceed the analogous deviations in the case of the metallic phase in class AII, see Table~\eqref{tab:lAII}. There are two reasons for this. First, in class D there exist two-loop corrections, while in class AII the corrections start from the four-loop order only. Second, the value of the resistance $t$ in the metallic phase of class AII is much smaller: the parameter $b$ was $0.0273$ in that case, while it is equal to $0.283$ for class D.  As a result, the corrections to one-loop formulas (and thus to the generalized parabolicity) turn out to be much larger in class D.

\begin{figure*}
	\centering
	\includegraphics[width=0.45\linewidth]{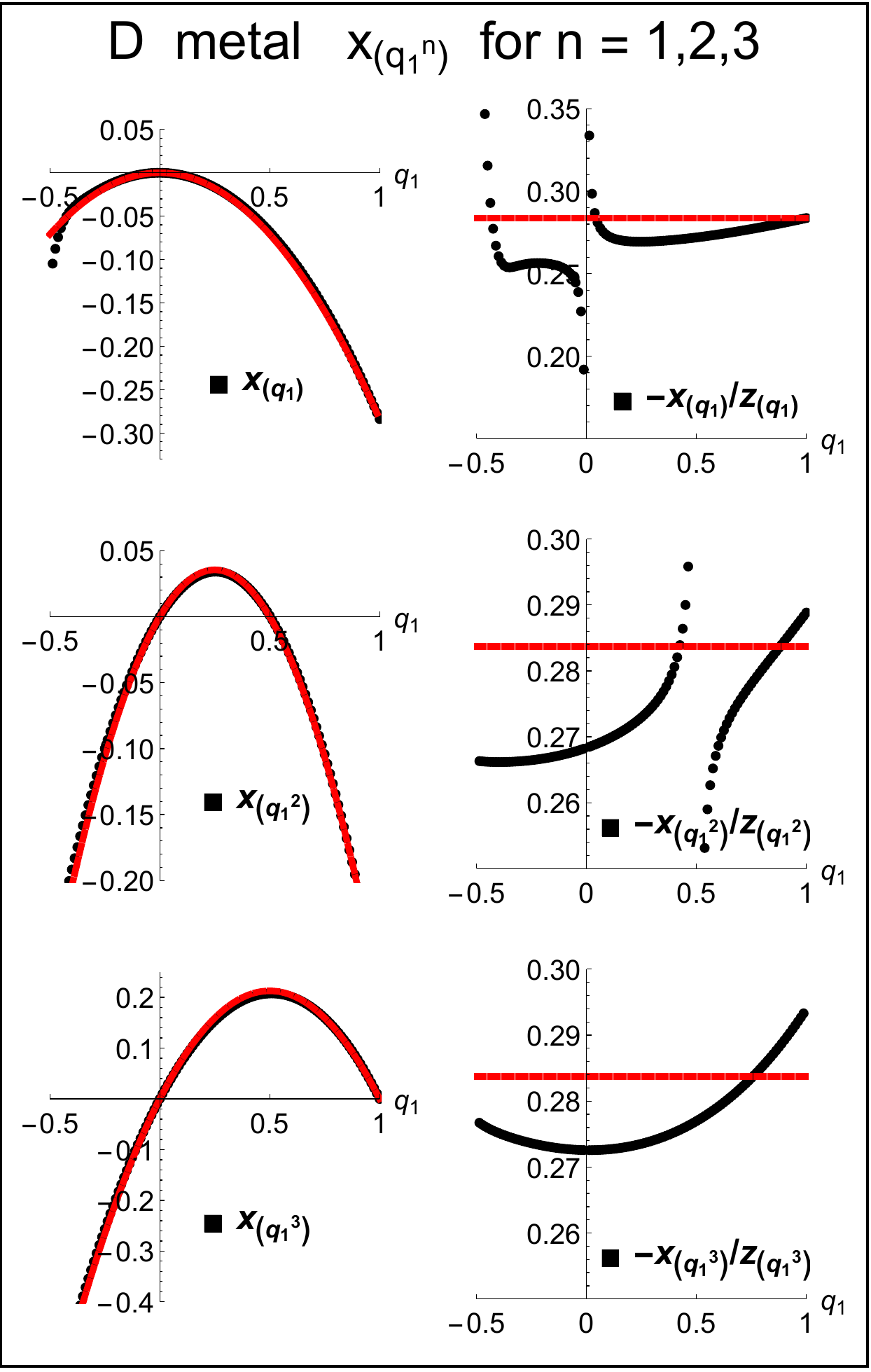}
	\includegraphics[width=0.45\linewidth]{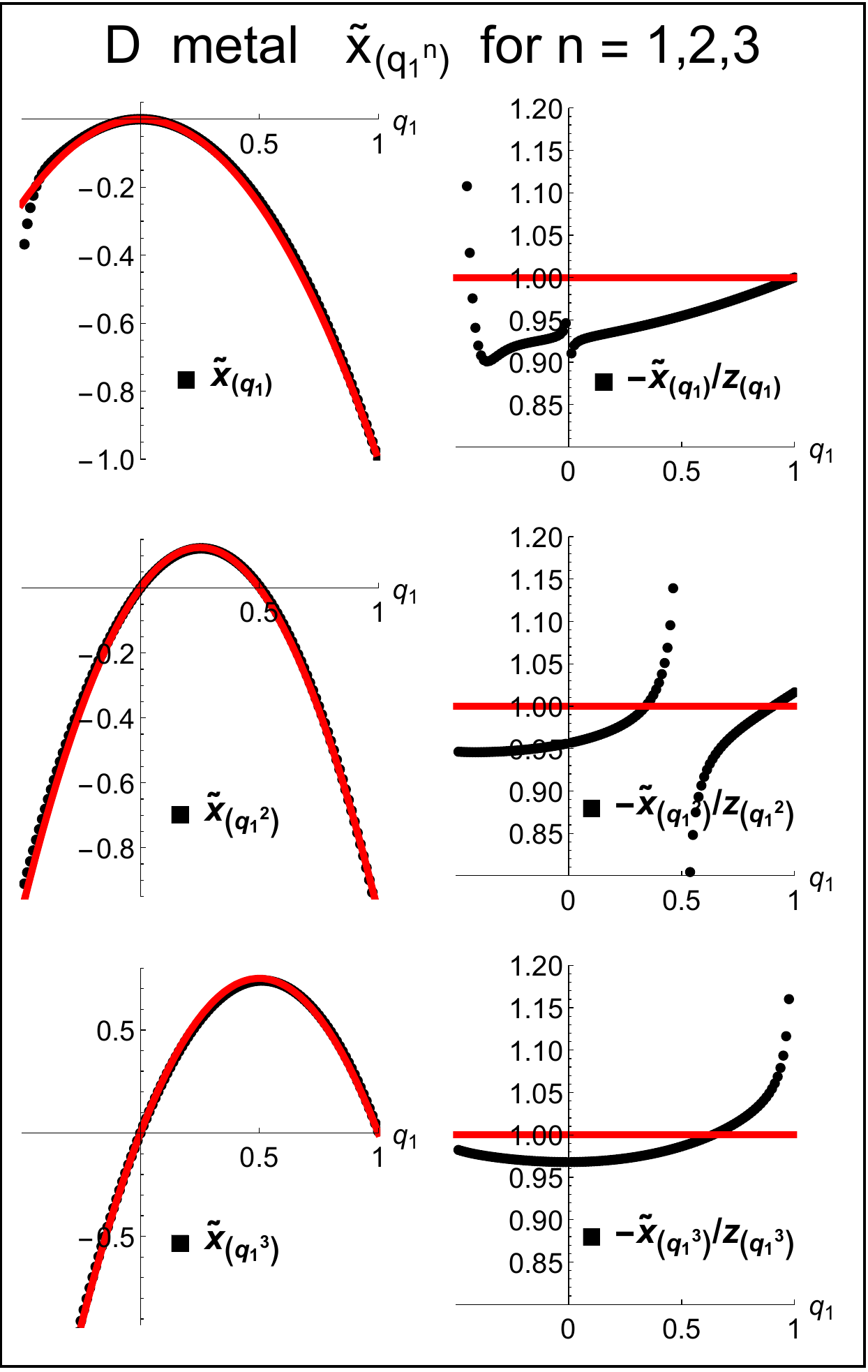}
	\caption{\textit{Left:} Exponents $x_{(q_1^n)}$ with $n=1, 2, 3$ for the metallic phase of class D (same parameters as in Fig.~\ref{fig:D-metal}). Data points in the left panels are  numerically obtained exponents $x_\lambda$. In the respective right panels, the same data are shown in the form $- x_\lambda / z_\lambda$, where $z_\lambda$ is the quadratic Casimir invariant.  The red  lines in all panels correspond to the generalized parabolicity~\eqref{eq:xqd} with $b=0.283$.
		It is seen that the generalized parabolicity holds to a good accuracy for not too large $q_1$.
		\textit{Right:} Same data analyzed according to Eq.~\eqref{eq:metal-scaling-P-lambda-L_3}.  Left panels show the exponents $\tilde{x}_\lambda \equiv \tilde{\Delta}_\lambda - q$, while right panels show the ratio $- \tilde{x}_\lambda / z_\lambda$.
	}
	\label{fig:D-metal}
\end{figure*}

In the right panel of Fig.~\ref{fig:D-metal-2} we show the same data for $L^{2q} \langle P_\lambda[\psi] \rangle$ in an alternative way, corresponding to the analytically predicted asymptotic behavior ~\eqref{eq:metal-scaling-P-lambda-L}. The one-loop formula ~\eqref{eq:metal-scaling-P-lambda-L} can be equivalently written as
\begin{equation}
L^{2q} \langle P_\lambda[\psi] (L) \rangle  \sim \left(1+ t(\ell) \ln \frac{L}{\ell} \right)^{z_\lambda - q },
\label{eq:metal-scaling-P-lambda-L_2}
\end{equation}
where we used the class-D values $\gamma=\alpha=1$ and $z_{(1)}=1$. In the present case of a not so weak disorder, extrapolation of Eq.~\eqref{eq:metal-scaling-P-lambda-L_2}
to the ultraviolet limit, $\ell \to a$, does not work, since the resistance $t$ blows up. We thus choose $\ell$ roughly in the middle of the range of system sizes that we consider. Specifically, we take $\ell = 180$. We now choose $t(\ell)$ to optimize the fit of the data with all $\lambda$ to  Eq.~\eqref{eq:metal-scaling-P-lambda-L_2}, which yields $t(\ell) = 0.263$.  As expected, the obtained value of $t(\ell)$ is close to $b=0.283$ found from the power-law fits.  Now we fix $t(\ell)= 0.263$ and fit the data for each $\lambda$ to the form analogous to Eq.~\eqref{eq:metal-scaling-P-lambda-L_2} but with an exponent  $\tilde{\Delta}_\lambda^{\rm metal}$ left as a fit parameter:
\begin{equation}
L^{2q} \langle P_\lambda[\psi] (L) \rangle  \sim \left(1+ t(\ell) \ln \frac{L}{\ell} \right)^{- \tilde{\Delta}_\lambda^{\rm metal}}.
\label{eq:metal-scaling-P-lambda-L_3}
\end{equation}
The resulting values of $\tilde{\Delta}_\lambda^{\rm metal}$ are also shown in Table~\ref{tab:lD}. One can see that, for all $\lambda$, the values of $\tilde{\Delta}_\lambda^{\rm metal}$ are rather close to $\Delta_\lambda^{\rm metal} / b$. This shows that the two types of fits (power-law and logarithmic) do not differ too much for the considered metallic system in the considered range of $L$.  Deviations of $\tilde{\Delta}_\lambda^{\rm metal}$ from the (integer) values $q - z_\lambda$ (equal to $\Delta_\lambda^{\rm para}$ from the last column of the Table, without factor $b$) is attributed to two-loop (and higher) corrections, as discussed above.

In Fig.~\ref{fig:D-metal}, we display the exponents $x_{(q_1^n)}$ with $n=1, 2, 3$ and continuously changing $q_1$ for the metallic phase of class D (same parameters as in Fig.~\ref{fig:D-metal-2}).  In the left half of the figure, we show exponents obtained by power-law fits analogous to left panel of Fig.~\ref{fig:D-metal-2}. We have converted $\Delta_\lambda$ into $x_\lambda$ by using $x_{(1)} = 0.283$. To the right of each panel with $x_\lambda$ data, we show the same data plotted as $-x_\lambda / z_\lambda$. We also include lines corresponding to the generalized parabolicity $x_\lambda = - b z_\lambda$, Eq.~\eqref{eq:xqd}, with $b= 0.283$.  It is seen that the generalized parabolicity holds to a good accuracy in the range of not too large $q_1$ considered in this figure.

In the right half of the figure, the same analysis is performed by plotting the data in the form~\eqref{eq:metal-scaling-P-lambda-L_3}. We show the corresponding exponents $\tilde{x}_\lambda \equiv \tilde{\Delta}_\lambda - q$ and, to the right of each plot, the ratio $- \tilde{x}_\lambda / z_\lambda$. If the one-loop approximation (yielding the generalized parabolicity) were exact, we would have $- \tilde{x}_\lambda / z_\lambda = 1$. We see again that deviations from the generalized parabolicity are relaitively small in this range of $q_1$.
Both ways of fitting the data (power law of $L$ and  Eq.~\eqref{eq:metal-scaling-P-lambda-L_3}) work rather well.

It is worth commenting on the apparent singularities in the plots for $- x_\lambda / z_\lambda$ and $- \tilde{x}_\lambda / z_\lambda$ near the points where $z_\lambda = 0$ ($x=0$ for $n=1$ and $x = 0.5$ for $n=2$). The Weyl symmetry predicts $x_\lambda$ vanishing in these points such that $- x_\lambda / z_\lambda$ has a finite limiting value (and similarly for $- \tilde{x}_\lambda / z_\lambda$). However, statistical errors in $x_\lambda$ and $ \tilde{x}_\lambda$ unavoidably violate this cancellation, leading to a spurious singularity seen in the figure.

\subsection{Metal-insulator transition}
\label{sec:D-MIT}

\begin{figure}
	\centering
	\includegraphics[width=0.43\textwidth]{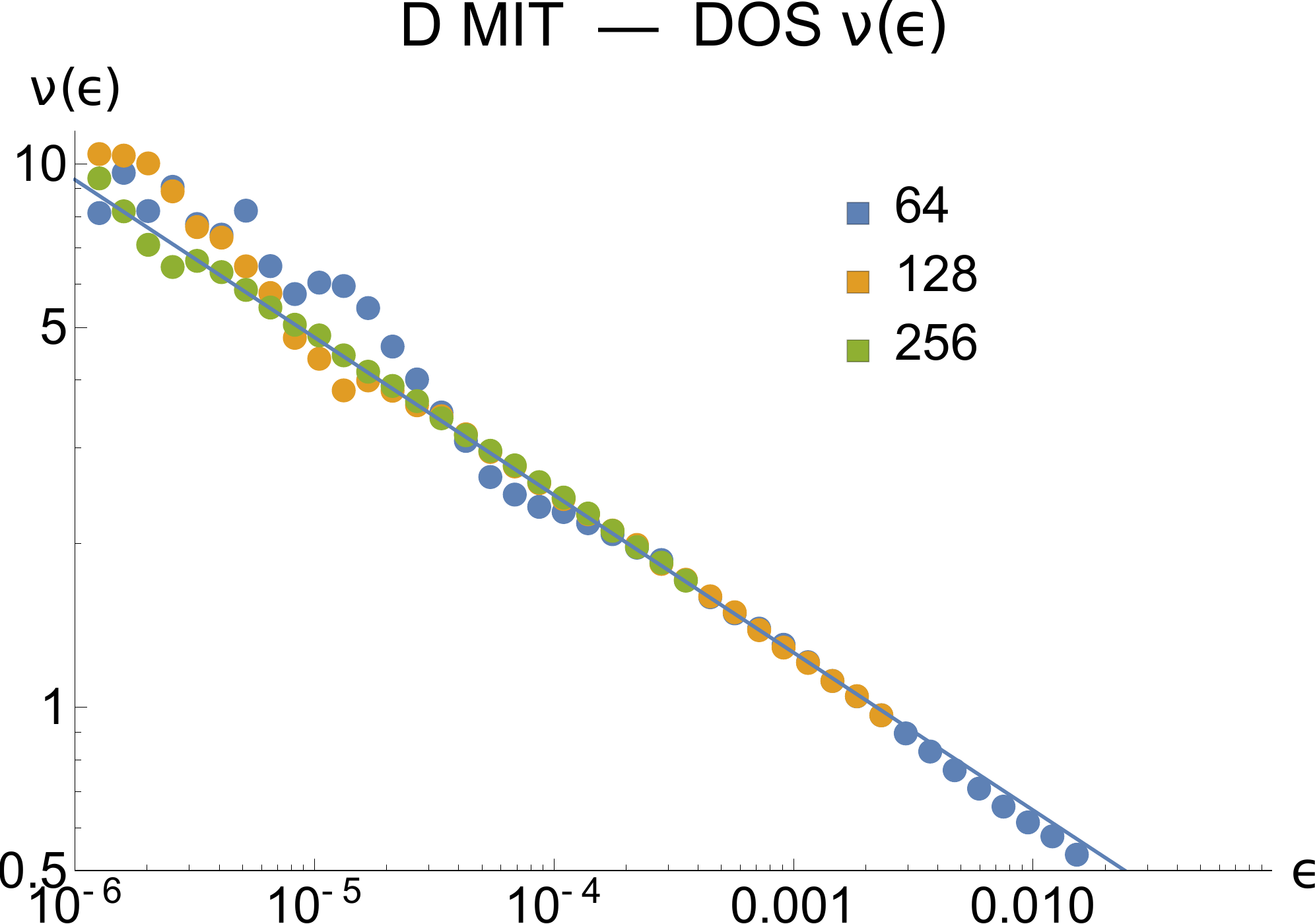}
	\includegraphics[width=0.43\textwidth]{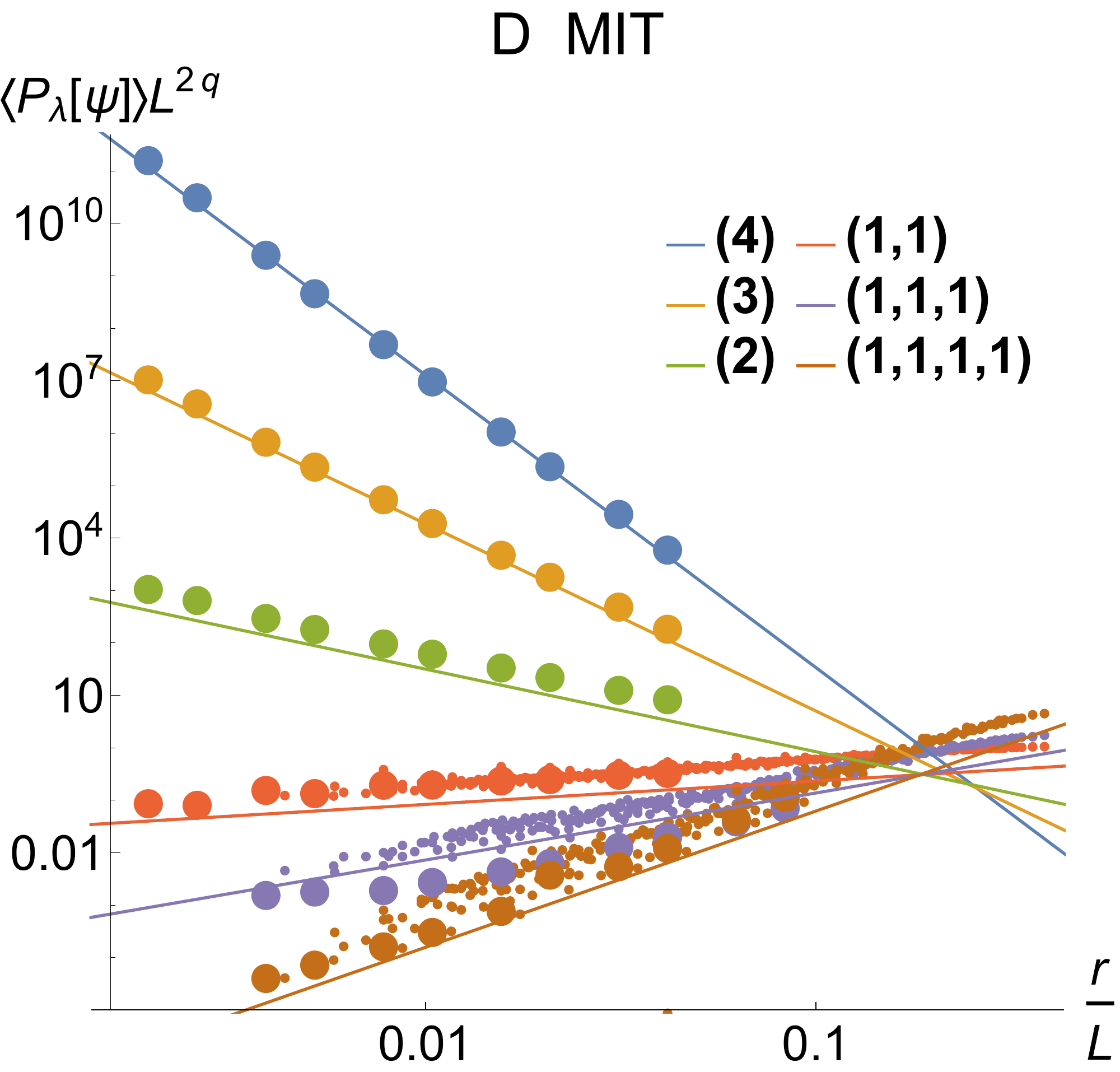}
	\caption{\textit{Upper panel:} Density of states $\bar{\nu}(\epsilon)$ in the metal-insulator transition of class D [Cho-Fisher network model, $\sin^2\alpha=0.19$, $p=0.19$]. The behavior exhibited by $\bar{\nu}(\epsilon)$ follows a power law $\epsilon^{\kappa}$ as expected.  The slope yields $\kappa = -0.30$, which translates into $x_{(1)}=-0.85$ by virtue of Eq.~\eqref{eq:exprel}.
		\textit{Lower panel:} Numerical determination of generalized multifractality at the metal-insulator transition of class D. The spinless pure-scaling combinations
		\eqref{eq:abelianA},~\eqref{eq:det} are computed, with averaging over the system area and $10^4$ realizations of disorder. The data are scaled with $r^{\Delta_{q_1}+ \ldots + \Delta_{q_n}}$, which yields an expected collapse as functions of $r/L$. For each $\lambda$, data points corresponding to the smallest $r \sim 1$ are highlighted as large dots, visualizing the $L$-dependence at a fixed $r$. The full lines are fits to these data points; the resulting exponents $\Delta^{\rm MIT}_\lambda$ are given in Table~\ref{tab:lD}.}
	\label{fig:D-MIT}
\end{figure}

We turn now to the analysis of the generalized multifractality at the metal-insulator transition point in the Cho-Fisher model ($\sin^2\alpha = 0.19$, $p=0.19$).
First, we determine the scaling of the average density of states $\bar{\nu}(\epsilon)$ in order to find the exponent $x_{(1)}$ relating the scaling of the wavefunction observables $\Delta_\lambda$ to the operator scaling dimensions $x_\lambda$. In the upper panel of Fig.~\ref{fig:D-MIT}, we show the numerically obtained  $\bar{\nu}(\epsilon)$, which is nicely fitted by  a power law $\bar{\nu}(\epsilon) \sim \epsilon^{\kappa}$ as expected. The slope gives the exponent $\kappa = -0.30 \pm 0.01 $, which yields  $x_{(1)}=-0.85 \pm 0.03$ according to Eq.~\eqref{eq:exprel}.

We proceed with the numerical analysis of the scaling of eigenfunction observables. We compute spinless combinations~\eqref{eq:abelianA},~\eqref{eq:det}, with averaging over the system area and $10^4$ realizations of disorder. The linear system sizes reach from $L=24$ to $L=512$.  We perform this for all Young diagrams $\lambda$ with $2\leq|\lambda|\leq 4$;  the resulting exponents $\Delta^{\rm MIT}_\lambda$ are given in Table~\ref{tab:lD}. The data for $\lambda = (q)$ and $(1^q)$ are shown in the lower panel of Fig.~\ref{fig:D-MIT}.

The numerical results confirm once again that Eqs.~\eqref{eq:abelianA},~\eqref{eq:det} yield properly the pure-scaling observables. This is particularly non-trivial at the Anderson transition of class D, since the $\sigma$-model domain walls play a central role here, as discussed above. At the same time, we did not include the domain walls when deriving the pure-scaling combination with the help of the Iwasawa decomposition (and also in the alternative approach using the one-loop RG). This can be explained in the following way.

{ At the intuitive physical level, the presence of domain walls does not affect the argumentation in Sec.~\ref{sec:pure_scaling} leading to the the eigenfunction pure-scaling observable construction, with wave-function Slater  determinants as building blocks. In a more technical language,} the domain walls (i.e. jumps between two connected components of the target space) respect the symmetry of the target space. Thus, they do not affect the form of the pure-scaling observables that are determined by this symmetry.  At the same time, the domain walls affect crucially the exponents characterizing the scaling of these observables. In particular, they lead to a strong breakdown of the Weyl symmetry relations. As an example, one of the relations~\eqref{eq:weyl_d} reads $x_{(1)} = x_{(1,1)}$. We find, however, $x_{(1)} = - 0.85$ and $x_{(1,1)} = \Delta_{(1,1)} + 2 x_{(1)} = - 1.26$, implying a clear violation of the Weyl symmetry.  Even more dramatic is the violation of the relation $x_{(1,1,1)}= 0 $, since we get $x_{(1,1,1)} = \Delta_{(1,1,1)} + 3 x_{(1)} = -1.37$.

{
	To shed more light on these points, it is instructive to recall the case of a quasi-one-dimensional (thick wire) geometry, which maps onto a one-dimensional (1D) $\sigma$-model. For Wigner-Dyson classes, the 1D $\sigma$-model approach was used, in particular, to study the conductance and its variance~\cite{zirnbauer1992super, mirlin1994conductance} as well as fluctuations and spatial correlations of eigenstates~\cite{fyodorov1994statistical}. The results are expressed in terms of a Fourier expansion over eigenfunctions of the Laplace operator on the $\sigma$-model manifold.
	
	An extension of the calculation of the conductance to several non-Wigner-Dyson classes, including class D, was carried out in Ref.~\cite{altland2015topology}. The eigenfunctions of the transfer-operator on the $\sigma$-model target space that enter the Fourier expansion in Ref.~\cite{altland2015topology} are, in the absence of jumps between two components of the manifold, the spherical functions ${\cal P}_{(q)}$ with $q=il$, where $l$ is real. The corresponding eigenvalues (which control the exponential decay rate of the associated contribution to the conductance with the length of the wire) are proportional to $- z_{(il)} = l^2$, satisfying the Weyl symmetry $l  \to -l$. Inclusion of the jumps induces a $2 \times 2$ structure of the transfer-operator in the space of the manifold components, with off-diagonal terms proportional to the domain wall fugacity $\chi$. This leads to a splitting of each eigenvalue.
	
	Importantly, the eigenfunctions remain the spherical functions~\cite{altland2015topology}, independently of the domain-wall fugacity, which is due to the fact that the domain-wall term reflects the symmetry. (Two eigenfunctions resulting from the splitting differ by a relative sign on two components of the manifold, as for a splitting of a degenerate two-level system.) At the same time, the eigenvalues get modified:  $ l^2 \mapsto l^2 \pm i \chi l$. The eigenvalues $l^2 + i \chi l$ that correspond to symmetric eigenfunctions enter the Fourier expansion for the conductance. This shift of the eigenvalues implies a breakdown of the Weyl symmetry $l  \to -l$ and is responsible for the localization in class-D wires. The situation in the case of class-DIII wires turns out to be very similar~\cite{altland2015topology}.
	
	While a systematic analysis of the effect of domain walls in $\sigma$-models of classes D and DIII in $d>1$ dimensions remains to be done, we expect that the above two important statements will be inherited from the 1D analysis and can be extended to the whole spectrum of generalized multifractality: (i) pure-scaling observables (e.g., eigenfunctions of RG) are independent of the domain-wall fugacity and (ii) eigenvalues are modified by the domain walls, leading to a breakdown of the Weyl symmetry.}

\begin{figure}
	\centering
	\includegraphics[width=.8\linewidth]{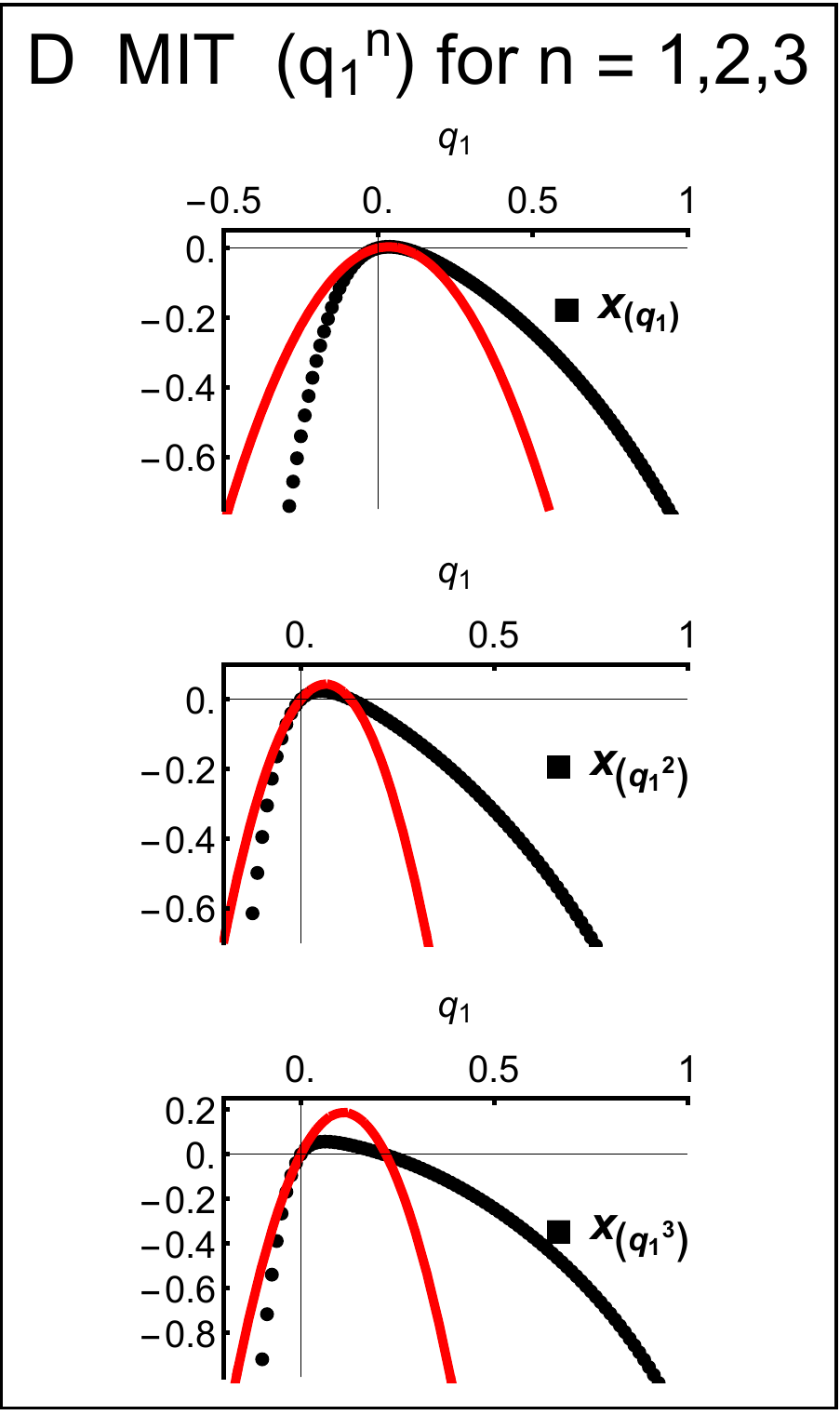}
	\caption{Exponents $x_{(q_1^n)}$ with $n=1, 2, 3$ for the metal-insulator transition of class D (same parameters as in Fig.~\ref{fig:D-MIT}). The red lines represent the ``parabolic approximations'' to the data (fixed by the derivative at $q_1 = 0$ and the root $x_{(q_1^n)}= 0$.  Clearly, the curves $x_{(q_1^n)}$ are strongly non-parabolic, implying violation of the local conformal invariance.  They also strongly violate the Weyl symmetries~\eqref{eq:weyl-D-q1n}, which is a manifestation of the effect of domain walls.
	}
	\label{fig:D-MIT-2}
\end{figure}

In Fig.~\ref{fig:D-MIT-2}, we show the exponents $x_{(q_1^n)}$ with a continuously changing (and relatively small) $q_1$ and $n=1$, 2, and 3. These plots serve as another illustration of a strong violation of the Weyl symmetry relations~\eqref{eq:weyl-D-q1n}. Furthermore, they demonstrate a strong violation of the generalized parabolicity, even in its weak form (assuming no Weyl symmetry)~\eqref{eq:Deltaq_CFT}. Indeed, within the general parabolicity, the dependence $x_{(q_1^n)}$ on $q_1$ should be parabolic, which is clearly not the case. Thus, we can rule out local conformal invariance also for this transition.

\section{Class DIII}
\label{sec:DIII}

\subsection{Model and generalities}

\begin{table*}
	\centering
	\begin{tabular}{cc||c|cc||c}
		& rep. $\lambda$& $\Delta_\lambda^{\rm MIT}$ & $\Delta_\lambda^{\rm metal}$ & $\displaystyle\frac{\Delta_\lambda^{\rm metal}}{b} $ & $\Delta^{\rm para}_\lambda $\\[5pt]
		\hline
		\hline
		&&&&&\\[-5pt]
		$q=2$ & (2) & $-1.365\pm 0.003$ &  $-0.285\pm 0.001$ & $-2.41\pm 0.01$ & $-2b$\\[3pt]
		& (1,1) & $1.31\pm 0.02$ &   $0.238\pm 0.001$    &$2.01\pm 0.01$  & $2b$
		\\[5pt]
		\hline
		&&&&&\\[-5pt]
		$q=3$ & (3)& $-3.097\pm 0.007$ &  $-0.949\pm 0.007$ & $-8.02\pm 0.06$ & $-6b$\\[3pt]
		& (2,1)& $-0.45\pm 0.03$ &  $-0.021\pm 0.001$ & $-0.17\pm 0.01$ & $0$\\[3pt]
		& (1,1,1)& $2.90\pm 0.01$ & $0.695\pm 0.001$  & $5.87\pm 0.01$ & $6b$
		\\[5pt]
		\hline
		&&&&&\\[-5pt]
		$q=4$ & (4) & $-4.93\pm 0.01$ &  $-1.97\pm 0.03$ & $-16.61\pm 0.25$  & $-12b$\\[3pt]
		& (3,1) & $-1.97\pm 0.04$ &  $-0.620\pm 0.007$  & $-5.23\pm 0.06$ & $-4b$\\[3pt]
		& (2,2) & $-1.16\pm 0.07$ &  $-0.05\pm 0.01$ & $-0.41\pm 0.08$ &   $0$\\[3pt]
		& (2,1,1) & $1.96\pm 0.06$ &  $0.455\pm 0.002$ & $3.84\pm 0.02$ &   $4b$\\[3pt]
		& (1,1,1,1) & $5.11\pm 0.02$ &  $1.366\pm 0.001$ & $11.53\pm 0.01$ & $12b$
	\end{tabular}
	\caption{
		Scaling exponents of generalized multifractality in class DIII for eigenstate observables with $q\equiv |\lambda|\leq 4$.
		The exponents $\Delta_\lambda$ shown in the table are related to the exponents $x_\lambda$ via $\Delta_\lambda = x_\lambda - qx_{(1)}$.
		The exponents $\Delta_\lambda^{\rm MIT}$ are found numerically from the helical network model at the transition point, $\alpha = 1.2$, $\theta = 0.5$, and $p = 0.175$.  The thermal-metal exponents $\Delta_\lambda^{\rm metal}$ are obtained from the same model with defect concentrated increased up to $p=0.5$. The last column displays the exponents $\Delta_\lambda^{\rm para} = b(q- z_\lambda)$ corresponding to the generalized parabolic spectrum, Eq.~\eqref{eq:xqdiii}, with a single parameter $b$. In the metallic phase, the exponents are quite close to the generalized parabolicity, as can be seen from the comparison of the column $\Delta_\lambda^{\rm metal} / b$ with  $\Delta_\lambda^{\rm para}$.
		Here $b= - x_{(1)}^{\rm metal} = 0.119$, which agrees well with the scaling of the density of states. The Weyl symmetry relations~\eqref{eq:weyl_diii} (that can be easily translated to $\Delta_\lambda$) are nicely satisfied in the metallic phase, which can be seen by inspection of $\Delta_\lambda^{\rm metal} / b$.  	
		The metal-insulator transition exponents $\Delta_\lambda^{\rm MIT}$, in combination with $x_{(1)}^{\rm MIT} = -0.44$ obtained from the fit of the density of states, substantially violate the Weyl symmetry, which is a manifestation of the effect of topological excitations (domain walls between two connected components) in the $\sigma$-model. 	
	}	
	\label{tab:lDIII}
\end{table*}

Superconductors with time-reversal invariance and spin-orbit interaction are in the Bogolyubov-de-Gennes class DIII, which is characterized by the particle-hole symmetry $P$ satisfying $P^2=1$ and the time-reversal symmetry $T$ with $T^2=-1$. As in class D, the density of states is critical, so that, in addition to determining $\Delta_\lambda$,  we need to study the scaling of the local density of states with energy to determine the exponent $x_{(1)}$.
Similarly to class D, the beta-function has a two-loop correction to the leading (one-loop) term:
\begin{align}
\dfrac{\mathrm{d} \ln t}{\mathrm{d} \ln \ell } = - 2t + 2t^2 + O (t^3),
\label{eq:DIII-beta}
\end{align}
with $t= 1/\pi g$.
The one-loop zeta-functions for scaling dimensions of the operators read  [see Eq.~\eqref{eq:one-loop-RG-C-lambda}]
\begin{align}
\dfrac{\mathrm{d} \ln \mathcal{C}_\lambda}{\mathrm{d} \ln \ell }=  z_\lambda t + O (t^2).
\label{eq:DIII-zeta}
\end{align}
As for class D, the target space of the $\sigma$-model for class DIII consists of two disjoint components. In full analogy with the above discussion of class D, the corresponding jumps (domain walls) are crucial for establishing localization and are expected to violate the Weyl symmetry at the metal-insulator transition in class DIII.

For our numerical analysis, we use the helical superconductor network model from Ref.~\cite{fulga2012thermal}.  Each link carries two counter-propagating Majorana modes that can be identified by their spin (helicity).  The scattering matrices $S$ and $S'$ at even and odd nodes read, respectively,
\begin{align}
S &= \begin{pmatrix}
0 & r & t\cos\theta & -t \sin\theta \\
-r & 0 & t\sin \theta & t\cos\theta \\
-t\cos\theta & -t \sin\theta& 0 &-r\\
t\sin \theta & -t\cos\theta &r &0\\
\end{pmatrix},\nonumber\\
S' &= \begin{pmatrix}
0 & -t\cos\theta & -r & -t \sin\theta \\
t\cos\theta &0 & t \sin\theta &-r\\
r & -t\sin \theta & 0 & t\cos\theta \\
t\sin \theta & r & -t\cos\theta &0\\
\end{pmatrix}.
\label{eq:net_diii}
\end{align}
The transmission and reflection amplitudes are $t=\sin\alpha$ and $r=\cos\alpha$, with the angle $\alpha$ drawn from the Cho-Fisher distribution~\eqref{eq:dcf}.
The angle $\theta$ describes the coupling between the two helical copies of the class-D Cho-Fisher network model. (At $\theta=0$, the copies decouple and one can study a crossover to class-D behavior.)  For the choice of parameters $\alpha = 1.2$, $\theta = 0.5$, and $p = 0.175$, the system is at the critical point of the metal-insulator transition~\cite{fulga2012thermal}. When the disorder concentration is increased up to its maximal value, $p=0.5$, the system is deeply in the thermal-metal phase. We will study these two points in the phase diagram to investigate the metal-insulator transition and the metallic phase, respectively.

\subsection{Metallic phase}
\label{sec:DIII-metallic}

\begin{figure*}
	\centering
	\includegraphics[width=0.48\textwidth]{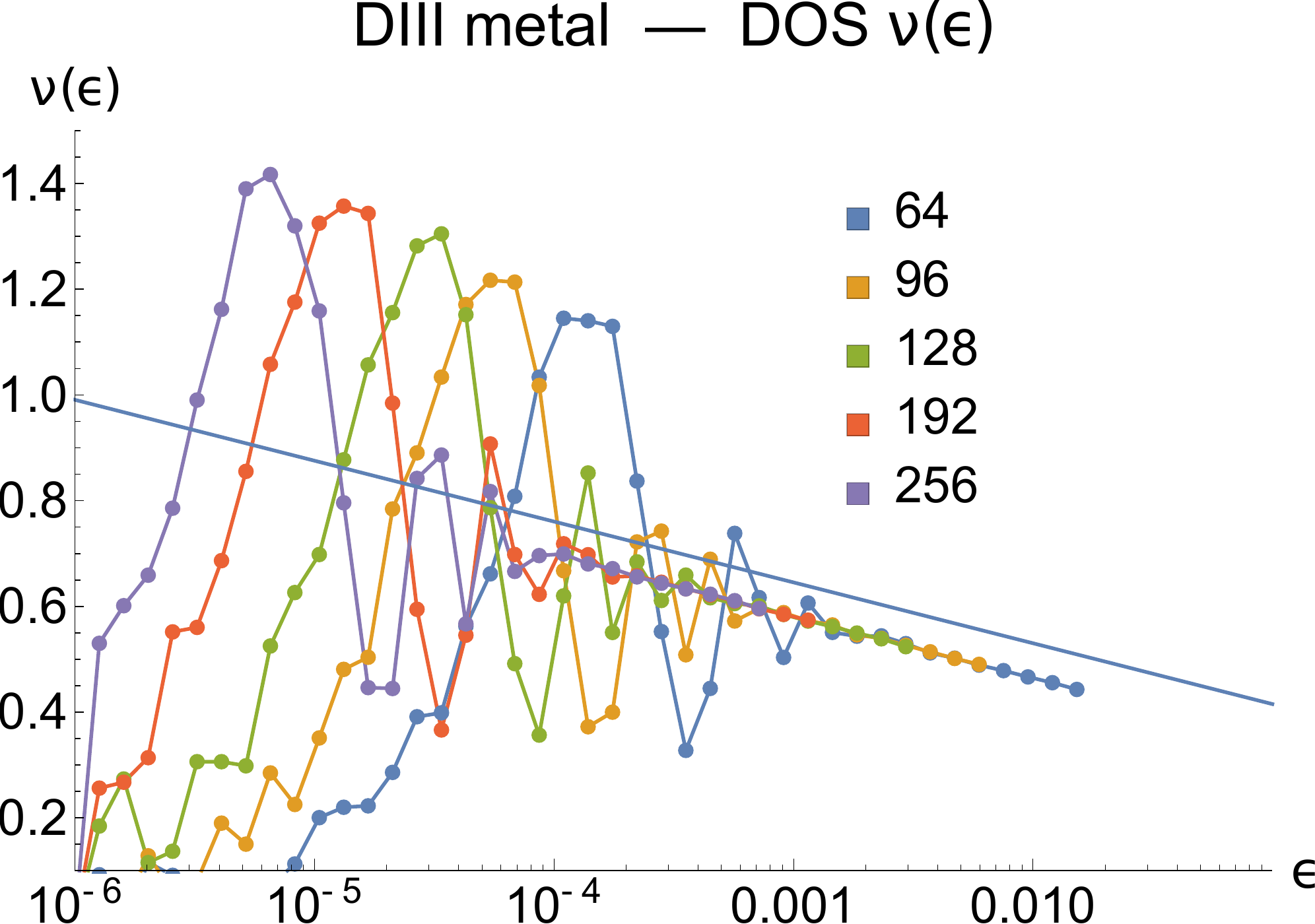}
	\hspace{0.05\textwidth}
	\includegraphics[width=0.4\textwidth]{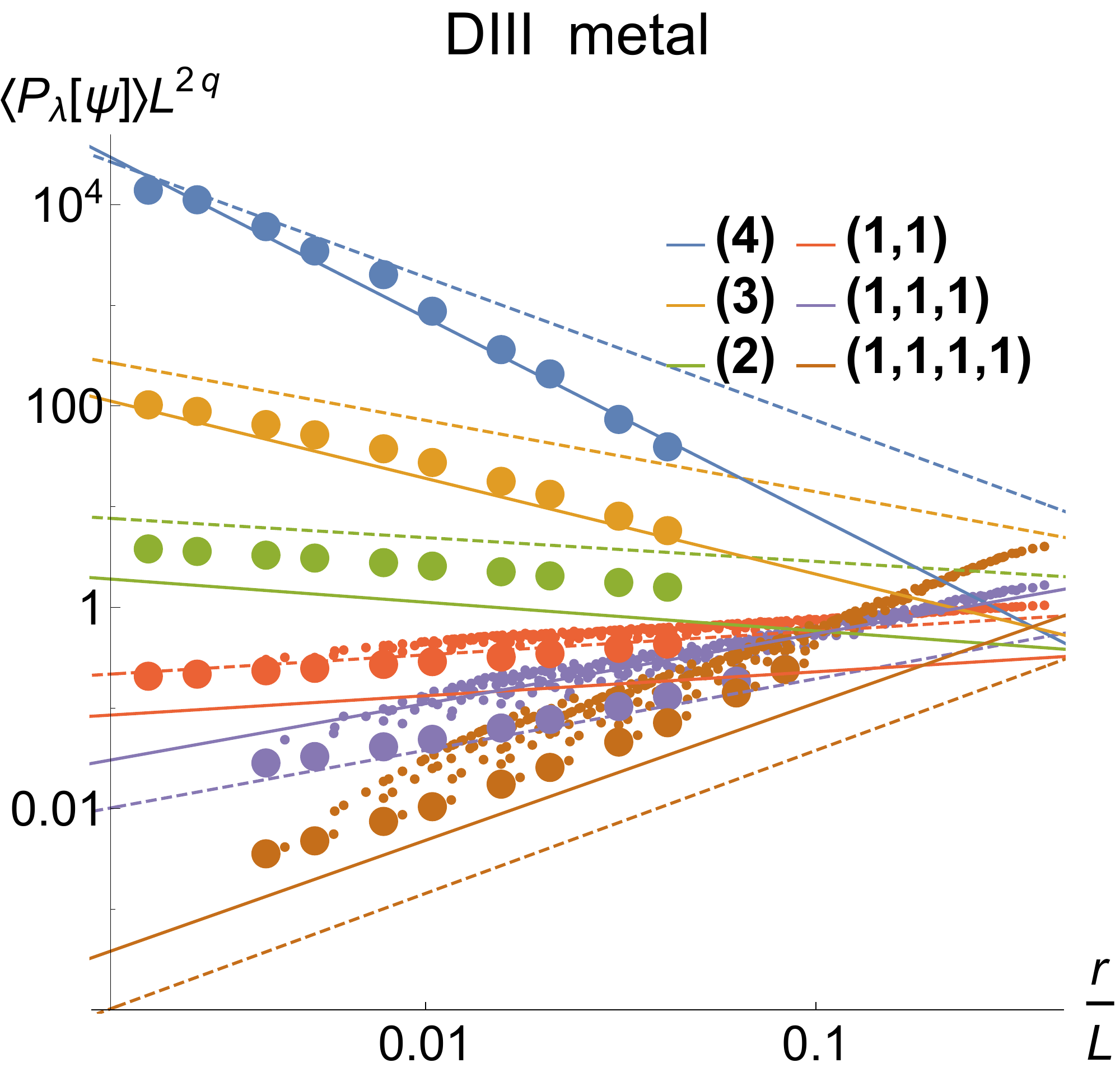}
	\caption{\textit{Left panel:} Density of states $\bar{\nu}(\epsilon)$ in the thermal-metal phase of class DIII (helical superconductor network model~\eqref{eq:net_diii}, at the scattering angles $\alpha=0.5$, $\theta = 1.2$ and the defect concentration $p=0.5$).  The density of states slowly increases with lowering energy, in consistency with the analytical prediction~\eqref{eq:one-loop-renorm-DOS-e}. A power-law fit to $\bar{\nu}(\epsilon) \sim \epsilon^{\kappa}$ yields the running exponent $\kappa = -0.06$, implying $x_{(1)} \approx - 0.12$.	{ At the lowest energies, the RMT oscillations of class DIII	\cite{altland1997nonstandard} are observed.}
		\textit{Right panel:} Numerical determination of generalized multifractality in the metallic phase of class DIII. The spinfull pure-scaling combinations~\eqref{eq:det-sp},~\eqref{eq:det} are computed, with averaging over the system area and $10^4$ realizations of disorder. The data are scaled with $r^{\Delta_{q_1}+ \ldots + \Delta_{q_n}}$, which yields an expected collapse as functions of $r/L$. For each $\lambda$, data points corresponding to the smallest $r \sim 1$ are highlighted as large dots, visualizing the $L$-dependence at a fixed $r$. The full lines are fits to these data points; the resulting exponents $x^{\rm metal}_\lambda$ are given in Table~\ref{tab:lDIII}. The dashed lines correspond to the generalized parabolic spectrum~\eqref{eq:xq} with $b=0.119$; see the column $x^{\rm para}_\lambda$ in  Table~\ref{tab:lDIII}.  The slopes of full and dashed lines are close for all $\lambda$, which means that the generalized parabolicity holds to a good accuracy in the metallic phase. Further, this value of $b$ matches $x_{(1)}$ extracted from the DOS scaling very well.
		Curvatures in the data is related to the fact that all exponents are in fact the running ones and reduce logarithmically with increasing $L$, see Sec.~\ref{sec:metal_scaling}
		and Eq.~\eqref{eq:metal-scaling-P-lambda-L_2-DIII}.}
	\label{fig:DIII-metal}
\end{figure*}

As for class D, we study numerically the polynomial observables up to the order $q=4$. The Weyl symmetry expected to hold in the metallic phase of class DIII, implies the following relations between the corresponding exponents:
\begin{align}
x_{(1,1)} &= 0,
&
x_{(2,2)} &= x_{(2)} = 0,
&
x_{(2,1,1)} &= 0.
\label{eq:weyl_diii}
\end{align}
The single-parameter generalized parabolic spectrum~\eqref{eq:xq} (which combines parabolicity with the Weyl symmetry) reads in class DIII:
\begin{align}
x_{(q_1,q_2,\ldots)}^{\rm para}&= b\left[ -q_1^2 + q_2(2-q_2) + q_3(4-q_3) +\ldots\right].
\label{eq:xqdiii}
\end{align}
As for other classes, we will also study numerically the exponents $x_\lambda$ for $\lambda = (q_1^n)$ with $q_1 = q/n$ and $n =1,2,3$. The Weyl symmetry for such exponents reads
\begin{align}
& x_{(q_1)} = x_{(- q_1)}, \quad x_{(q_1, q_1)} = x_{(1 - q_1, 1-q_1)},
\nonumber \\
& x_{(q_1, q_1, q_1)} = x_{(2- q_1, 2-q_1, 2-q_1)}.
\label{eq:weyl-DIII-q1n}
\end{align}

In the left panel of Fig.~\ref{fig:DIII-metal}, we show the numerically determined density of states $\bar{\nu}(\epsilon)$.
The analytical prediction is given by Eq.~\eqref{eq:one-loop-renorm-DOS-e}. In the available range of energies, the density of states looks linear as a function of $\ln (1/\epsilon)$. This is because the slope [i.e., the variation of  $\bar{\nu}(\epsilon)$] is rather small, which means a small resistance $t$. To observe clearly the asymptotic $\ln^{1/2} (1/\epsilon)$ behavior, one would need to proceed to much smaller energies, which would require unrealistically large system sizes. In the considered range of energies, the density of states can be also fitted very well to a power law $\bar{\nu}(\epsilon) \sim \epsilon^{\kappa}$ with $\kappa = -0.06$, which translates into the exponent $x_{(1)} \approx - 0.12$. As has been extensively discussed above, all exponents $x_\lambda$ in the metallic phase are in fact running ones; we determine their values corresponding to the available range of system sizes.

{ At the lowest energies, the density of states exhibits RMT oscillations, as expected for a metallic system of class DIII~\cite{altland1997nonstandard}. Note that we deal here with a model of Cho-Fisher-type, i.e., with an even number of defects, so that the corresponding RMT ensemble is of DIII-even type~\cite{ivanov2002supersymmetric}.
}

The right panel of Fig.~\ref{fig:DIII-metal} displays the scaling of eigenfunction observables. We have computed the observables~\eqref{eq:det-sp},~\eqref{eq:det} corresponding to the spinful situation, with averaging over the system area and $10^4$ realizations of disorder.  As discussed in Sec.~\ref{sec:metal_scaling}, the asymptotic behavior of the pure scaling observables in the thermal-metal phase of class DIII has the form
\begin{equation}
L^{2q} \langle P_\lambda[\psi] (L) \rangle  \sim \left(1+ 2t(\ell) \ln \frac{L}{\ell} \right)^{\frac12 (z_\lambda - q) }.
\label{eq:metal-scaling-P-lambda-L_2-DIII}
\end{equation}
As we have also discussed, for sufficiently small $t(\ell)$ and in a restricted range of system sizes $L$, this can be approximated by a power law,
\begin{equation}
L^{2q} \langle P_\lambda[\psi] (L) \rangle  \sim L^{-\Delta_\lambda},
\qquad \Delta_\lambda = - t(\ell) (z_\lambda - q),
\label{eq:metal-scaling-P-lambda-L_3-DIII}
\end{equation}
which is the generalized parabolicity~\eqref{eq:xqdiii} with $b= t(\ell)$. We have seen in the case of class D that fits of both types (power law and ``logarithmic'') work well
(see Fig.~\ref{fig:D-metal-2}) and yield very close values of the exponents (see columns $\Delta_\lambda^{\rm metal}/b$ and $\tilde{\Delta}_\lambda^{\rm metal}$ in Table~\ref{tab:lD}). In the DIII metallic phase, the parameter $b$ (determined by the running resistance $t(\ell)$) characterizing the strength of multifractality turns out to be substantially smaller ($b= - x_{(1)} = 0.12$) than in the case of class D, where we had $b=0.3$. (Of course, we mean the specific points in the phase diagram that we consider.)
Thus, the approximation of  Eq.~\eqref{eq:metal-scaling-P-lambda-L_2-DIII}  by Eq.~\eqref{eq:metal-scaling-P-lambda-L_3-DIII} should work still better than an analogous approximation in class D, and the exponents obtained in both ways should be even closer. For this reason, we restrict ourselves to the conventional log-log representation as shown in the right panel of Fig.~\ref{fig:DIII-metal}, with straight-line fits corresponding to power-law dependence on $L$.

The numerics confirms that the spinful eigenstates combinations~\eqref{eq:det-sp},~\eqref{eq:det} are correct pure-scaling observables for class DIII.
The obtained exponents are presented in Table~\ref{tab:lDIII} as $\Delta_\lambda^{\rm metal}$. They satisfy very well the Weyl symmetry relations~\eqref{eq:weyl_diii} in agreement with analytical predictions. The exponents are sufficiently close to the single-parameter generalized parabolic form~\eqref{eq:xqdiii} with $b= - x_{(1)} = 0.119$, as can be seen by comparing the columns $\Delta_\lambda^{\rm metal} / b$ and $\Delta_\lambda^{\rm para}$ in Table~\ref{tab:lDIII}, as well as full and dashed lines in the right panel of Fig.~\ref{fig:DIII-metal}. We observe, however, deviations that can be attributed to the two-loop corrections.   It is also worth mentioning that the data points in the right panel of Fig.~\ref{fig:DIII-metal} show small but noticeable curvature: the slopes have a tendency to become smaller when $L$ increases. This is in full consistency with analytical expectations: the exponents are in fact the running ones and reduce logarithmically with increasing $L$, see Sec.~\ref{sec:metal_scaling} and Eq.~\eqref{eq:metal-scaling-P-lambda-L_2-DIII}.

\begin{figure}
	\centering
	\includegraphics[width=.9\linewidth]{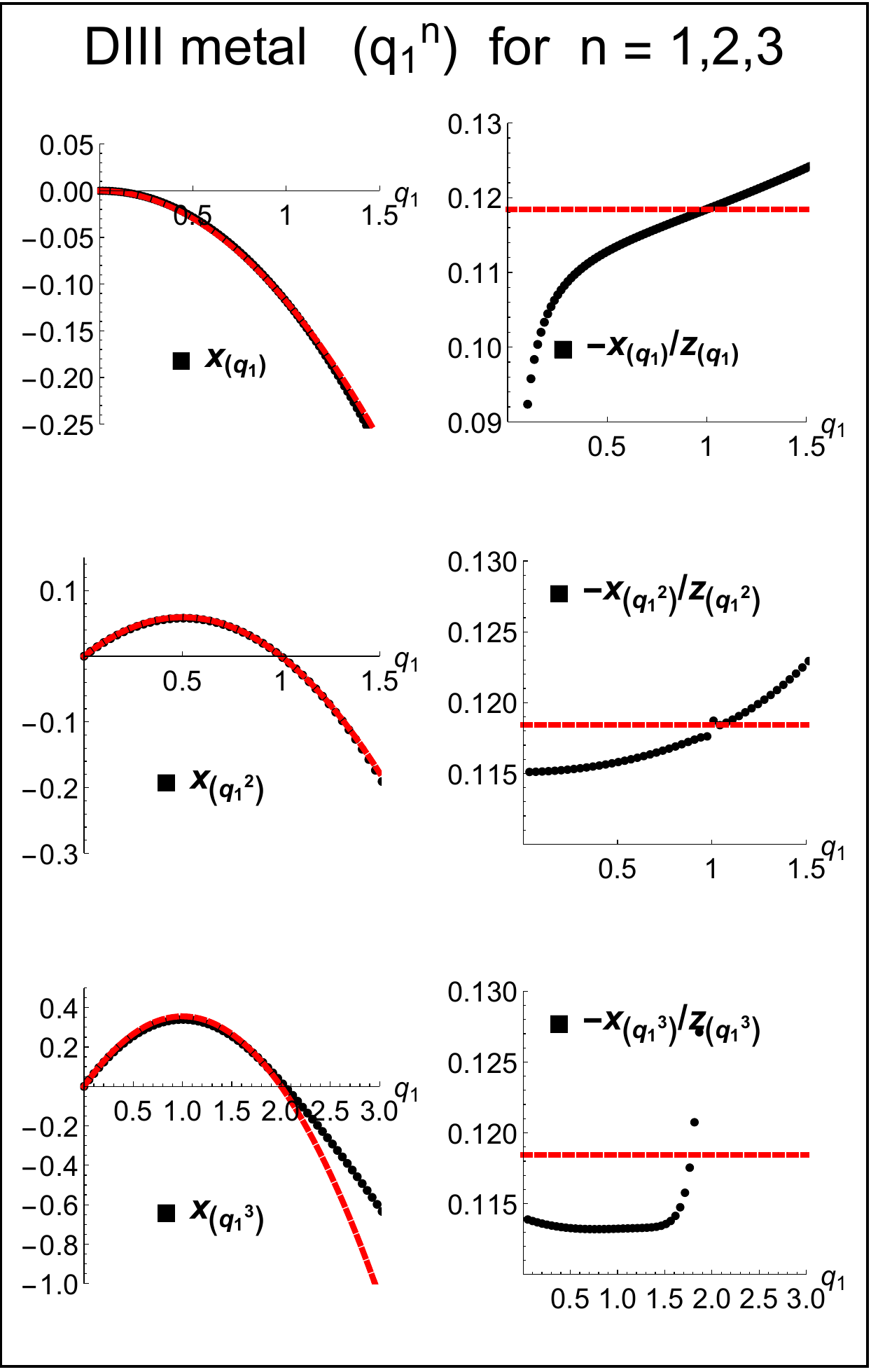}
	\caption{ Numerically obtained exponents $x_{(q_1^n)}$ with $n=1, 2, 3$ for the metallic phase of class DIII (same parameters as in Fig.~\ref{fig:DIII-metal}). In the left panels, data points are exponents $x_\lambda$; in the right panels, the same data are shown as a ratio $- x_\lambda / z_\lambda$, where $z_\lambda$ is the quadratic Casimir invariant.  The red lines correspond to generalized parabolicity~\eqref{eq:xqdiii} with $b=0.119$. Both the
		Weyl symmetry~\eqref{eq:weyl-DIII-q1n} and the  generalized parabolicity~\eqref{eq:xqdiii} hold very accurately. At large $q$, deviations related to insufficient ensemble averaging are observed.}
	\label{fig:DIII-metal-2}
\end{figure}

In Fig.~\ref{fig:DIII-metal-2}, we present the exponents $x_{(q_1^n)}$ with $n=1, 2, 3$ for the metallic phase in class DIII. In the left panels, data points are  numerically obtained exponents $x_\lambda$; in the right panels, the same data are shown in the form $- x_\lambda / z_\lambda$, where $z_\lambda$ is the quadratic Casimir invariant.  The red lines in all panels correspond to the single-parameter generalized parabolicity~\eqref{eq:xqdiii} with $b=0.119$. It is seen that the
Weyl symmetry~\eqref{eq:weyl-DIII-q1n} and, moreover, the  generalized parabolicity~\eqref{eq:xqdiii}, hold to a high accuracy.

\subsection{Metal-insulator transition}
\label{sec:DIII-MIT}

\begin{figure*}
	\centering
	\includegraphics[width=0.48\textwidth]{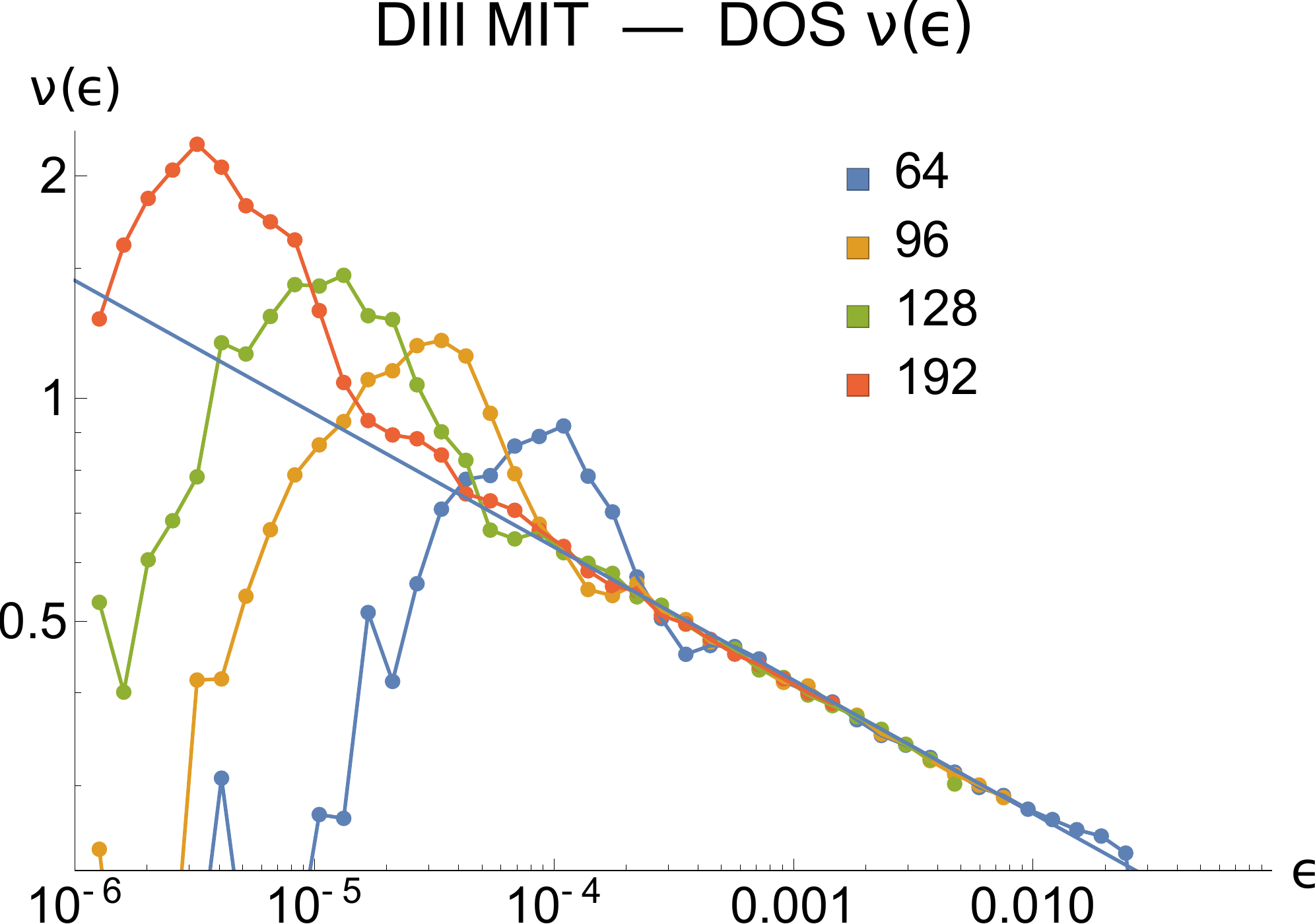}
	\includegraphics[width=0.48\textwidth]{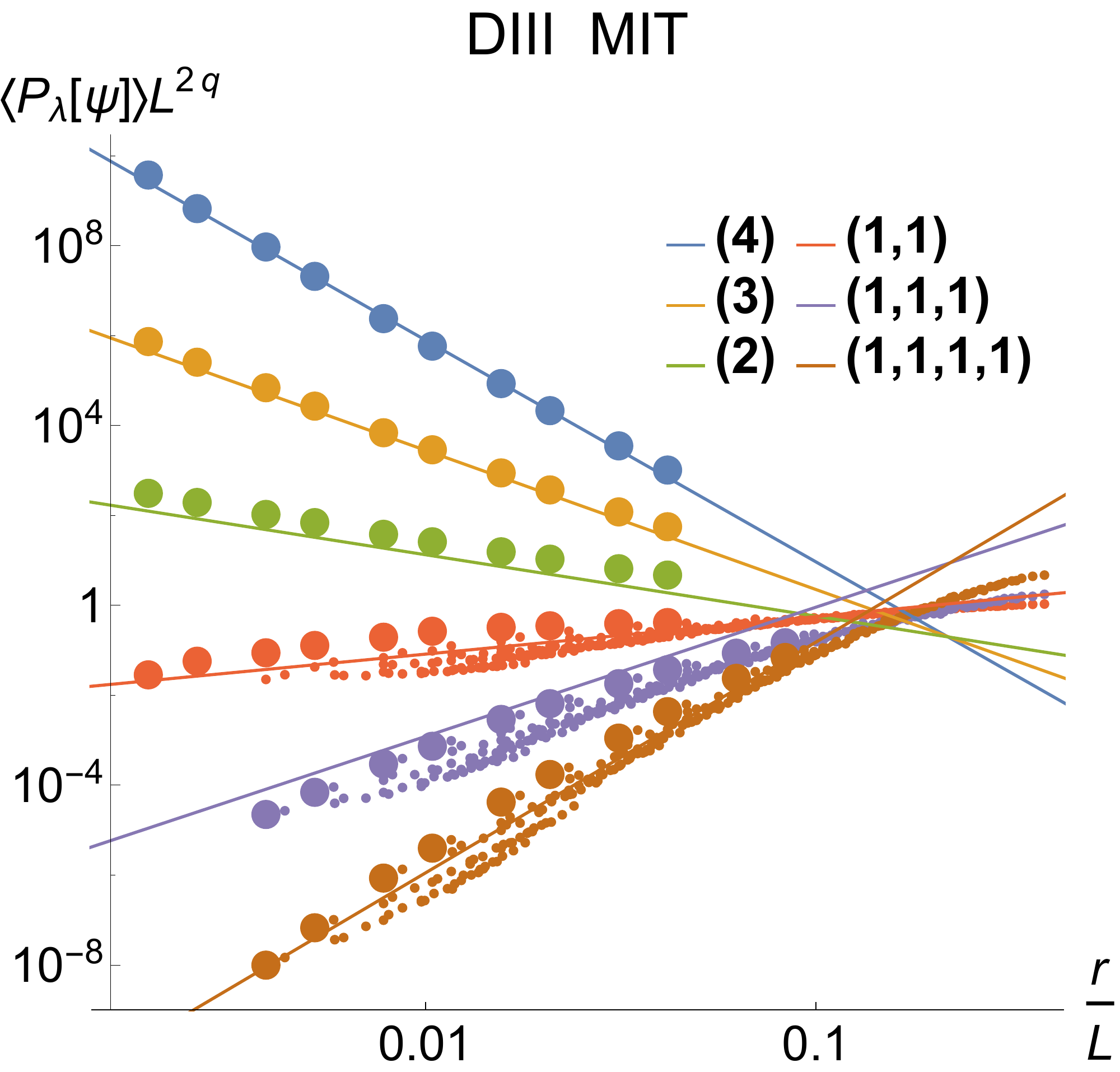}
	\caption{\textit{Left panel:} Density of states $\bar{\nu}(\epsilon)$ at the metal-insulator transition of class DIII (helical-superconductor network model~\eqref{eq:net_diii}, at scattering angles $\alpha=0.5$ and $\theta=1.2$, and defect concentration $p=0.175$).
		The density of states exhibits a power-law scaling $\bar{\nu}(\epsilon) \propto \epsilon^{-\kappa}$ with $\kappa = -0.18$, which translates into  $x_{(1)}=-0.44$ according to Eq.~\eqref{eq:exprel}.
		\textit{Right panel:} Numerical determination of generalized multifractality at the metal-insulator transition of class DIII. The spinfull pure-scaling combinations~\eqref{eq:abelianA},~\eqref{eq:det-sp} are computed, with averaging over the system area and $10^4$ realizations of disorder. The data are scaled with $r^{\Delta_{q_1}+ \ldots + \Delta_{q_n}}$, which yields an expected collapse as functions of $r/L$. For each $\lambda$, data points corresponding to the smallest $r \sim 1$ are highlighted as large dots, visualizing the $L$-dependence at a fixed $r$. The full lines are fits to these data points; the resulting exponents $\Delta^{\rm MIT}_\lambda$ are given in Table~\ref{tab:lDIII}.}
	\label{fig:DIII-MIT-2}
\end{figure*}

\begin{figure}
	\centering
	\includegraphics[width=\linewidth]{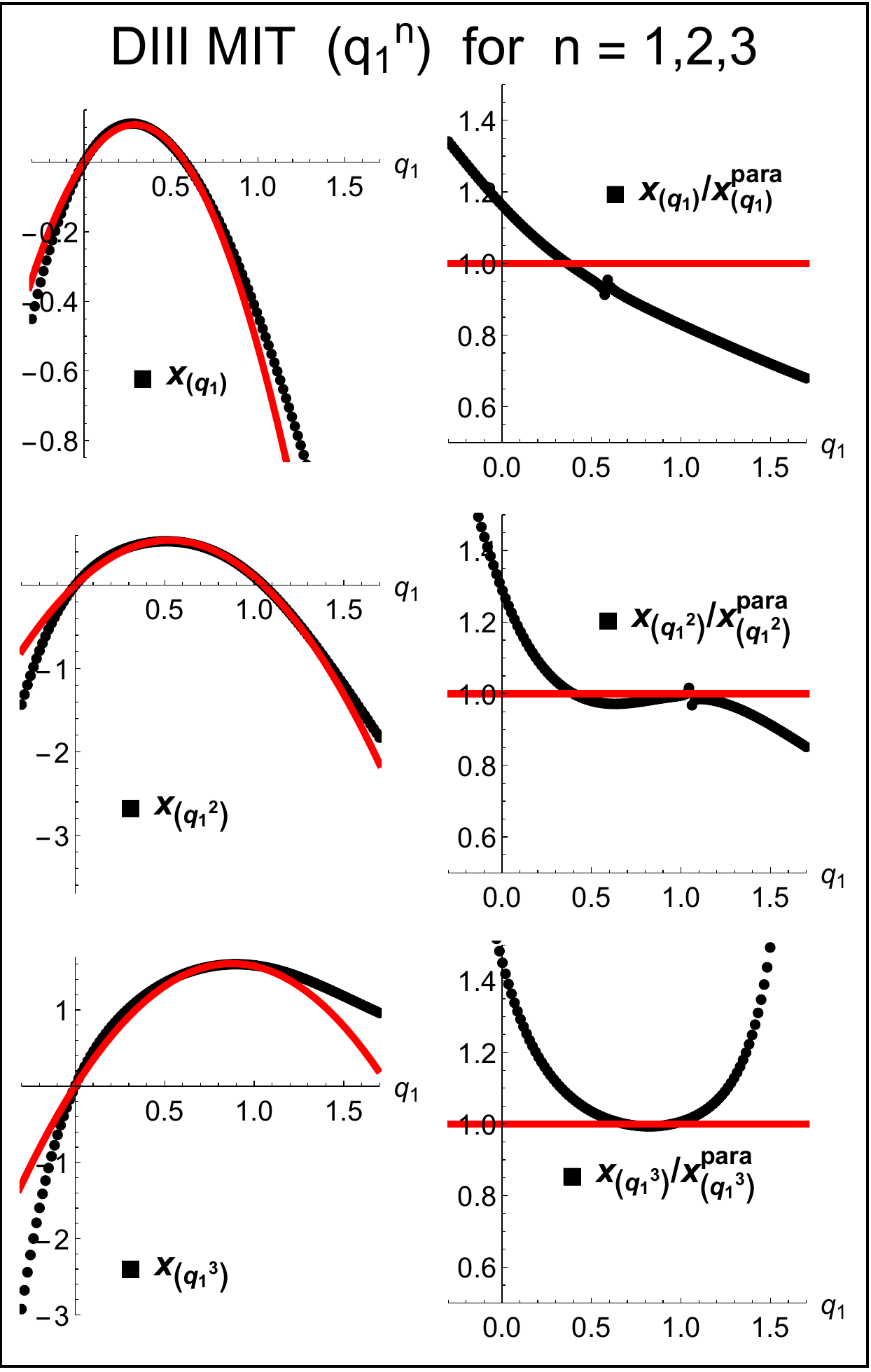}
	\caption{Exponents $x_{(q_1^n)}$ with $n=1,2,3$ for the metal-insulator transition of class DIII (same parameters as in Fig.~\ref{fig:DIII-MIT-2}).
		The Weyl symmetry	\eqref{eq:weyl-DIII-q1n} is manifestly violated. At the same time, the data are described sufficiently well by a many-parameter parabolic formula~\eqref{eq:Deltaq_CFT}. Still, there are clear deviations, as demonstrated in the right panels where the ratio of $x_\lambda$ to the parabolic fit~\eqref{eq:Deltaq_CFT} (see parameters in the text) is shown.  }
	\label{fig:DIII-MIT}
\end{figure}

Finally, we study the class-DIII network model at the metal-insulator transition point. The corresponding density of states $\bar{\nu}(\epsilon)$ is shown in the left panel of
Fig.~\ref{fig:DIII-MIT-2}.  It exhibits a power-law scaling $\bar{\nu}(\epsilon) \propto \epsilon^{\kappa}$ as expected at criticality. The numerically obtained exponent is $\kappa = -0.180 \pm 0.005$, which yields $x_{(1)}=-0.440 \pm 0.015$ by virtue of Eq.~\eqref{eq:exprel}. In the right panel of Fig.~\ref{fig:DIII-MIT-2}, we present the data for the eigenfunction observables corresponding to $\lambda = (q)$ and $(1^q)$  with $q=2,3,4$.  The extracted exponents for all polynomial observables with $q \le 4$ are presented as $\Delta_\lambda^{\rm MIT}$
in Table~\ref{tab:lDIII}.  In addition, we show in Fig.~\ref{fig:DIII-MIT} numerical results for the exponents $x_{(q_1^n)}$  with $n=1,2,3$ and continuously changing $q_1$.  Violation of the Weyl symmetry~\eqref{eq:weyl-DIII-q1n} is evident from this data. This is expected (in full analogy with class D), in view of the domain walls between the two connected components of the $\sigma$-model target space. At the same time, the form of $x_{(q_1)}$,  $x_{(q_1,q_1)}$, and  $x_{(q_1,q_1,q_1)}$ turns out to be not so far from a parabolic one; see parabolic fits by red lines. We recall that, since the Weyl symmetry does not hold at the metal-insulator transition point of class DIII, a parabolic spectrum may have the general, multi-parametric form~\eqref{eq:Deltaq_CFT}. Such an unrestricted parabolic fit yields the approximation $x^{\rm para}_{(q_1)} = 1.27q_1(0.5829-q_1) $,  $x^{\rm para}_{(q_1,q_1)} =  2.05q_1(1.0364-q_1) $, and $x^{\rm para}_{(q_1,q_1,q_1)} =  2.1 q_1 (1.75 - q_1)$.  While it approximates quite well the numerically obtained spectrum, there are sizeable deviations. They can be best seen in the right panels, where the ratio $x_\lambda / x_\lambda^{\rm para}$ is plotted. The deviations for the generalized parabolicity imply violation of the conformal invariance also at the class-DIII metal-insulator transition. It is worth noticing, however, that the deviations from parabolicity are much less dramatic than for the metal-insulator transition in class D studied above, as can be seen by a comparison of Fig.~\ref{fig:DIII-MIT} with Fig.~\ref{fig:D-MIT-2}.

\section{Summary}
\label{sec:summary}

In this paper, we have studied the generalized multifractality in 2D disordered systems. While our main focus was on the symmetry classes AII, D, and DIII, which exhibit 2D metal-insulator transitions, some of our results are more general, extending to all symmetry classes.  Our key findings are as follows:

\begin{enumerate}
	
	\item  We have performed a derivation of pure-scaling operators in terms of the $\sigma$-model field theory and their translation to the language of observables constructed from eigenfunctions of the Hamiltonian. Analyzing the composite operators in the $\sigma$-model field theories, we have used two complementary approaches: the Iwasawa decomposition (that we have carried out for classes AII, D, and DIII) and the one-loop RG (that we have worked out for all ten symmetry classes). We have shown that the ten symmetry classes can be subdivided in two groups: ``spinless'' (A, AI, AIII, BDI, and D) and ``spinful'' (AII, CII, C, CI, DIII). This subdivision is quite transparent physically: the spinful classes are characterized either by Kramers degeneracy due to time-reversal invariance $T$ with $T^2=-1$ or by similar ``near-degeneracy'' due to the particle-hole symmetry $P$ satisfying $P^2=-1$, or by both of them. The eigenfunction pure-scaling observables [which are classified according to representations $\lambda= (q_1, \ldots, q_n)$] are given in both cases (spinless and spinful) by Eq.~\eqref{eq:abelianA}. At the same, the building blocks of this construction have two distinct forms:
	they are given by Eq.~\eqref{eq:det} for spinless symmetry classes and by Eq.~\eqref{eq:det-sp} for spinful classes.
	
	\item  We have carried out extensive numerical simulations of the generalized multifractality in classes AII, D, DIII, using the Ando model for class AII and suitable network models for classes D and DIII. We have studied metal-insulator transition points as well as metallic phases in these models.  The results fully confirm that the spinful construction~\eqref{eq:abelianA}, ~\eqref{eq:det-sp} yields properly the pure-scaling observables for classes AII and DIII, while the observables in class D are correctly given by the spinless form, Eqs.~\eqref{eq:abelianA} and~\eqref{eq:det-sp}. What adds an interesting twist to this result is the fact that the localization in these three classes is crucially associated with topological excitations in the $\sigma$ model: vortices in class AII and domain walls in classes D and DIII.  The situation is particularly non-trivial for classes D and DIII, where the $\sigma$-model target spaces consist of two disjoint components. While the derivation (based on Iwasawa decomposition or one-loop RG) does not incorporate the associated jumps between the components, these jumps do not affect the symmetry analysis leading to Eqs.~\eqref{eq:abelianA},~\eqref{eq:det}, and~\eqref{eq:det-sp} for the pure-scaling eigenfunction observables.
	
	\item In the metallic phases, the scaling exponents should be viewed as running ones; they experience a slow (logarithmic) renormalization towards zero with increasing $L$, as is also observed in the numerics.  In all three classes, the numerically obtained exponents in the metallic phase satisfy well the Weyl symmetry, as expected. Furthermore, in class AII, the exponents are described by the single-parameter generalized parabolicity~\eqref{eq:xq} with an excellent accuracy. At the same time, in classes D and DIII, sizeable deviations from the generalized parabolicity are observed. This is in agreement with analytical expectations: the generalized parabolicity~\eqref{eq:xq}  is exact on the one-loop level but is in general violated by higher-loop corrections. In class AII, these corrections are particularly small since they start from the four-loop order and because of smallness of the resistance $t$.
	
	\item At the metal-insulator transition in class AII, the exponents nicely satisfy the Weyl symmetry, in agreement with the analytical prediction. At the same time, the generalized parabolicity~\eqref{eq:xq} is strongly violated, which implies violation of local conformal invariance at this critical point.
	
	\item At the critical points of metal-insulator transitions in classes D and DIII, the Weyl-symmetry relations do not hold. This result is again in agreement with analytical predictions, since the $\sigma$-model domain walls are expected to lead to a breakdown of the Weyl symmetry.  Furthermore, the numerically found generalized-multifractality exponents do not obey the generalized parabolicity even in its weak form~\eqref{eq:Deltaq_CFT}, with deviations being especially strong in class D. This implies that the local conformal invariance does not hold at metal-insulator transitions in classes D and DIII as well.
	
\end{enumerate}

The results of this work demonstrate that the violation of the generalized parabolicity---and thus of local conformal invariance---that was found (both analytically and numerically) for the SQH transition in our previous works~\cite{karcher2021generalized, karcher2022generalized} is in fact a quite general feature shared by many localization critical points in 2D disordered systems.

The generalized multifractality explored here is an important hallmark of quantum disordered systems, and the corresponding set of exponents constitutes a ``fingerprint'' of the critical point (or a ``nearly critical point'', as in the case of metallic phases in 2D systems).  There is a broad variety of interesting potential extensions of this work, including systems of different spatial dimensionalities, symmetry classes, and topologies, interacting systems, as well as surfaces of disordered systems. Before closing, we briefly discuss a few prospective research directions:

\begin{enumerate}
	
	\item  The generalized multifractality can be studied also in 2D systems of classes A, AI, C, CI in the regime of weak localization. On the analytical side, the formulas analogous to those in Sec.~\ref{sec:metal_scaling} will hold, with straightforward modifications corresponding to a replacement of weak antilocalization by weak localization. An essential difference is that, in this situation, the generalized multifractality holds only on scales below the localization length. However, for a sufficiently small bare resistance, the localization length is exponentially large and cannot be realistically reached.
	
	\item  The classification of pure-scaling observables does not depend on the system dimensionality. In particular, our results pave the way to an investigation of the generalized multifractality at localization transitions in three-dimensional systems.
	
	\item It was shown~\cite{rodriguez2011multifractal} that the multifractal analysis can be a very useful tool for locating the Anderson transition, determining the critical exponent of the localization length, and establishing universality. An extension of such an analysis to the generalized multifractality may be very advantageous.
	
	\item Three chiral classes (AIII, BDI, and CII) are special in the sense that pure-scaling observables for them are labeled not by a single representation $\lambda$ but rather by a pair $(\lambda, \overline{\lambda})$.  In terms of eigenfunction observables, $\lambda$ should correspond to one sublattice, and $\overline{\lambda}$  to another one.
	Furthermore, in 2D geometry, these classes possess critical-metal phases and transitions between these phases and insulating phases. Numerical studies of the generalized multifractality in these phases and critical points would be of much interest.
	
	\item In classes AIII, DIII, and CI, there exist critical points that emerge on surfaces of topological superconductors (or, alternatively, in models of disordered Dirac fermions). These critical points are described by Wess-Zumino-Novikov-Witten models and are expected to exhibit the spectrum of generalized-multifractality exponents satisfying the single-parameter generalized parabolicity. It would be very interesting to verify this numerically.

\end{enumerate}

\section{Acknowledgments}

We thank I. Burmistrov for useful discussions. J.F.K. and A.D.M. acknowledge support by the Deutsche Forschungsgemeinschaft (DFG) via the grant MI 658/14-1.

\pagebreak

\appendix

\section{The Iwasawa construction}
\label{app:Iwasawa}

In this appendix, we describe the construction of pure-scaling $\sigma$-model observables $\mathcal{P}_\lambda(Q)$ based on the Iwasawa decomposition. The pure-scaling observables obtained in this way satisfy the abelian fusion. The construction explicitly demonstrates the difference between ``spinless'' and ``spinful'' symmetry classes. Upon ``translation'', these composite operators yield pure-scaling eigenfunction observables as presented in Sec.~\ref{sec:pure_scaling} of the paper.

\subsection{Generalities}

The Iwasawa construction has already been presented for class A  by two of us and M. Zirnbauer in Ref.~\cite{gruzberg2013classification} and for class C by the present authors and N. Charles in Ref.~\cite{karcher2021generalized}, so here we only provide basic steps. Further details relevant to the three symmetry classes, AII, D, and DIII, studied in this paper, will be presented in the subsequent sections.

The analysis of the generalized multifractality in class A in Ref.~\cite{gruzberg2013classification} was done using the supersymmetry approach to disordered systems~\cite{efetov1983supersymmetry, efetov1997supersymmetry}. It is sufficient for our purposes in this paper to work within the bosonic sectors of the relevant $\sigma$-models. For some quantities this requires to take the limit $n \to 0$, where $n$ is the number of bosonic replicas. The bosonic $\sigma$-model target spaces have the form $M_B = G/K$ where $G$ is a real non-compact group and $K$ is its maximal compact subgroup. As we explained in Ref.~\cite{gruzberg2013classification}, the pure-scaling operators $\mathcal{P}_\lambda(Q)$ are joint eigenfunctions of the $G$-invariant differential operators on $G/K$, also known as the Laplace-Casimir operators. The Iwawasa decomposition allows us to construct the desired eigenfunctions as the $N$-radial spherical functions on $G/K$.

We begin with the Cartan decomposition
\begin{align}
\mfg= \mfk \oplus \mfp
\label{Cartan-decomposition-real}
\end{align}
of the Lie algebra of $G$, $\mfg = \text{Lie}(G)$, into a maximal compact subalgebra $\mfk$ and the complementary subspace $\mfp$. The two parts of the Cartan decomposition are the $+1$ and $-1$ eigenspaces of a Cartan involution (a Lie algebra automorphism that squares to the identity) $\theta$. If we write an element $Z \in \mfg$ as $Z = X + Y$ where $X \in \mfk$ and $Y \in \mfp$, then
$\theta(X + Y) = X - Y$.
The parts of the Cartan decomposition satisfy the commutation relations
\begin{align}
[\mfk, \mfk] &\subseteq \mfk,
&
[\mfk, \mfp] &\subseteq \mfp,
&
[\mfp, \mfp] &\subseteq \mfk.
\label{comm-rel-Cartan-real}
\end{align}

Then we choose a maximal Abelian subspace $\mfa \subset \mfp$ and consider the adjoint action of elements $H \in \mfa$ on $\mfg$. The eigenvectors $E_\alpha$ of this action satisfy
\begin{align}
[H, E_\alpha] = \alpha(H) E_\alpha
\end{align}
and are called restricted root vectors, and the eigenvalues $\alpha$ are called restricted roots. The dimension $m_\alpha$ of the restricted root space
$\mfg_\alpha = \text{span}\, \{E_\alpha\}$
is called the multiplicity of the restricted root $\alpha$, and can be bigger that 1. Restricted roots are linear functions on $\mfa$, and lie in the space $\mfa^*$ dual to $\mfa$. The dimension $n$ of both $\mfa$ and $\mfa^*$ is the {\it rank} of the symmetric space $G/K$. This is what we earlier called the number of bosonic replicas. Basis elements of $\mfa$ will be denoted by $H_k$, so that a generic element $H \in \mfa$ is $H = \sum_{k=1}^n h_k H_k$.
The dual basis in $\mfa^*$ is defined as elements $x_i$ such that $x_i(H) = h_i$ ($i = 1,\ldots, n$). In terms of this basis the restricted roots will be of three types:
\begin{align}
\pm \alpha_{ij} &= \pm(x_i - x_j),
\quad
\pm\beta_{ij} = \pm(x_i + x_j), & i < j,
\nonumber \\
\pm\gamma_i &= \pm 2x_i.
\label{roots}
\end{align}
The roots $\pm \alpha_{ij}$ and $\pm \beta_{ij}$ are ordinary roots with multiplicities $m_{o,\alpha}$ and $m_{o,\beta}$, respectively, while the roots $\pm \gamma_{i}$ are long roots with multiplicity $m_l$. Short roots $\pm x_i$ will not appear in the context of the $\sigma$-models that we consider. The multiplicities of the roots are known in all ten symmetry classes, see, for examples, the books~\cite{Helgason-Differential-1978, onishchik1990}. In the chiral classes there are no $\beta$-roots, so $m_{o,\beta} = 0$. In the remaining seven classes $m_{o,\alpha} = m_{o,\beta}$. In what follows, we will compute these multiplicites for the three classes AII, D, and DIII that are in the main focus of this paper. They turn out nozero, so the root system~\eqref{roots} will be $C_n$ in the usual Cartan notation.

A system of positive restricted roots is defined by choosing some hyperplane through the origin of $\mfa^*$ which divides $\mfa^*$ in two halves, and then defining one of these halves as positive. We will always choose $\alpha_{ij}$, $\beta_{ij}$, and $\gamma_i$ as the positive roots. The Weyl vector $\rho$  is defined as the half-sum of positive restricted roots accounting for their multiplicities. In the replica limit $n \to 0$ this gives
\begin{align}
\rho &= \lim_{n \to 0} \frac{1}{2} \sum_{\alpha > 0} m_\alpha \alpha = \sum_{i} c_i x_i,
\nonumber \\
c_i &= \frac{m_{o,\alpha} - m_{o,\beta}}{2} + m_l - m_{o,\alpha} i.
\label{Weyl-vector}
\end{align}
Positive restricted roots generate the nilpotent Lie algebra
$\mfn = \sum_{\alpha > 0} \mfg_\alpha$.
The Iwasawa decomposition at the Lie algebra level is
\begin{align}
\mfg= \mfk\oplus \mfa \oplus \mfn
\label{Iwasawa-decomposition-algebra}
\end{align}

Exponentiation of Eq.~\eqref{Iwasawa-decomposition-algebra} gives the global form of the Iwasawa decomposition
\begin{align}
G = NAK,
\label{Iwasawa-decomposition-group}
\end{align}
which allows us to represent any element $g \in G$ in the form $g = nak$, with $n \in N = e^\mfn$, $a \in A = e^\mfa$, and $k \in K = e^{\mfk}$. This factorization is unique once the system of positive restricted roots is fixed, and provides a very useful parametrization of the target space $G/K$. An element $a \in A$ is fully specified by $n$ real numbers $x_i(\ln a)$, which play the role of radial coordinates on $G/K$. For simplicity, we will denote these radial coordinates simply by $x_i$. Thus $x_i$ may now have two different meanings: either its original meaning as a basis element in $\mfa^*$, or the new one as an $N$-radial function $x_i(\ln a)$ on $G/K$. It should be clear from the context which of the two meanings is being used.

Using the radial coordinates, the joint $N$-radial eigenfunctions of the Laplace-Casimir operators on $G/K$ take a very simple exponential form
\begin{align}
\phi_\mu(Q) = e^{(\mu + \rho)(\ln a)},
\label{plane-wave-1}
\end{align}
where $a$ is the $a$-factor in the Iwasawa decomposition of $g$ in $Q = g \Lambda g^{-1}$, and
$\mu = \sum_{i} \mu_i x_i$
is a weight vector in $\mfa^*$ with arbitrary real or even complex components $\mu_i$. We will also use the notation
\begin{align}
\lambda &= -(\mu + \rho)/2,
&
q_i &= -(\mu_i + c_i)/2,
\end{align}
in which the exponential functions~\eqref{plane-wave-1} become
\begin{align}
\phi_\lambda \equiv \phi_{(q_1,q_2,\ldots,q_n)} = \exp \Big(\!-\!2 \sum_{i} q_i x_i \Big).
\label{plane-wave-2}
\end{align}

To construct the exponential $N$-radial eigenfunctions explicitly as combinations of matrix elements of $Q$, we use the key fact that there exists a choice of basis in which elements of $\mfa$ and $a \in A$ are diagonal matrices, while elements of $\mfn$ are strictly upper triangular, and elements $n \in N$ are upper triangular with units on the diagonal. This has immediate consequences for the matrix $Q \Lambda$: since elements of $K$ commute with $\Lambda$, the Iwasawa decomposition $g = nak$ leads to $Q \Lambda = n a^2 \Lambda n^{-1} \Lambda$, which is a product of an upper triangular, a diagonal, and a lower triangular matrices. In this form the lower principal minors of the advanced-advanced (AA)  block of $Q \Lambda$ are simply products of diagonal elements of $a^2$, which are exponentials of the radial coordinates $x_i$ on $G/K$. These minors are basic $N$-radial spherical functions on $G/K$ which can be raised to arbitrary powers and multiplied to produce the most general exponential functions~\eqref{plane-wave-1}. A great advantage of this construction is that is directly gives the general positive scaling operators that can be raised to arbitrary powers and satisfy the Abelian fusion rules.

Let us now present elements of the Iwasawa construction that are the same for all symmetry classes. The groups $G$ and $K$ will act in the space
\begin{align}
\mathbb{C}^{4n} &= \mathbb{C}^2 \otimes \mathbb{C}^2 \otimes \mathbb{C}^n,
\label{tensor-product-space-3}
\end{align}
where the factors in the tensor product correspond in this order to advanced-retarded, spin, and replica spaces. We will use the standard Pauli matrices $\sigma_i$ including the identity matrix $\sigma_0$. These act in either of the two first factors in Eq.~\eqref{tensor-product-space-3}, and we introduce short-hand notations for various tensor products
\begin{align}
\Sigma_{i} &\equiv \sigma_i \otimes I_n,
\qquad
\sigma_{jk} \equiv \sigma_j \otimes \sigma_k,
\nonumber \\
\Sigma_{jk} &\equiv \sigma_{jk} \otimes I_n
= \sigma_j \otimes \sigma_k \otimes I_n.
\label{Sigma-ij-definition}
\end{align}
For example $\Sigma_{00} = I_{4n}$, and $\Sigma_{30} = \Lambda$, the usual $\Lambda$ matrix from the sigma model. In symmetry classes with broken spin symmetry we can omit the second factor in the space~\eqref{tensor-product-space-3}, resulting in
\begin{align}
\mathbb{C}^{2n} &= \mathbb{C}^2 \otimes \mathbb{C}^n.
\label{tensor-product-space-2}
\end{align}

We will use a standard notation for the matrix units: $E_{ij}$ is the matrix with $1$ in the $i$-th row and $j$-th column, all other entries being zero. The symmetric and anti-symmetric combinations of matrix units are denoted as
\begin{align}
E^+_{ij} & = E_{ij} + E_{ji},
& i &\leqslant j,
&
E^-_{ij} & = E_{ij} - E_{ji},
& i &< j.
\label{E^pm}
\end{align}

Another common element in the constructions below is a basis rotation in the spaces~\eqref{tensor-product-space-3} or~\eqref{tensor-product-space-2}. This will be facilitated by the unitary matrix
\begin{align}
U &= (\sigma_0 + i \sigma_1 + i \sigma_2 + i \sigma_3)/2
\label{U-Iwasawa}
\end{align}
that cyclically permutes the Pauli matrices: $U^{-1} \sigma_j U = \sigma_{j+1}$, where the addition of one in the index is understood modulo 3.

\subsection{Class D}

In this section we present details of the Iwasawa construction for class D which is the simplest of the three classes considered in this paper. In class D we have $M_B = \text{Sp}(2n, \mathbb{R})/\text{U}(n)$. Elements $g \in \text{Sp}(2n, \mathbb{R})$
are complex matrices that act in the space~\eqref{tensor-product-space-2} and
\begin{align}
g^T \Sigma_2 g &= \Sigma_2,
&
g^\dagger \Sigma_3 g &= \Sigma_3.
\label{Sp-definition}
\end{align}
(This definition is related to a more common one where $\hat{g}$ are real matrices satisfying $\hat{g}^T \Sigma_2 \hat{g} = \Sigma_2$ by the unitary transformation $g = U_\text{D}^{-1} \hat{g} U_\text{D}$ with $U_\text{D}$ from Eq.~\eqref{U-D}.) For elements of the Lie algebra $Z = \ln g\in \mfg$ we have
\begin{align}
Z^T \Sigma_2 + \Sigma_2 Z &= 0,
&
Z^\dagger \Sigma_3 + \Sigma_3 Z &= 0,
\end{align}
In terms of $n \times n$ blocks in the replica space, we have
\begin{align}
Z = \begin{pmatrix} X & Y \\ Y^* & X^* \end{pmatrix},
\end{align}
where $X$ is skew-Hermitian and $Y$ is symmetric. In this form it is very easy to identify the subalgebra u$(n)$ as the one with $Y = 0$.

The Cartan involution is
\begin{align}
\theta(Z) = - Z^\dagger = \Sigma_3 Z \Sigma_3,
\label{Cartan-involution-D}
\end{align}
and its eigenspaces are characterised as follows: $Z \in \mfk$ if $Y = 0$, and $Z \in \mfp$ if $X = 0$. We have two groups of generators in both $\mfk$ and $\mfp$:
\begin{align}
X^{(1)}_{ij} &= \sigma_0 \otimes E^-_{ij},
&
X^{(2)}_{ij} &= i\sigma_3 \otimes E^+_{ij},
\nonumber \\
Y^{(1)}_{ij} &= \sigma_1 \otimes E^+_{ij},
&
Y^{(2)}_{ij} &= \sigma_2 \otimes E^+_{ij}.
\label{p-generators-D}
\end{align}


We choose the maximal Abelian subspace $\mfa \subset \mfp$ as:
\begin{align}
\mfa &= \text{span}\left\{ H_k = \sigma_1 \otimes E_{kk}
= Y^{(1)}_{kk}/2 \right\}.
\end{align}
Straightforward computations show that the system of restricted roots is given by Eq.~\eqref{roots} with multiplicities $m_o = 1$ and $m_l = 1$. The positive restricted root vectors are
\begin{align}
E_{\alpha_{ij}} &= X^{(1)}_{ij} + Y^{(1)}_{ij},
&
E_{\beta_{ij}} &= X^{(2)}_{ij} + Y^{(2)}_{ij},
\nonumber \\
E_{\gamma_i} &= X^{(2)}_{ii} + Y^{(2)}_{ii}.
\label{positive-RRV-D}
\end{align}


Next, we perform a unitary transformation of basis in the space~\eqref{tensor-product-space-3} that makes the generators of $\mfa$ diagonal, and the generators of $\mfn$ strictly upper-triangular. Using the matrix $U$ in Eq.~\eqref{U-Iwasawa}, we define
\begin{align}
U_\text{D} &= U \otimes I_n.
\label{U-D}
\end{align}
The unitary transformation $M \to U_\text{D} M U_\text{D}^{-1}$ makes the elements of $\mfa$ diagonal and rotates the $\Lambda$ matrix:
\begin{align}
U_\text{D} H_k U_\text{D}^{-1} &= \sigma_3 \otimes E_{kk},
&
U_\text{D} \Lambda U_\text{D}^{-1} &= \Sigma_2.
\end{align}
In this basis the positive restricted root vectors are upper-triangular in the retarded-advanced (RA) space. An additional permutation $\pi$ that reverses the order of the basis in the AA sector accomplishes the necessary upper triangularization. The permutation is given by
\begin{align}
\pi(n+i) = 2n + 1 - i.
\label{permutation-pi-D}
\end{align}
Let $\Pi$ be the permutation matrix with elements $\Pi_{ij} = \delta_{\pi(i), j}$, then the unitary transformation
\begin{align}
\tilde{M} &= \Pi^{-1} U_\text{D} M U_\text{D}^{-1} \Pi
\label{tilde-M-D}
\end{align}
leads to
\begin{align}
\tilde{\Lambda} &= \sigma_{2} \otimes \mathcal{I}_n,
\end{align}
where $\mathcal{I}_n$ is the $n \times n$ matrix with units on the ``anti-diagonal'', that is, $(\mathcal{I}_n)_{ij} = \delta_{i, n + 1 - j}$. It is easy to show that in the new basis the positive restricted root vectors $\tilde{E}$ are strictly upper-triangular.

Let us exploit consequences of the Iwasawa decomposition of $G$ and the permutation $\pi$ for the sigma model field $Q$. In the original basis we write $g = nak$ with $n \in N$, $a \in A$, and $k \in K$, and then
\begin{align}
\mathcal{Q} &\equiv Q \Lambda
= n a^2 \Lambda n^{-1} \Lambda.
\end{align}
Here we used $k \Lambda k^{-1} = \Lambda$ and $a \Lambda a^{-1} = a^2
\Lambda$, which is a special case of second condition in Eq.~\eqref{Sp-definition} for a diagonal matrix $a \in G$. Now we perform the permutation $P_\pi$. Using the notation~\eqref{tilde-M-D}, we get
\begin{align}
\tilde{\mathcal{Q}} = \tilde{n} \tilde{a}^2 \tilde{\Lambda} \tilde{n}^{-1} \tilde{\Lambda}.
\label{rotated-Q}
\end{align}
As should be clear from the previous discussion, the matrices $\tilde{n}$ and $\tilde{n}^{-1}$ are upper-triangular with units on the diagonals, while $\tilde{a}$ is diagonal:
\begin{align}
\tilde{a} = \text{diag} (e^{x_1}, \ldots, e^{x_n},  e^{-x_n},\ldots, e^{-x_1}).
\label{eq:iwasawa-d-tilde-a}
\end{align}
Conjugation by $\tilde{\Lambda}$ converts $\tilde{n}^{-1}$ into $\tilde{\Lambda} \tilde{n}^{-1} \tilde{\Lambda}$ which is {\it lower-triangular} with units on the diagonal. This results in the following structure of the lower-right $m \times m$ submatrix of the AA block of the matrix $\tilde{\mathcal{Q}}$ for any $m \leq n$:
\begin{align}
\tilde{\mathcal{Q}}_{AA}^{(m)} =
\begin{pmatrix}
1 & \ldots & * \\
\vdots & \ddots & \vdots \\
0 & \ldots & 1
\end{pmatrix}
\begin{pmatrix}
e^{-2x_{m}} & \ldots & 0 \\
\vdots & \ddots & \vdots \\
0 & \ldots & e^{-2x_1}
\end{pmatrix}
\begin{pmatrix}
1 & \ldots & 0 \\
\vdots & \ddots & \vdots \\
* & \ldots & 1
\end{pmatrix}.
\label{Q-tilde-AA}
\end{align}
Determinants of these submatrices give the basic positive $N$-radial eigenfunctions
\begin{align}
d_{m} = \phi_{(1^m)}
= \exp \Big(- 2 \sum_{i=1}^{m} x_i \Big).
\label{basic-d-2m}
\end{align}
We can form the most general $N$-radial eigenfunctions as products
\begin{align}
\phi_{(q_1,\ldots,q_n)} = d_1^{q_1 - q_2} d_2^{q_2 - q_3}
\ldots d_{n-1}^{q_{n-1} - q_n} d_{n}^{q_n},
\label{d-product}
\end{align}
where we may take $q_i$ to be arbitrary complex numbers.

The resulting form of the Iwasawa construction for class D as given by Eqs.~\eqref{eq:iwasawa-d-tilde-a}--\eqref{d-product} is fully analogous to that in class A, Ref.~\cite{gruzberg2013classification}. This demonstrates that class D is one of ``spinless'' symmetry classes. Upon translation to the language of eigenfunctions, one obtains the pure-scaling observables in the form~\eqref{eq:det},~\eqref{eq:abelianA}. We will see below that the situation is different for classes DIII and AII, for which a distinct (although bearing much similarity), spinful construction emerges out of the Iwasawa decomposition.

\subsection{Class DIII}

In this section we present details of the Iwasawa construction for class DIII. In this class we have $M_B = \text{Sp}(2n, \mathbb{C})/\text{Sp}(2n)$. The group $\text{Sp}(2n, \mathbb{C})$ can be realized as the group of complex $4n \times 4n$ matrices that act in the space~\eqref{tensor-product-space-3} and satisfy the following constraints:
\begin{align}
g^\dagger \Sigma_{30} g &= \Sigma_{30},
&
g^T \Sigma_{32} g &= \Sigma_{32},
&
g^T \Sigma_{20} g &= \Sigma_{20}.
\end{align}
(This definition is equivalent to a more common one where $\hat{g}$ are $2n \times 2n$ complex matrices acting in the space~\eqref{tensor-product-space-2} and satisfying $\hat{g}^T \Sigma_2 \hat{g} = \Sigma_2$.) For elements of the Lie algebra $Z = \ln g\in \mfg$ we have
\begin{align}
Z^\dagger \Sigma_{30} + \Sigma_{30} Z &= 0,
\qquad
Z^T \Sigma_{32} + \Sigma_{32} Z = 0,
\nonumber \\
Z^T \Sigma_{20} + \Sigma_{20} Z &= 0.
\label{hat-Z-conditions-DIII}
\end{align}
This leads to the following structure in the RA space:
\begin{align}
Z &= \begin{pmatrix*}[c] X & Y \\ Y^* & X^* \end{pmatrix*},
\end{align}
where the blocks can be split further:
\begin{align}
X &= \begin{pmatrix*}[r] X_1 & X_2 \\ -X_2^* & X_1^* \end{pmatrix*},
&
Y &= \begin{pmatrix*}[r] Y_2 & Y_1 \\ -Y_1^* & Y_2^* \end{pmatrix*}.
\end{align}
Here all sub-blocks are complex $n \times n$ matrices: $X_1$ and $Y_1$ skew-Hermitian, and $X_2$ and $Y_2$ symmetric.

We use the Cartan involution
\begin{align}
\theta(Z) = - Z^\dagger = \Sigma_{30} Z \Sigma_{30},
\label{Cartan-involution-AII}
\end{align}
whose eigenspaces are characterised as follows: $Z \in \mfk$ if $Y = 0$, and $Z \in \mfp$ if $X = 0$. This gives four groups of generators for each eigenspace:
\begin{align}
X^{(0)}_{ij} &= \sigma_{00} \otimes E^-_{ij},
&
Y^{(0)}_{ij} &= \sigma_{10} \otimes E^+_{ij},
\nonumber \\
X^{(1)}_{ij} &= i \sigma_{31} \otimes E^+_{ij},
&
Y^{(1)}_{ij} &= \sigma_{21} \otimes E^+_{ij},
\nonumber \\
X^{(2)}_{ij} &= i \sigma_{02} \otimes E^+_{ij},
&
Y^{(2)}_{ij} &= i \sigma_{12} \otimes E^-_{ij},
\nonumber \\
X^{(3)}_{ij} &= i \sigma_{33} \otimes E^+_{ij},
&
Y^{(3)}_{ij} &= \sigma_{23} \otimes E^+_{ij}.
\label{p-generators-DIII}
\end{align}

We choose the maximal Abelian subspace $\mfa \subset \mfp$ as:
\begin{align}
\mfa &= \text{span}\left\{ H_k = \sigma_{10} \otimes E_{kk}
= Y^{(0)}_{kk}/2 \right\}.
\end{align}
Straightforward computations show that the system of restricted roots is given by Eq.~\eqref{roots} with multiplicities $m_o = 2$ and $m_l = 2$. The positive restricted root vectors are
\begin{align}
E_{\alpha_{ij}}^{(1)} &= X^{(0)}_{ij} + Y^{(0)}_{ij},
&
E_{\alpha_{ij}}^{(2)} &= X^{(2)}_{ij} + Y^{(2)}_{ij},
\nonumber \\
E_{\beta_{ij}}^{(1)} &= X^{(1)}_{ij} + Y^{(1)}_{ij},
&
E_{\beta_{ij}}^{(2)} &= X^{(3)}_{ij} + Y^{(3)}_{ij},
\nonumber \\
E_{\gamma_i}^{(1)} &= X^{(1)}_{ii} + Y^{(1)}_{ii},
&
E_{\gamma_i}^{(2)} &= X^{(3)}_{ii} + Y^{(3)}_{ii}.
\label{positive-RRV-DIII}
\end{align}

The unitary transformation that makes the generators of $\mfa$ diagonal and the generators of $\mfn$ strictly upper-triangular is accompished with the help of the matrix
\begin{align}
U_\text{DIII} &= U \otimes \sigma_0 \otimes I_n.
\label{U-DIII}
\end{align}
We also need the permutation matrix $\Pi_1$ with elements $(\Pi_1)_{ij} = \delta_{\pi_1(i), j}$ where the permutation $\pi_1$ of the basis of the space~\eqref{tensor-product-space-3} can be described as follows: for $i \in 1, \ldots, n$, we have
\begin{align}
\pi_1(i) &= 2i-1,
&
\pi_1(2n+i) &= 4n + 2 - 2i,
\nonumber \\
\pi_1(n+i) &= 2i,
&
\pi_1(3n+i) &= 4n + 1 - 2i.
\label{permutation-pi-DIII}
\end{align}
The unitary transformation
\begin{align}
\tilde{M} &= \Pi_1^{-1} U_\text{DIII} M U_\text{DIII}^{-1} \Pi_1
\label{tilde-M-D}
\end{align}
rotates the $\Lambda$ matrix to
\begin{align}
\tilde{\Lambda} &= \sigma_{2} \otimes \mathcal{I}_{2n},
\end{align}
makes the elements of $\mfa$ diagonal, and the positive restricted root vectors $\tilde{E}$ strictly upper-triangular.

The subsequent construction is almost verbatim as in class D, except that each entry in the diagonal matrix $\tilde{a} \in A$ is now repeated twice (the doubling is a manifestation of the Kramers degeneracy):
\begin{align}
\tilde{a} = \text{diag} (e^{x_1} \sigma_0, \ldots, e^{x_n} \sigma_0,
e^{- x_n} \sigma_0, e^{-x_1} \sigma_0).
\label{tilde-a-DIII}
\end{align}
The structure of $\tilde{\mathcal{Q}}_{AA}^{(2m)}$, the lower-right $2m \times 2m$ submatrix of the $AA$ block of the matrix $\tilde{\mathcal{Q}}$ for any $m \leq n$ is the same as in Eq.~\eqref{Q-tilde-AA}, except that now all entries are understood as $2 \times 2$ matrices, with the blocks on the diagonals proportional to the identity matrix $\sigma_0$. Determinants of $\tilde{\mathcal{Q}}_{AA}^{(2m)}$ give the basic positive $N$-radial eigenfunctions
\begin{align}
d_{2m} = \exp \Big(- 4 \sum_{i=1}^{m} x_i \Big).
\label{basic-d-2m}
\end{align}
We can form the most general $N$-radial eigenfunctions as products
\begin{align}
\phi_{(q_1,\ldots,q_n)} = d_2^{(q_1 - q_2)/2} d_4^{(q_2 - q_3)/2}
\ldots d_{2n}^{q_n/2},
\label{d-product-DIII}
\end{align}
where we may take $q_i$ to be arbitrary complex numbers. It is easy to see that the product~\eqref{d-product-DIII} is the same as the exponential eigenfunction~\eqref{plane-wave-2}, while the basic function $d_{2m}$ is $\phi_{(2,2,\ldots)}$ with $m$ twos in the subscript.

Notice that the doubling of the diagonal entries $e^{-2x_i}$ for each $i$ in Eq.~\eqref{tilde-a-DIII} compelled us to take determinants of sub-matrices of even size and raise the resulting functions to powers written as $(q_i - q_{i+1})/2$. In the Iwasawa formalism it is also possible to obtain directly the basic solutions $\varphi_{(1^m)}$. Using definitions above, it is straightforward to show that the matrix $\tilde{\mathcal{Q}}_{AA}^{(2m)} (I_{m} \otimes i\sigma_2)$ is anti-symmetric, and that its Pfaffian gives the basic $N$-radial eigenfunction
\begin{align}
p_m \equiv \phi_{(1^m)} = \exp \Big(-2 \sum_{i=1}^{m} x_i \Big)
= \text{Pf} \, [\tilde{\mathcal{Q}}_{AA}^{(2m)} (I_{m} \otimes i\sigma_2)].
\end{align}
The general $N$-radial functions are then obtained as products of powers of $p_m$:
\begin{align}
\phi_{(q_1,\ldots,q_n)} = p_1^{q_1 - q_2} p_2^{q_2 - q_3}
\ldots p_{n-1}^{q_{n-1} - q_n} p_n^{q_n}.
\label{p-product-DIII}
\end{align}

The resulting form of the Iwasawa construction for class DIII as given by Eqs.~\eqref{tilde-a-DIII}--\eqref{p-product-DIII} is fully analogous to that in class C, Ref.~\cite{karcher2021generalized}. This shows that class DIII is one of ``spinful'' symmetry classes. Upon translation into the language of eigenfunctions, one obtains the pure-scaling observables as given by Eqs.~\eqref{eq:abelianA} and~\eqref{eq:det-sp}.

\subsection{Class AII}

In this section we present details of the Iwasawa construction for class AII, which is very similar to that for class DIII above. In class AII we have $M_B = G/K$ where $G = \text{Sp}(2n,2n)$, $K = \text{Sp}(2n) \times \text{Sp}(2n)$. The group $\text{Sp}(2n,2n)$ is the subgroup of the complex symplectic group $\text{Sp}(4n,\mathbb{C})$ that preserves the symmetric bilinear from with the matrix $\Sigma_{30}$ in the space~\eqref{tensor-product-space-3}:
\begin{align}
g^T \Sigma_{20} g &= \Sigma_{20},
&
g^\dagger \Sigma_{30} g &= \Sigma_{30}.
\label{Sp-definition-Helgason}
\end{align}
For elements of the Lie algebra $Z = \ln g\in \mfg$ we have
\begin{align}
Z^T \Sigma_{20} + \Sigma_{20} Z &= 0,
&
Z^\dagger \Sigma_{30} + \Sigma_{30} Z &= 0.
\label{sp-definition-Helgason}
\end{align}
In terms of $n \times n$ blocks in the replica space we have the following structure:
\begin{align}
Z &= \begin{pmatrix*}[r]
Z_{11} & Z_{12} & Z_{13} & Z_{14} \\
-Z_{12}^* & Z_{11}^* & Z_{14}^* & - Z_{13}^* \\
Z_{13}^\dagger & Z_{14}^T & Z_{33} & Z_{34} \\
Z_{14}^\dagger & - Z_{13}^T & -Z_{34}^* & Z_{33}^*
\end{pmatrix*},
\label{Z-sp-2n-blocks}
\end{align}
where $Z_{11}$ and $Z_{33}$ are skew-Hermitian, $Z_{12}$ and $Z_{34}$ are complex symmetric,
and $Z_{13}$ and $Z_{14}$ are arbitrary complex matrices.

The Cartan involution is the same as in class DIII:
\begin{align}
\theta(Z) = - Z^\dagger = \Sigma_{30} Z \Sigma_{30}.
\label{Cartan-involution-AII}
\end{align}
Its eigenspaces are characterised as follows: $Z \in \mfk$ if $Z_{13} = Z_{14} = 0$, and $Z \in \mfp$ if $Z_{11} = Z_{12} = Z_{33} = Z_{34} = 0$. The eight groups of generators of $\mfk$ are
\begin{align}
X^{00}_{ij} &\equiv \sigma_{00} \otimes E^-_{ij},
&
X^{30}_{ij} &\equiv \sigma_{30} \otimes E^-_{ij},
\nonumber \\
X^{01}_{ij} &\equiv i\sigma_{01} \otimes E^+_{ij},
&
X^{31}_{ij} &\equiv i\sigma_{31} \otimes E^+_{ij},
\nonumber \\
X^{02}_{ij} &\equiv i\sigma_{02} \otimes E^+_{ij},
&
X^{32}_{ij} &\equiv i\sigma_{32} \otimes E^+_{ij},
\nonumber \\
X^{03}_{ij} &\equiv i\sigma_{03} \otimes E^+_{ij},
&
X^{33}_{ij} &\equiv i\sigma_{33} \otimes E^+_{ij}.
\label{k-generators-AII}
\end{align}
and the eight groups of generators of $\mfp$ are
\begin{align}
Y^{20}_{ij} &= \sigma_{20} \otimes E^+_{ij},
&
Y^{10}_{ij} &= i\sigma_{10} \otimes E^-_{ij},
\nonumber \\
Y^{21}_{ij} &= i\sigma_{21} \otimes E^-_{ij},
&
Y^{11}_{ij} &= \sigma_{11} \otimes E^+_{ij},
\nonumber \\
Y^{22}_{ij} &= i\sigma_{22} \otimes E^-_{ij},
&
Y^{12}_{ij} &= \sigma_{12} \otimes E^+_{ij},
\nonumber \\
Y^{23}_{ij} &= i\sigma_{23} \otimes E^-_{ij},
&
Y^{13}_{ij} &= \sigma_{13} \otimes E^+_{ij}.
\label{p-generators-AII}
\end{align}


We choose the maximal Abelian subspace $\mfa \subset \mfp$ as
\begin{align}
\mfa &= \text{span}\left\{ H_k = \sigma_{20} \otimes E_{kk}
= Y^{20}_{kk}/2 \right\}.
\end{align}
Straightforward computations show that the system of restricted roots is given by Eq.~\eqref{roots} with multiplicities $m_o = 4$, $m_l = 3$.
The positive restricted root vectors are
\begin{align}
E_{\alpha_{ij}}^{(1)} &= X^{00}_{ij} + Y^{20}_{ij},
&
E_{\alpha_{ij}}^{(2)} &= X^{01}_{ij} + Y^{21}_{ij},
\nonumber \\
E_{\alpha_{ij}}^{(3)} &= X^{02}_{ij} + Y^{22}_{ij},
&
E_{\alpha_{ij}}^{(4)} &= X^{03}_{ij} + Y^{23}_{ij},
\nonumber \\
E_{\beta_{ij}}^{(1)} &= X^{30}_{ij} + Y^{10}_{ij},
&
E_{\beta_{ij}}^{(2)} &= X^{31}_{ij} - Y^{11}_{ij},
\nonumber \\
E_{\beta_{ij}}^{(3)} &= X^{32}_{ij} - Y^{12}_{ij},
&
E_{\beta_{ij}}^{(4)} &= X^{33}_{ij} - Y^{13}_{ij},
\nonumber \\
E_{\gamma_i}^{(1)} &= X^{31}_{ii} - Y^{11}_{ii},
&
E_{\gamma_i}^{(2)} &= X^{32}_{ii} - Y^{12}_{ii},
\nonumber \\
E_{\gamma_i}^{(3)} &= X^{33}_{ii} - Y^{13}_{ii}.
\end{align}


The upper-triangularization is accomplished with the help of the same matrices $U_\text{DIII}$~\eqref{U-DIII} and $\Pi_1$ that we used in class DIII. The unitary transformation
\begin{align}
\tilde{M} &= \Pi_1^{-1} U_\text{DIII}^{-1} M U_\text{DIII} \Pi_1
\label{tilde-M-AII}
\end{align}
rotates the $\Lambda$ matrix to
\begin{align}
\tilde{\Lambda} &= \sigma_{11} \otimes \mathcal{I}_n,
\end{align}
makes the elements of $\mfa$ diagonal, and the positive restricted root vectors $\tilde{E}$ strictly upper-triangular.

The rest of the construction is identical to that in class DIII. Therefore, we obtain the same construction with each matrix element of the diagonal matrix $\tilde{a}$ repeated twice, yielding Eqs.~\eqref{tilde-a-DIII}--\eqref{p-product-DIII}. In terms of eigenfunction observables, we thus get the spinful construction, Eqs.~\eqref{eq:abelianA} and~\eqref{eq:det-sp}.

\section{One-loop $\sigma$-model RG and $K$-invariant eigenfunctions for all ten symmetry classes}
\label{appendix:rg}

In this appendix, we present the one-loop RG analysis for the $\sigma$-model, which allows us to determine the $K$-invariant polynomial composite operators
${\cal P}_\lambda (Q)$ as eigenfunctions of the RG. While in the rest of the paper we focus on three classes AII, D, and DIII, here we present the results for all ten symmetry classes. This allows us to demonstrate relations between different symmetry classes and, in particular, their splitting in two groups (``spinless'' and ``spinful''), in full agreement with physical considerations based on the presence or absence of Kramers degeneracy and Kramers-like near-degeneracy (see Sec.~\ref{sec:pure_scaling}) and with the Iwasawa construction (Appendix~\ref{app:Iwasawa}).

For classes A and C, this RG analysis was presented in Ref.~\cite{karcher2021generalized}; the corresponding main results (with some sign typos corrected) are also collected in Appendix B of Ref.~\cite{karcher2022generalized}. Below we use the same notations as in Refs.~\cite{karcher2021generalized,karcher2022generalized}.

Let us first comment on the notion of $K$-invariant (or, equivalently, $K$-radial) eigenfunctions, and their difference from (and relation to) the $N$-radial eigenfunctions resulting from the Iwasawa decomposition. This was explained in Ref.~\cite{gruzberg2013classification} and is brielfy reiterated here. The $Q$-matrix is given by $Q= g \Lambda g^{-1}$, with $g \in G$.  We use the Cartan decomposition, $G=KAK$, where $A$ is the maximal Abelian subgroup in $G/K$ (the same as in the Iwasawa construction, Appendix~\ref{app:Iwasawa}), which implies that any element $g$ of $G$ can be presented in the form $g = k_1 a k_2$, with $k_1, k_2 \in K$ and $a \in A$. This yields $Q=k_1 a \Lambda a^{-1} k_1^{-1}$. An operator ${\cal P}(Q)$ is called $K$-invariant if it satisfies ${\cal P}(Q) = {\cal P}(kQk^{-1})$ for any $k \in K$. Clearly, one has then ${\cal P}(Q) = {\cal P}(a \Lambda a^{-1})$, i.e.,  a $K$-invariant operator depends only on the coordinates parametrizing $a \in A$ (called $K$-radial coordinates). For any irreducible representation $\lambda$, one can (uniquely) construct a $K$-radial representative, which is done as follows. Let $\tilde{\cal P}_\lambda(Q)$ be any function belonging to $\lambda$.  Then by symmetry $\tilde{\cal P}_\lambda(kQk^{-1})$ also belongs to $\lambda$.  Integrating with the Haar measure over the group $K$, we obtain the sought  $K$-invariant function:  ${\cal P}_\lambda(Q) = \int_K d\mu(k) \tilde{\cal P}_\lambda(kQk^{-1})$. (Notice that for some choices of $\tilde{\cal P}_\lambda(Q)$ this integral over $K$ may vanish.) In particular, $\tilde{\cal P}_\lambda(Q)$ here can be the $N$-radial function $\phi_\lambda(Q)$ resulting from the Iwasawa construction.

The $K$-invariant eigenfunctions ${\cal P}_\lambda (Q)$ are known as zonal spherical functions. In the case of conventional spherical functions $Y_{lm}$ on the sphere $S^2 = {\rm SU(2) / U(1)}$, these are the $m=0$ spherical harmonics, which do not depend on the azimuthal angle $\phi$ (Legendre polynomilas of $\cos \theta$). For comparison, the $N$-radial eigenfunctions (or, equivalently, the highest-weight vectors) are in this case the $m=l$ harmonics.

We consider $K$-invariant, polynomial-in-$Q$ functions on the $\sigma$-model target space $G/K$. For a given order $q$ of the polynomials, it is convenient to introduce a basis in the corresponding linear space, with basis function labeled by integer partitions $\lambda=(q_1,\ldots,q_n)$  of $q$ with $q_1\geq\ldots\geq q_n$, where $q_i$ are positive integers and $ \sum_i q_i =q$. Elements of this basis have the form
\begin{align}
{\cal O}_\lambda \equiv {\cal O}_{(q_1,\ldots, q_n)} &= \prod_{i=1}^n \mathrm{tr} \left[(\Lambda Q)^{q_i}\right].
\label{eq:o_lambda}
\end{align}
As is clear from Eq.~\eqref{eq:o_lambda}, the integers $q_i$ here are the lengths of $n$ cycles of $\Lambda Q$-strings over which the traces are taken.

For $q=2$, there are two basis operators, with $\lambda = (1,1)$ and (2).  The one-loop RG conserves the order $q$. Since it is a linear operation, it works as a matrix in the corresponding two-dimensional space. To derive this matrix, it is convenient to use the background-field formalism. The field $g \in G$ is split into fast $g_f$ and slow $g_s$ components, $g= g_s g_f$. The fast field is expanded as $g_f = e^{- {\cal X}} = 1 - {\cal X} + \frac12 {\cal X}^2 + \ldots$, with ${\cal X}\Lambda = -\Lambda {\cal X} $,
which yields
\begin{align}
Q &= g \Lambda g^{-1} = g_s g_f \Lambda g_f^{-1} g_s^{-1}
\nonumber \\
&= Q_s + 2 g_s \Lambda {\cal X} g_s^{-1} + 2 g_s \Lambda {\cal X}^2 g_s^{-1} + \ldots,
\label{eq:Q-RG-expansion}
\end{align}
where $Q_s = g_s \Lambda g_s^{-1}$. For the one-loop RG, the terms beyond the ${\cal X}^2$ order are not needed. To perform the RG transformation on an operator ${\cal O}(Q)$, we expand  ${\cal O}(Q)$ up to the second order in ${\cal X}$ using Eq.~\eqref{eq:Q-RG-expansion},
${\cal O} = {\cal O}^{(0)} + {\cal O}^{(1)} [{\cal X}]  + \frac12 {\cal O}^{(2)} [{\cal X}]+ \ldots$,
and then perform the Gaussian averaging over the fast fields ${\cal X}$.
The corresponding Gaussian action $S_f[{\cal X}]$ is obtained by an expansion of the $\sigma$-model action up to the quadratic order; it depends on the symmetry class under consideration. We denote by $\delta {\cal O}$ the result of the averaging of the contribution of second order in ${\cal X}$ over the fast mode with the action $S_f[{\cal X}]$, i.e.,
$\delta {\cal O} = \frac12 \langle {\cal O}^{(2)} [{\cal X}] \rangle_{S_f[{\cal X}]}$. This $\delta {\cal O}$ is a result of one-loop renormalization of the composite operator ${\cal O}(Q)$. By construction, $\delta {\cal O}$  depends on the slow field $Q_s$; we denote this field again by $Q$.

The one-loop RG procedure is  presented in detail in Ref.~\cite{karcher2021generalized} for classes A and C; the derivation proceeds in the same way for other symmetry classes. The only difference is in the action $S_f$ that determines the $2 \times 2$ RG matrix $M_2$ (see below), which should be derived separately for each $G/K$. We omit the corresponding details and only present the results for all ten symmetry classes.

Performing the RG for $q=2$ operators, we get
\begin{align}
\delta
\begin{pmatrix}
{\rm tr}(AQ) {\rm tr}(BQ)\\
{\rm tr}(AQ BQ)
\end{pmatrix}
\!\!
&= 2I_f \, M_2
\begin{pmatrix}
{\rm tr}(AQ) {\rm tr}(BQ)\\
{\rm tr}(AQ BQ)
\end{pmatrix}\!,
\nonumber \\
M_2 &=
\begin{pmatrix*}[c]
c &\ 2a \\
1 &\ b + c
\end{pmatrix*} \!.
\label{eq:RG-M2}
\end{align}
Here $A$ and $B$ are arbitrary fixed matrices (strictly speaking, they should satisfy ${\rm tr} A = {\rm tr} B= {\rm tr} AB = 0$; otherwise additional, $Q$-independent terms appear upon RG transformation). Setting $A= B = \Lambda$ yields the RG flow for $q=2$ $K$-invariant operators. The constant $I_f$ is the one-loop integral; its value plays no role for determination of eigenfunctions that we are interested in. The RG matrix $M_2$ is determined by three constants, $a$, $b$, and $c$, as shown in Eq.~\eqref{eq:RG-M2}. The values of these parameters in different symmetry classes are presented in Table~\ref{M2-coeff}.

\begin{table}
	\begin{tabular}{c|ccc}
		class & \ $a $ \ & \ $b $ \ & \ $c $ \\
		\hline
		\hline \\[-0.3cm]
		A & 1/2 & 0 & 0 \\
		AI & 1 & 1 & 0 \\
		AII & 1 & -1 & 0 \\
		\hline \\[-0.3cm]
		D & 1 & 1 & 2 \\
		C & 1 & -1 & -2 \\
		DIII & 2 & 0 & 2 \\
		CI & 2 & 0 & -2 \\
		\hline \\[-0.3cm]
		AIII & 1/2 & 0 & 0 \\
		BDI & 1 & 1 & 1 \\
		CII & 1 & -1 & -1
	\end{tabular}
	\caption{Coefficients of the matrix $M_2$ in Eq.~\eqref{eq:RG-M2} and consequently of different terms in the differential operator~\eqref{eq:diff} for all symmetry classes. }
	\label{M2-coeff}
\end{table}

We briefly comment on the origin of contributions proportional to 1, $a$, $b$, and $c$ in Eq.~\eqref{eq:RG-M2}. Consider first the terms coming from the renormalization of  ${\rm tr}(AQ) {\rm tr}(BQ)$.  When two ${\cal X}$ fields that enter the contraction are taken from two different traces, one obtains a single trace ${\rm tr}(AQ BQ)$, with a coefficient $a$. When both ${\cal X}$ fields are taken from the same $Q$ matrix, the original structure ${\rm tr}(AQ) {\rm tr}(BQ)$ is reproduced with a coefficient $c$. Now we turn to the terms coming from the renormalization of a single trace ${\rm tr}(AQ BQ)$.  When two ${\cal X}$ fields are taken from different $Q$ fields, one gets the same structure (with a coefficient $b$) or splits the trace into two traces, ${\rm tr}(AQ) {\rm tr}(BQ)$, with a coefficient 1. Finally, if both ${\cal X}$ fields are taken from the same $Q$ field, one reproduce the original structure with a coefficient $c$.

Remarkably, the rules encoded in the matrix $M_2$, Eq.~\eqref{eq:RG-M2}, are sufficient to extend the RG onto polynomial operators of any degree $q$, as was shown in Ref.~\cite{karcher2021generalized}. A convenient formalism to extract the action of RG on a polynomial operator of any degree is as follows~\cite{karcher2022generalized}. We identify $K$-invariant basis operators ${\cal O}_\lambda$, Eq.~\eqref{eq:o_lambda}, with polynomials in variables $w_k$ in the following way. We rewrite $\lambda = (1^{m_1}, 2^{m_2},\dots, k^{m_k}, \ldots)$  in terms of cycle lengths $k$ and multiplicities $m_k$. Then  we associate the monomial $W_\lambda = \prod_k w_k^{m_k}$ to ${\cal O}_\lambda$. It is easy to see that the degree $q$ of the operator ${\cal O}_\lambda$ is in this notations $q = \sum_ k km_k$.  A generic $K$-invariant operator ${\cal O}(Q)$ maps onto a linear combination $W$ of such monomials. The action of one-loop RG can now be presented as $\delta W = 2I_f \mathcal{D} W$, where $\mathcal{D}$  is the following differential operator:
\begin{align}
\mathcal{D}&= \sum_{j< i} jw_{i-j} w_j \partial_i + a \sum_{i,j} ijw_{i+j} \partial_i\partial_j
\nonumber \\
& \quad + b \sum_i \dfrac{i(i-1)}{2}w_i\partial_i + c \sum_i iw_i\partial_i,
\label{eq:diff}
\end{align}
with $\partial_i \equiv \partial/\partial w_i$.

We briefly comment on the four terms in the RG operator ${\cal D}$ (proportional to unity, $a$, $b$, and $c$, respectively).
\begin{itemize}
	\item[(i)]
	The first term in Eq.~\eqref{eq:diff} (proportional to unity) describes cutting a cycle of length $i$  into two cycles of length $j$ and $i-j$.  Here, the derivative removes one factor $w_i$ and yields a factor $m_i$, corresponding to the fact that this can happen to any of the $m_i$ cycles of length $i$.
	The multiplication by $w_j w_{i-j}$ corresponds to the appearance of two cycles with the lengths $j$ and $i-j$. In total, there are $i=j+(i-j)$ realizations of such a cut.
	\item[(ii)]
	The second term (quadratic with respect to the derivatives, proportional to $a$)  describes the fusion of cycles of length $i$ and $j$ into a cycle of length $i+j$. Here, the derivatives remove one cycle of length $i$ and one of length $j$, while the multiplication by $w_{i+j}$ adds one cycle of the corresponding length. In total, there are $ij$ channels for this process: the first fast field can come from each of the $i$ $Q$-fields in the cycle of length $i$,
	and the second one from each of the $j$ $Q$-fields in the cycle of length $j$.
	\item[(iii)]
	The third term (proportional to $b$) originates from contractions of fast fields coming from distinct $Q$ fields in a cycle of length $i$ and preserving this cycle. This terms affects  only diagonal entries of matrices $M_q$ defined below.
	\item[(iv)]
	Finally, the last term in Eq.~\eqref{eq:diff} (proportional to $c$) results from contractions between fast fields coming from the same $Q$.  It preserves the structure of the monomial $W_\lambda$, multiplying it by $\sum_i i m_i =q$, i.e., by the total number of $Q$ fields. This term is associated with the renormalization of the average density of states, providing a contribution $q x_{(1)}$ to the exponents $x_\lambda$.  For any given $q$, it yields a contribution proportional to unit matrix to the RG matrices $M_q$ and thus does not influence their eigenvectors (which are $K$-invariant scaling operators).
	
\end{itemize}

Let us comment on peculiarities of three chiral classes, AIII, BDI, and CII. Models of these classes conventionally emerge when one studies tight-binding models with sublattice structure (two sublattices A and B, with all non-zero matrix elements of the Hamiltonian corresponding to hopping between them). For these classes, irreducible representations are labeled not simply by $\lambda$ but rather by a pair of Young diagrams $(\lambda,\bar{\lambda})$  ~\cite{gade1991the}, with $\lambda$ corresponding to observables on one sublattice and $\bar{\lambda}$ on the other sublattice~\cite{gade1993anderson}.   At the level of polynomials $W$ introduced above, we introduce a second, independent set of variables $\bar{w}_k$; the operators are now represented by a linear combination of monomials $W_\lambda(\{w_k\}) W_{\bar{\lambda}}(\{\bar{w}_k\})$. The RG operator~\eqref{eq:diff} is extended to $\mathcal{D} + \bar{\mathcal{D}}$, where $\bar{\mathcal{D}}$ has the same structure as $\mathcal{D}$, with a replacement $w_k \mapsto \bar{w}_k$. In addition, there is a contribution to the one-loop RG operator that originates from the U(1) sector of chiral-class $\sigma$-models~\cite{gade1991the}. This contribution does not affect the eigenfunctions but shifts the eigenvalues by a term proportional to $(q-\bar{q})^2$, where $q = |\lambda|$ and $\bar{q} = |\bar{\lambda}|$.

It is easy to verify that the differential operator~\eqref{eq:diff} preserves the degree $q = \sum_{k} k m_k$ of a composite operator. We can therefore restrict it to a sector of the theory with a given $q=|\lambda|$. This yields
\begin{align}
\mathcal{D} \sum_\lambda a_\lambda W_\lambda &= \sum_{\lambda,\mu} a_\lambda  (M_q)_{\lambda,\mu} W_\mu,
\label{eq:rg_operator_action}
\end{align}
with matrices $M_q$ describing the renormalization of operators of degree $q$.  For $q=2$, we are of course back to the matrix $M_2$ given in Eq.~\ref{eq:RG-M2}. For higher values of $q$ (i.e., $q=3, 4$) the matrices $M_q$ can be straightforwardly obtained numerically for each symmetry class.
If we consider~\eqref{eq:rg_operator_action} as equations describing the action of the RG operators $\mathcal{D}$ on vectors of the coefficients $a_\lambda$, this action is clearly characterized by the transposed matrix $M_q^T$.  Once  the matrices $M_q^T$ are found with this procedure, one can determine their eigenvectors that yield the sought
$K$-invariant pure-scaling composite operators ${\cal P}_\lambda (Q)$.  To assign the Young label $\lambda$ (with $|\lambda| = q$) to each of the eigenvectors, we use the fact that the corresponding eigenvalues are identical to the eigenvalues $z_\lambda$ of the Laplacian on the $\sigma$-model manifold~\cite{karcher2021generalized, Friedan-Nonlinear-1980, Friedan-Nonlinear-1985}.

Inspecting the Table~\ref{M2-coeff}, we observe that the symmetry classes split into four groups according to pairs of values $(a,b)$. (We recall the coefficient $c$ does not affect the eigenvectors.) We discuss now the results for eigenvectors in each of these groups.

For classes A and AIII we have $a=\frac12$ and $b=0$.  The results for the eigenvectors
${\cal P}_\lambda^{\rm A} = {\cal P}_\lambda^{\rm AIII}$ for $q = 2$, 3, and 4 are as follows (the analysis for class A was carried out in Ref.~\cite{karcher2021generalized}):

\onecolumngrid

\begin{align}
\begin{pmatrix}
\mathcal{P}_{(1,1)}^A[Q]\\
\mathcal{P}_{(2)}^A[Q]
\end{pmatrix}
=
\begin{pmatrix*}[r]
1 & -1 \\
1 & 1
\end{pmatrix*}
\begin{pmatrix}
\mathrm{tr}(\Lambda Q)
\mathrm{tr}(\Lambda Q)\\
\mathrm{tr}(\Lambda Q\Lambda Q)
\end{pmatrix},
&&
\begin{pmatrix}
\mathcal{P}_{(1,1,1)}^A[Q]\\
\mathcal{P}_{(2,1)}^A[Q]\\
\mathcal{P}_{(3)}^A[Q]
\end{pmatrix}
=
\begin{pmatrix*}[r]
1 & -3 & 2 \\
1 & 0 & -1 \\
1 & 3 & 2
\end{pmatrix*}
\begin{pmatrix}
\mathrm{tr}(\Lambda Q)
\mathrm{tr}(\Lambda Q)
\mathrm{tr}(\Lambda Q)\\
\mathrm{tr}(\Lambda Q\Lambda Q)
\mathrm{tr}(\Lambda Q)\\
\mathrm{tr}(\Lambda Q\Lambda Q\Lambda Q)
\end{pmatrix},
\nonumber\\
\begin{pmatrix}
\mathcal{P}_{(1,1,1,1)}^A[Q]\\
\mathcal{P}_{(2,1,1)}^A[Q]\\
\mathcal{P}_{(2,2)}^A[Q]\\
\mathcal{P}_{(3,1)}^A[Q]\\
\mathcal{P}_{(4)}^A[Q]
\end{pmatrix}
=
\begin{pmatrix*}[r]
1 & -6 & 3 & 8 & -6 \\
1 & -2 & -1 & 0 & 2 \\
1 & 0 & 3 & -4 & 0 \\
1 & 2 & -1 & 0 & -2 \\
1 & 6 & 3 & 8 & 6
\end{pmatrix*}
\begin{pmatrix}
\mathrm{tr}(\Lambda Q)
\mathrm{tr}(\Lambda Q)
\mathrm{tr}(\Lambda Q)
\mathrm{tr}(\Lambda Q)\\
\mathrm{tr}(\Lambda Q\Lambda Q)
\mathrm{tr}(\Lambda Q)
\mathrm{tr}(\Lambda Q)\\
\mathrm{tr}(\Lambda Q\Lambda Q)\mathrm{tr}(\Lambda Q\Lambda Q)\\
\mathrm{tr}(\Lambda Q\Lambda Q\Lambda Q)
\mathrm{tr}(\Lambda Q)\\
\mathrm{tr}(\Lambda Q\Lambda Q\Lambda Q\Lambda Q)
\end{pmatrix}.\hspace{-6cm}
\label{eq:rg_A_result}
\end{align}

\twocolumngrid
\noindent
Our main interest is in the first row of each of the matrices in Eq.~\eqref{eq:rg_A_result} (and of analogous matrices for other symmetry classes below)  which yields the most antysimmetrized observable $(1^q)$. As we know (see Sec.~\ref{sec:pure_scaling} and Appendix~\ref{app:Iwasawa}), observables from these representations serve as building blocks for the construction of generic pure-scaling observables. It can be shown~\cite{karcher2021generalized} that the entries in the first row of the matrices are given by $(-1)^{q-l(\lambda)}d_\lambda$, where $d_\lambda$ is the number of elements in the permutation group $S_q$ that have the cycle shape $\lambda$, and $l(\lambda)$ is the number of cycles in $\lambda$.

We turn now to the second group that includes classes D, AI, and BDI, with $a=b=1$.  The results for eigenvectors ${\cal P}_\lambda^{\rm D} = {\cal P}_\lambda^{\rm AI} = {\cal P}_\lambda^{\rm BDI}$ read (the case of class AI was considered in Ref.~\cite{burmistrov2016mesoscopic})
\onecolumngrid
\begin{align}
\begin{pmatrix}
\mathcal{P}_{(1,1)}^{\rm D}[Q]\\
\mathcal{P}_{(2)}^{\rm D}[Q]
\end{pmatrix}
=
\begin{pmatrix*}[r]
1 & -1 \\
1 & 2
\end{pmatrix*}
\begin{pmatrix}
\mathrm{tr}(\Lambda Q)
\mathrm{tr}(\Lambda Q)\\
\mathrm{tr}(\Lambda Q\Lambda Q)
\end{pmatrix},
&&
\begin{pmatrix}
\mathcal{P}_{(1,1,1)}^{\rm D}[Q]\\
\mathcal{P}_{(2,1)}^{\rm D}[Q]\\
\mathcal{P}_{(3)}^{\rm D}[Q]
\end{pmatrix}
=
\begin{pmatrix*}[r]
1 & -3 & 2 \\
1 & 1 & -2 \\
1 & 6 & 8
\end{pmatrix*}
\begin{pmatrix}
\mathrm{tr}(\Lambda Q)
\mathrm{tr}(\Lambda Q)
\mathrm{tr}(\Lambda Q)\\
\mathrm{tr}(\Lambda Q\Lambda Q)
\mathrm{tr}(\Lambda Q)\\
\mathrm{tr}(\Lambda Q\Lambda Q\Lambda Q)
\end{pmatrix},
\nonumber\\
\begin{pmatrix}
\mathcal{P}_{(1,1,1,1)}^{\rm D}[Q]\\
\mathcal{P}_{(2,1,1)}^{\rm D}[Q]\\
\mathcal{P}_{(2,2)}^{\rm D}[Q]\\
\mathcal{P}_{(3,1)}^{\rm D}[Q]\\
\mathcal{P}_{(4)}^{\rm D}[Q]
\end{pmatrix}
=
\begin{pmatrix*}[r]
1 & -6 & 3 & 8 & -6 \\
1 & -1 & -2 & -2 & 4 \\
1 & 2 & 7 & -8 & -2 \\
1 & 5 & -2 & 4 & -8 \\
1 & 12 & 12 & 32 & 48
\end{pmatrix*}
\begin{pmatrix}
\mathrm{tr}(\Lambda Q)
\mathrm{tr}(\Lambda Q)
\mathrm{tr}(\Lambda Q)
\mathrm{tr}(\Lambda Q)\\
\mathrm{tr}(\Lambda Q\Lambda Q)
\mathrm{tr}(\Lambda Q)
\mathrm{tr}(\Lambda Q)\\
\mathrm{tr}(\Lambda Q\Lambda Q)\mathrm{tr}(\Lambda Q\Lambda Q)\\
\mathrm{tr}(\Lambda Q\Lambda Q\Lambda Q)
\mathrm{tr}(\Lambda Q)\\
\mathrm{tr}(\Lambda Q\Lambda Q\Lambda Q\Lambda Q)
\end{pmatrix}.
\hspace{-6cm}
\label{eq:rg_D_result}
\end{align}

\twocolumngrid
\noindent
Comparing Eq.~\eqref{eq:rg_D_result} with~\eqref{eq:rg_A_result}, we see that the first lines determining the $(1^q)$ observables are identically  the same. This demonstrates that the five classes A, AIII, D, AI, and BDI, all belong to the same ``spinless'' category. This is in full agreement with the physical arguments presented in Sec.~\ref{sec:pure_scaling} and with the calculations using the Iwasawa decomposition performed for class A in Ref.~\cite{gruzberg2013classification} and for class D in Appendix~\ref{app:Iwasawa}. The pure-scaling wave-function observables for this group of classes are given by Eqs.~\eqref{eq:abelianA} and~\eqref{eq:det}.

We turn now to the remaining five classes (which are ``spinful'' as explained in Sec.~\ref{sec:pure_scaling}), beginning with classes AII, C, and CII (for which $a=1$, $b= -1$).
The eigenvectors ${\cal P}_\lambda^{\rm AII} = {\cal P}_\lambda^{\rm C} = {\cal P}_\lambda^{\rm CII}$ of the RG transformation (and thus, the pure-scaling operators) are now found to be (the case of class C was explored in Refs.~\cite{karcher2021generalized, karcher2022generalized})
\onecolumngrid
\begin{align}
\begin{pmatrix}
\mathcal{P}_{(1,1)}^{\rm AII}[Q]\\
\mathcal{P}_{(2)}^{\rm AII}[Q]
\end{pmatrix}
=
\begin{pmatrix*}[r]
1 & -2 \\
1 & 1
\end{pmatrix*}
\begin{pmatrix}
\mathrm{tr}(\Lambda Q)
\mathrm{tr}(\Lambda Q)\\
\mathrm{tr}(\Lambda Q\Lambda Q)
\end{pmatrix},
&&
\begin{pmatrix}
\mathcal{P}_{(1,1,1)}^{\rm AII}[Q]\\
\mathcal{P}_{(2,1)}^{\rm AII}[Q]\\
\mathcal{P}_{(3)}^{\rm AII}[Q]
\end{pmatrix}
=
\begin{pmatrix*}[r]
1 & -6 & 8 \\
1 & -1 & -2 \\
1 & 3 & 2
\end{pmatrix*}
\begin{pmatrix}
\mathrm{tr}(\Lambda Q)
\mathrm{tr}(\Lambda Q)
\mathrm{tr}(\Lambda Q)\\
\mathrm{tr}(\Lambda Q\Lambda Q)
\mathrm{tr}(\Lambda Q)\\
\mathrm{tr}(\Lambda Q\Lambda Q\Lambda Q)
\end{pmatrix},
\nonumber\\
\begin{pmatrix}
\mathcal{P}_{(1,1,1,1)}^{\rm AII}[Q]\\
\mathcal{P}_{(2,1,1)}^{\rm AII}[Q]\\
\mathcal{P}_{(2,2)}^{\rm AII}[Q]\\
\mathcal{P}_{(3,1)}^{\rm AII}[Q]\\
\mathcal{P}_{(4)}^{\rm AII}[Q]
\end{pmatrix}
=
\begin{pmatrix*}[r]
1 & -12 & 12 & 32 & -48 \\
1 & -5 & -2 & 4 & 8 \\
1 & -2 & 7 & -8 & 2 \\
1 & 1 & -2 & -2 & -4 \\
1 & 6 & 3 & 8 & 6
\end{pmatrix*}
\begin{pmatrix}
\mathrm{tr}(\Lambda Q)
\mathrm{tr}(\Lambda Q)
\mathrm{tr}(\Lambda Q)
\mathrm{tr}(\Lambda Q)\\
\mathrm{tr}(\Lambda Q\Lambda Q)
\mathrm{tr}(\Lambda Q)
\mathrm{tr}(\Lambda Q)\\
\mathrm{tr}(\Lambda Q\Lambda Q)\mathrm{tr}(\Lambda Q\Lambda Q)\\
\mathrm{tr}(\Lambda Q\Lambda Q\Lambda Q)
\mathrm{tr}(\Lambda Q)\\
\mathrm{tr}(\Lambda Q\Lambda Q\Lambda Q\Lambda Q)
\end{pmatrix}.
\hspace{-6cm}
\label{eq:rg_AII_result}
\end{align}

\twocolumngrid
\noindent
It is well known that there is a duality within the pairs of classes AI $\leftrightarrow$ AII,  C$\leftrightarrow$D, and BDI $\leftrightarrow$ CII. This duality becomes manifest if one compares  Eqs.~\eqref{eq:rg_D_result} and~\eqref{eq:rg_AII_result}.
Specifically, one sees that the first (second, etc.) row of the matrix $(\mathcal{P}_{\lambda}^{\rm C}[Q])_\mu$  [Eq.~\eqref{eq:rg_AII_result}]  coincides with the last (respectively, second last, etc.) row of the matrix $(\mathcal{P}_{\lambda}^{\rm D}[Q])_\mu$ [Eq.~\eqref{eq:rg_D_result}] multiplied by $(-1)^{l(\mu)}$, where $\mu$ is the column label. This means that
\begin{align}
(\mathcal{P}_{\lambda}^{\rm D}[Q])_\mu &= (-1)^{l(\mu)} (\mathcal{P}_{\lambda^T}^{\rm C}[Q])_\mu,
\label{eq:duality}
\end{align}
where $\lambda^T$ is the conjugate Young diagram (obtained from $\lambda = (q_1, \ldots, q_n)$ by interchanging rows with columns, i.e., by reflection with respect to the diagonal).

%
%

Finally, for classes DIII and CI (with $a=2$ and $b=0$), we obtain the following results for the eigenoperators ${\cal P}_\lambda^{\rm DIII} = {\cal P}_\lambda^{\rm CI} $:

\onecolumngrid

\begin{align}
\begin{pmatrix}
\mathcal{P}_{(1,1)}^{\rm DIII}[Q]\\
\mathcal{P}_{(2)}^{\rm DIII}[Q]
\end{pmatrix}
=
\begin{pmatrix*}[r]
1 & -1 \\
1 & 1
\end{pmatrix*}
\begin{pmatrix}
\mathrm{tr}(\Lambda Q)
\mathrm{tr}(\Lambda Q)\\
2\mathrm{tr}(\Lambda Q\Lambda Q)
\end{pmatrix},
&&
\begin{pmatrix}
\mathcal{P}_{(1,1,1)}^{\rm DIII}[Q]\\
\mathcal{P}_{(2,1)}^{\rm DIII}[Q]\\
\mathcal{P}_{(3)}^{\rm DIII}[Q]
\end{pmatrix}
=
\begin{pmatrix*}[r]
1 & -3 & 2 \\
1 & 0 & -1 \\
1 & 3 & 2
\end{pmatrix*}
\begin{pmatrix}
\mathrm{tr}(\Lambda Q)
\mathrm{tr}(\Lambda Q)
\mathrm{tr}(\Lambda Q)\\
2\mathrm{tr}(\Lambda Q\Lambda Q)
\mathrm{tr}(\Lambda Q)\\
4\mathrm{tr}(\Lambda Q\Lambda Q\Lambda Q)
\end{pmatrix},
\nonumber\\
\begin{pmatrix}
\mathcal{P}_{(1,1,1,1)}^{\rm DIII}[Q]\\
\mathcal{P}_{(2,1,1)}^{\rm DIII}[Q]\\
\mathcal{P}_{(2,2)}^{\rm DIII}[Q]\\
\mathcal{P}_{(3,1)}^{\rm DIII}[Q]\\
\mathcal{P}_{(4)}^{\rm DIII}[Q]
\end{pmatrix}
=
\begin{pmatrix*}[r]
1 & -6 & 3 & 8 & -6 \\
1 & -2 & -1 & 0 & 2 \\
1 & 0 & 3 & -4 & 0 \\
1 & 2 & -1 & 0 & -2 \\
1 & 6 & 3 & 8 & 6
\end{pmatrix*}
\begin{pmatrix}
\mathrm{tr}(\Lambda Q)
\mathrm{tr}(\Lambda Q)
\mathrm{tr}(\Lambda Q)
\mathrm{tr}(\Lambda Q)\\
2\mathrm{tr}(\Lambda Q\Lambda Q)
\mathrm{tr}(\Lambda Q)
\mathrm{tr}(\Lambda Q)\\
4\mathrm{tr}(\Lambda Q\Lambda Q)\mathrm{tr}(\Lambda Q\Lambda Q)\\
4\mathrm{tr}(\Lambda Q\Lambda Q\Lambda Q)
\mathrm{tr}(\Lambda Q)\\
8\mathrm{tr}(\Lambda Q\Lambda Q\Lambda Q\Lambda Q)
\end{pmatrix}.\hspace{-6cm}
\label{eq:rg_DIII_result}
\end{align}

\twocolumngrid
\noindent
Note that we have chosen to include the factor $2^{q-l(\lambda)}$ in the vector of basis operators in the right-hand side of Eq.~\eqref{eq:rg_DIII_result}. This makes the matrices in
\eqref{eq:rg_DIII_result} identical to those in~\eqref{eq:rg_A_result}. Thus,
\begin{equation}
(\mathcal{P}^{\rm DIII}_\lambda[Q])_\mu = 2^{q-l(\mu)}(\mathcal{P}^{\rm A}_\lambda[Q])_\mu.
\end{equation}
For $\lambda=(1^q)$ (first lines of the matrices), this means that $\mathcal{P}^{\rm DIII}_{(1^q)} = \mathcal{P}^{\rm AII}_{(1^q)}$. Thus, for all five classes AII, C, CII, DIII, and CI, the $(1^q)$ eigenoperators have the same form, conforming that they belong to the same  ``spinful'' category. This is in perfect agreement with
physical arguments in Sec.~\ref{sec:pure_scaling} and with the calculations using the Iwasawa decomposition performed for class C in Ref.~\cite{karcher2021generalized}  and for classes AII and  DIII in Appendix~\ref{app:Iwasawa}.  The pure-scaling wave-function observables for this group of classes are given by Eqs.~\eqref{eq:abelianA} and~\eqref{eq:det-sp}.

\pagebreak

\bibliography{gener-MF}

\end{document}